\documentclass[ALICE,manyauthors]{cernphprep}
\usepackage[english]{babel}
\usepackage{graphicx}
\usepackage{verbatim}
\usepackage{amsfonts}
\usepackage{makeidx}
\usepackage{color}
\usepackage{amsmath}
\usepackage{lineno}
\usepackage{epsfig}
\usepackage{epstopdf}
\usepackage{hyperref}
\usepackage{cite}
\usepackage{multirow}
\usepackage{subfigure}

\usepackage[comma,square,numbers,sort&compress]{natbib}

\newcommand{\sqrts}{\sqrt{s}}

\newcommand{\GeV}{\mathrm{GeV}}
\newcommand{\TeV}{\mathrm{TeV}}

\newcommand{\mev}{\mathrm{MeV}}
\newcommand{\gev}{\mathrm{GeV}}
\newcommand{\gevc}{\mathrm{GeV}/c}
\newcommand{\mevc}{\mathrm{MeV}/c}
\newcommand{\tev}{\mathrm{TeV}}

\newcommand{\mum}{\mathrm{\mu m}}

\newcommand{\mub}{\mathrm{\mu b}}

\newcommand{\pt}{p_{\rm T}}

\newcommand{\DtoKpi}{{\rm D}^0 \to {\rm K}^-\pi^+}
\newcommand{\DtoKpipi}{{\rm D}^+\to {\rm K}^-\pi^+\pi^+}
\newcommand{\DstartoDpi}{{\rm D}^{*+} \to {\rm D}^0 \pi^+}

\newcommand{\DstartoKpipi}{{\rm D}^{*+} \to {\rm D}^0 \pi^+ \to {\rm K}^- \pi^+ \pi^+}
\newcommand{\DstoPhiPi}{{\rm D^{+}_{s}\to \phi\pi^+\to K^{+} K^{-} \pi^{+}}}
\newcommand{\Dzero}{{\rm D^0}}
\newcommand{\Ds}{{\rm D^+_s}}
\newcommand{\Dzerobar}{\overline{{\rm D^0}}}
\newcommand{\Dstar}{{\rm D^{*+}}}

\newcommand{\Dplus}{{\rm D^+}}

\newcommand{\Raa}{R_{\rm AA}}

\newcommand{\kt}{k_{\rm T}}
\newcommand{\fprompt}{f_{\rm prompt}}
\newcommand{\meanpt}{{\langle p_{ {\mathrm T} } \rangle}}

\begin{document}

\begin{titlepage}

\PHyear{2019}
\PHnumber{004}                 
\PHdate{19 January}              

\title{Measurement of $\Dzero$, $\Dplus$, $\Dstar$ and $\Ds$ production in pp collisions at 
$\mathbf{\sqrt{{\textit s}}~=~5.02~TeV}$ with ALICE}
\Collaboration{ALICE Collaboration}
                     
\ShortTitle{D-meson production in pp collisions at $\sqrts=5.02~\tev$}

\begin{abstract} 
The measurements of the production of prompt $\Dzero$, $\Dplus$, $\Dstar$, and $\Ds$ mesons in proton--proton (pp) collisions at $\sqrts=5.02~\tev$ with the ALICE detector at the Large Hadron Collider (LHC) are reported. D mesons were reconstructed at mid-rapidity ($|y|<0.5$) via their hadronic decay channels $\DtoKpi$, $\DtoKpipi$, $\DstartoKpipi$, $\DstoPhiPi$, and their charge conjugates. The production cross sections were measured in the transverse momentum interval $0<\pt<36~\gevc$ for $\Dzero$, $1<\pt<36~\gevc$ for $\Dplus$ and $\Dstar$, and in $2<\pt<24~\gevc$ for $\Ds$ mesons. Thanks to the higher integrated luminosity, an analysis in finer $\pt$ bins with respect to the previous measurements at $\sqrt{s}=7~\tev$ was performed, allowing for a more detailed description of the cross-section $\pt$ shape. The measured $\pt$-differential production cross sections are compared to the results at $\sqrts=7$~TeV and to four different perturbative QCD calculations. Its rapidity dependence is also tested combining the ALICE and LHCb measurements in pp collisions at $\sqrts=5.02~\tev$. This measurement will allow for a more accurate determination of the nuclear modification factor in p--Pb and Pb--Pb collisions performed at the same nucleon-nucleon centre-of-mass energy.
\end{abstract}

\end{titlepage}
\setcounter{page}{2}

\newpage 

\section{Introduction}
\label{sec:intro}
The study of the production of hadrons containing heavy quarks, i.e. charm and beauty, in proton--proton (pp) collisions at LHC energies is a sensitive test of Quantum Chromodynamics (QCD) calculations with the factorisation approach.
In this scheme, the transverse momentum ($\pt$) differential production cross sections of hadrons containing charm or beauty quarks are calculated as a convolution of three terms: (i) the parton distribution functions (PDFs) of the incoming protons, (ii) the partonic scattering cross section, calculated as a perturbative series in powers of the strong coupling constant $\alpha_{\rm s}$, and (iii) the fragmentation function, which parametrises the non-perturbative evolution of a heavy quark into a given species of heavy-flavour hadron.
Factorisation is implemented in terms of the squared momentum transfer $Q^2$ (collinear factorisation)~\cite{Collins:1989gx} or of the partonic transverse momentum $\kt$~\cite{Catani:1990eg}.  At LHC energies, calculations based on collinear factorisation are available in the general-mass variable-flavour-number scheme, GM-VFNS~\cite{Kniehl:2004fy,Kniehl:2005mk,Kniehl:2012ti,Helenius:2018uul}, and in the fixed order plus next-to-leading logarithms approach, FONLL~\cite{Cacciari:1998it,Cacciari:2012ny}, both of them having next-to-leading order (NLO) accuracy with all-order resummation of next-to-leading logarithms.
Within the $\kt$-factorisation framework, heavy-flavour production cross-section calculations exist only at leading order (LO) approximation in $\alpha_{\rm s}$~\cite{Luszczak:2008je,Maciula:2013wg,Catani:1990eg}. 
All these calculations describe within uncertainties the production cross sections of D and B mesons measured in pp and ${\rm p\overline{p}}$ collisions in different kinematic regions at centre-of-mass energies from 0.2 to 13~TeV (see e.g. Ref.~\cite{Andronic:2015wma} and references therein). In the case of charm production, the uncertainties on the theoretical predictions, which are dominated by the choice of the scales of the perturbative calculation (e.g. the factorisation and renormalisation scales), are significantly larger than the uncertainties on the measured data points~\cite{Adamczyk:2012af,Acosta:2003ax,Aad:2015zix,ALICE:2011aa,Abelev:2012vra,Abelev:2012xe, Abelev:2012tca,Adam:2016ich,Sirunyan:2017xss,Aaij:2013mga,Aaij:2016jht,Aaij:2015bpa}. However, as pointed out in Ref.~\cite{Cacciari:2015fta}, in the ratios of cross sections at different LHC energies and in different rapidity intervals the uncertainty due to choice of the factorisation and renormalisation scales becomes subdominant with respect to the uncertainty on the PDFs, thus making the measurement sensitive to the gluon PDF at small Bjorken-$x$ values. A precise measurement of the D-meson production cross sections down to $\pt=0$ can therefore provide important constraints to perturbative QCD (pQCD) calculations and to low-$x$ gluon PDFs. 
 Furthermore, D-meson measurements in pp collisions represent an essential reference for the study of effects induced by cold and hot strongly-interacting matter in the case of proton--nucleus and nucleus--nucleus collisions (see e.g. the recent reviews~\cite{Andronic:2015wma,Prino:2016cni,Aarts:2016hap}). 

In this article, the measurements of the $\pt$-differential production cross sections of prompt $\rm D^0$, $\rm D^+$, $\rm D^{*+}$, and $\rm D_s^+$ mesons (as average of particles and anti-particles) in pp collisions at the centre-of-mass energy $\sqrt s=5.02~\tev$ are reported together with their ratios. The measurements are performed at mid-rapidity ($|y|<0.5$) in the transverse momentum
intervals $0<\pt<36~\gev/c$ for $\rm D^0$ mesons, $1<\pt<36~\gev/c$ for $\rm D^+$ and $\rm D^{*+}$ mesons, and $2<\pt<24~\gev/c$ for $\rm D^+_s$ mesons. 
The $\pt$-integrated D-meson production cross sections per unit of rapidity is also reported for each D-meson species. 
The ratios of the $\rm D^0$, $\rm D^+$, and $\rm D^{*+}$-meson production cross sections measured at $\sqrt s=7~\tev$~\cite{Acharya:2017jgo} and $\sqrt s=5.02~\tev$ are presented as well, and compared to FONLL calculations. Finally, the ratios of $\rm D^0$-meson production cross sections at mid- and forward rapidity are also reported, using the measurements done at forward rapidity by the LHCb collaboration in pp collisions at $\sqrts=5.02~\tev$~\cite{Aaij:2016jht}.


\section{Experimental apparatus and data sample}
\label{sec:detector}
The ALICE experimental apparatus is composed of a set of detectors for particle reconstruction and identification at mid-rapidity, embedded in a 
large solenoidal magnet that provides a $B = 0.5~\mathrm{T}$ field parallel to 
the beams. It also includes a forward muon spectrometer and various forward and backward detectors for triggering and event characterisation. A complete description and an overview of their typical performance in pp, p--Pb, and Pb--Pb collisions is presented in Refs.~\cite{Aamodt:2008zz,Abelev:2014ffa}. 

The tracking and particle identification capabilities of the ALICE central barrel detectors were exploited to reconstruct the D-meson decay products at mid-rapidity. 
The Inner Tracking System (ITS), consisting of six cylindrical layers of silicon detectors, is used to track charged particles 
and to reconstruct primary and secondary vertices. The Time Projection Chamber (TPC) provides track reconstruction with up to 159 three-dimensional space points per track, as well as particle identification via the measurement 
of their specific ionisation energy loss d$E$/d$x$. The particle identification capabilities of the TPC are complemented by 
the Time-Of-Flight detector (TOF), which is used to measure the flight time of the charged particles from the interaction point. These detectors cover the pseudorapidity interval $|\eta| < 0.9$.
The V0 detector, composed of two arrays of 32
scintillators each, covering the pseudorapidity ranges $-3.7 < \eta < -1.7$ and $2.8 < \eta < 5.1$, provides the minimum-bias (MB) trigger  used to collect the data sample. 
In addition, the timing information of the two V0 arrays
and the correlation between the number of hits and track segments 
in the two innermost layers of the ITS, consisting of Silicon Pixel Detectors (SPD), was used for an offline event selection, in order to remove background due to the interaction between one of the beams and the residual gas present in the beam vacuum tube.
In order to maintain a uniform acceptance in pseudorapidity, collision vertices were required to be within $\pm10~\rm{cm}$ from the centre of the detector in the beam-line direction.
The pile-up events (less than 1\%) were rejected by detecting multiple primary vertices using track segments defined with the SPD layers. 
After the aforementioned selections, the data sample used for the analysis consists of about 990 million MB events, corresponding to an integrated luminosity $L_{\rm int} = (19.3 \pm 0.4)$~nb$^{-1}$, collected during
the 2017 pp run at $\sqrts = 5.02$~TeV.

\section{Data analysis}
\label{sec:analysis}
\subsection{Analysis with D-meson decay vertex reconstruction}
\label{subsec:vertexing}
The D mesons and their charge conjugates were reconstructed via the decay channels $\DtoKpi$ (with branching ratio, ${\rm BR}=3.89\pm 0.04\%$), $\DtoKpipi$ (${\rm BR}=8.98\pm 0.28\%$), $\DstartoKpipi$ (${\rm BR}=2.63\pm 0.03\%$), and $\DstoPhiPi$ (${\rm BR}=2.27\pm 0.08\%$)~\cite{Tanabashi:2018oca}. The analysis was based on the reconstruction of decay vertices displaced from the interaction vertex, exploiting the separation of a few hundred $\mu$m induced by the weak decays of $\Dzero$, $\Dplus$, and $\Ds$ mesons ($c\tau \simeq 123$, 312, and 150 $\mu$m, respectively~\cite{Tanabashi:2018oca}). The $\Dzero$, $\Dplus$, and $\Ds$ candidates were built combining pairs or triplets of tracks with the proper charge, each with $|\eta| < 0.8$, $\pt > 0.3~\gev/c$, at least 70 associated TPC space points, $\chi^{2}/{\rm ndf} < 2$ in the TPC (where ndf is the number of degrees of freedom involved in the 
track fit procedure), and at least one hit in either of the two layers of the SPD. The $\Dstar$ candidates were defined by the combination of $\Dzero$ candidates with tracks reconstructed with at least two points in the ITS, including at least one in the SPD, and $\pt >$ 80 $\mevc$. 
As a consequence of these track selection criteria, the acceptance for D mesons decreases rapidly for $|y|>0.5$ at low $\pt$ and for $|y|>0.8$ for $\pt>5~\gevc$. Therefore, only D-meson candidates within a fiducial acceptance region, $|y| < y_{\rm fid}(\pt)$, were selected. The $y_{\mathrm{fid}}(\pt)$ factor was defined as a second-order polynomial function, increasing from 0.5 to 0.8 in the transverse momentum range $0 < \pt < 5~\gevc$, and a constant term, $y_{\mathrm{fid}}=0.8$, for $\pt > 5~\gevc$.

In order to reduce the combinatorial background and to increase the signal-over-background ratio ($S/B$), 
geometrical selections on the $\Dzero$, $\Dplus$, and $\Ds$-meson decay topology were applied. In the $\DstartoDpi$
case, the decay vertex cannot be resolved from the primary vertex and
geometrical selections were applied on the secondary vertex topology of the produced $\Dzero$ mesons. The selection requirements, tuned to provide a large statistical significance for the signal and to keep the selection efficiency as high as 
possible, were mainly based on the displacement of the tracks from the primary vertex ($d_0$), the distance between 
the D-meson decay vertex and the primary vertex (decay length, $L$), and the pointing of the reconstructed D-meson momentum to the primary vertex. 
Additional selection criteria, already introduced in Refs.~\cite{Acharya:2017jgo,Acharya:2018hre}, were applied to $\Dplus$ and $\Ds$ candidates. 
These selections reject both combinatorial background and D mesons from beauty-hadron decays (selection efficiency reduced by 50\% at high $\pt$), denoted as ``feed-down'' in the following. For the $\Ds$-candidate selection, one of the two pairs of opposite-sign tracks was required have a reconstructed ${\rm K^+K^-}$ invariant mass within $\pm 10~\mev /c^{2}$ with respect to the PDG world average of the $\phi$ meson~\cite{Tanabashi:2018oca}.

Further reduction of the combinatorial background was obtained by applying particle identification (PID) to the decay tracks, except for the soft-pion track coming from $\DstartoDpi$ decays. 
Pions and kaons were identified requiring compatibility with the respective particle hypothesis within three standard deviations ($3\,\sigma$) between the measured and the expected signals for both the TPC $\mathrm{d}E/\mathrm{d}x$ and the time-of-flight. Tracks without TOF hits were identified using only the TPC information with a $3\,\sigma$ selection, except for the decay products of $\Ds$ candidates with $\pt<6~\gevc$, for which a $2\,\sigma$ selection was needed to suppress the larger fraction of combinatorial background in this mode. 

\begin{figure}[!th]
\begin{center}
\includegraphics[width=0.49\textwidth]{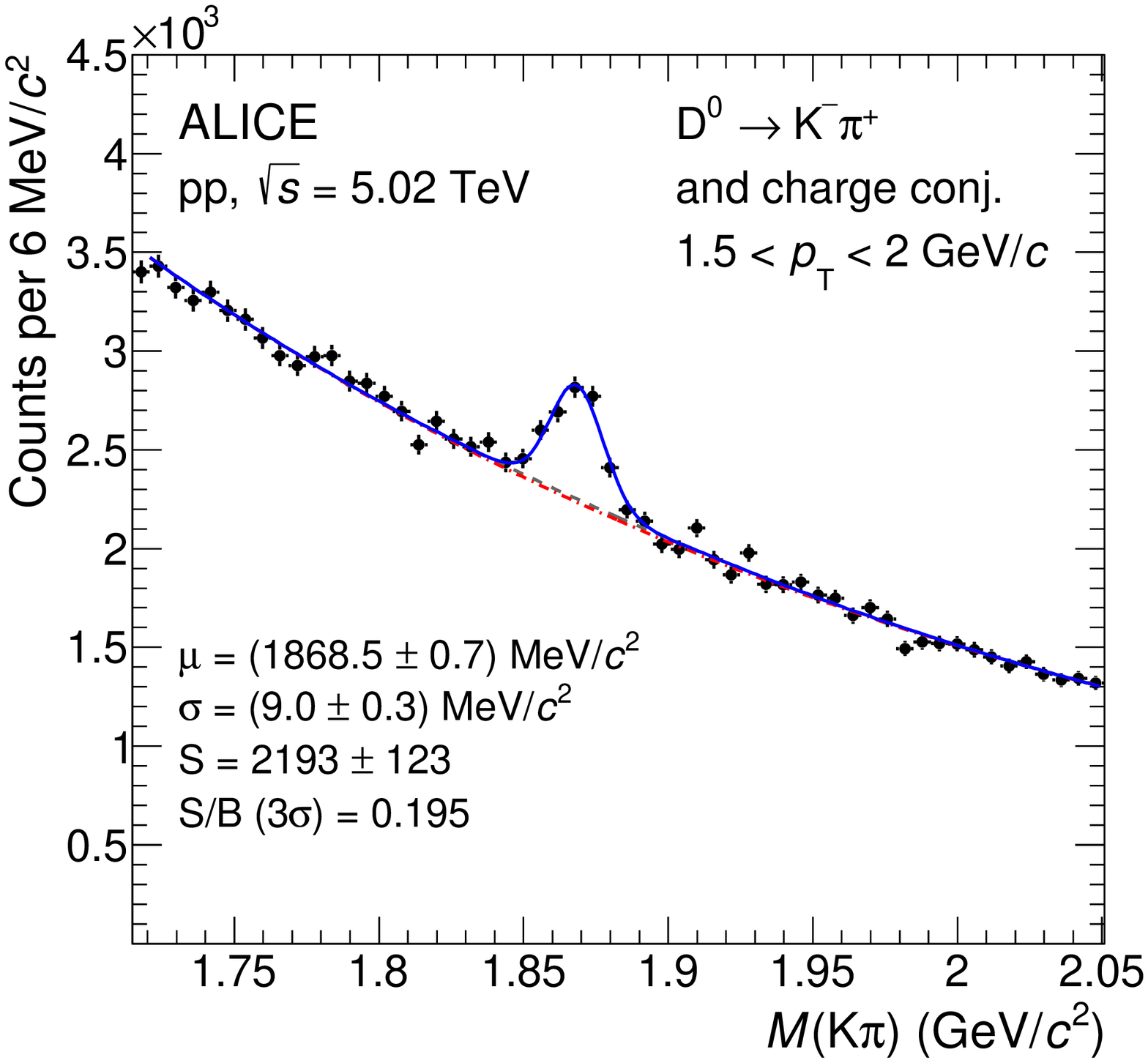}
\includegraphics[width=0.49\textwidth]{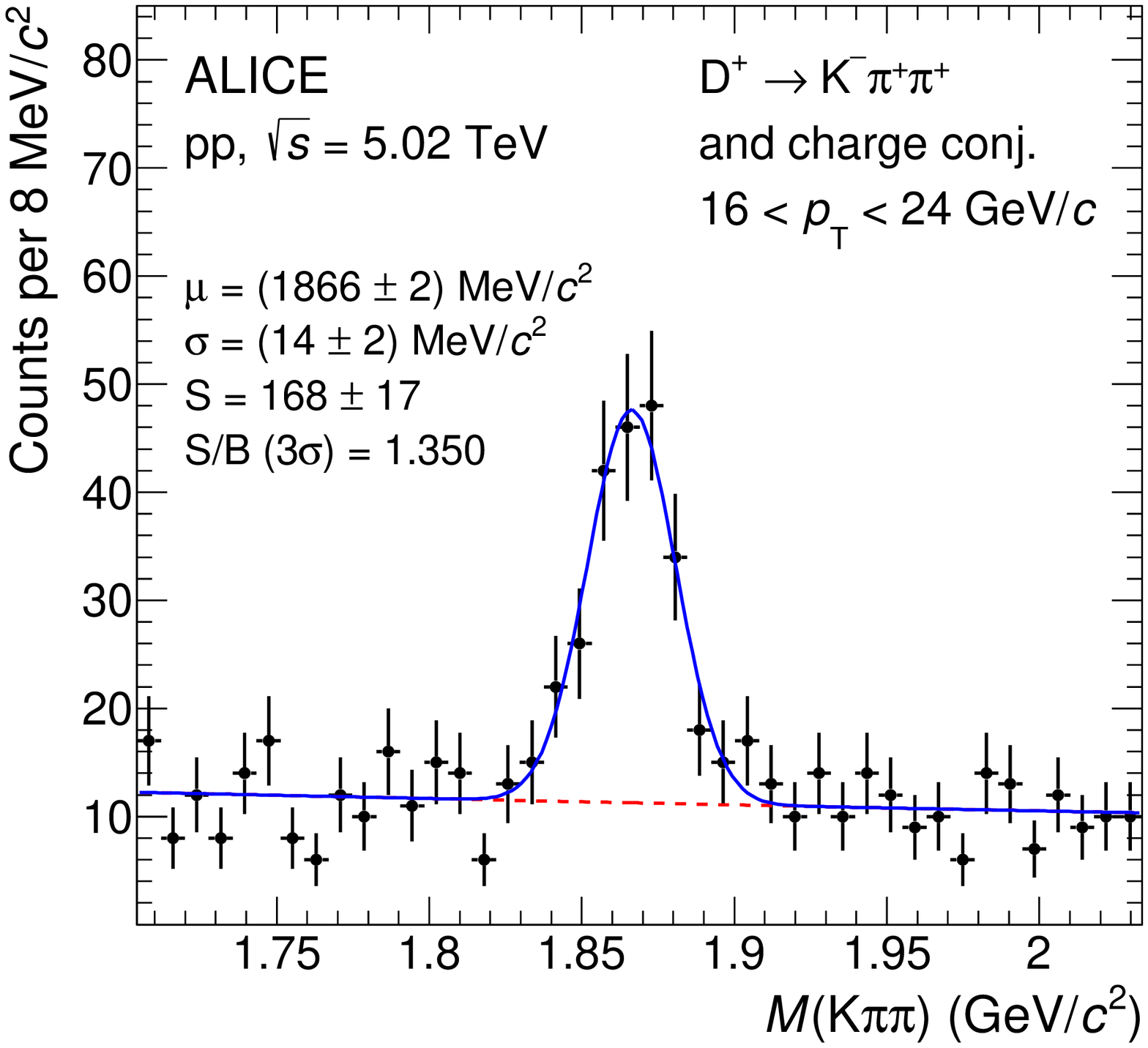}
\includegraphics[width=0.49\textwidth]{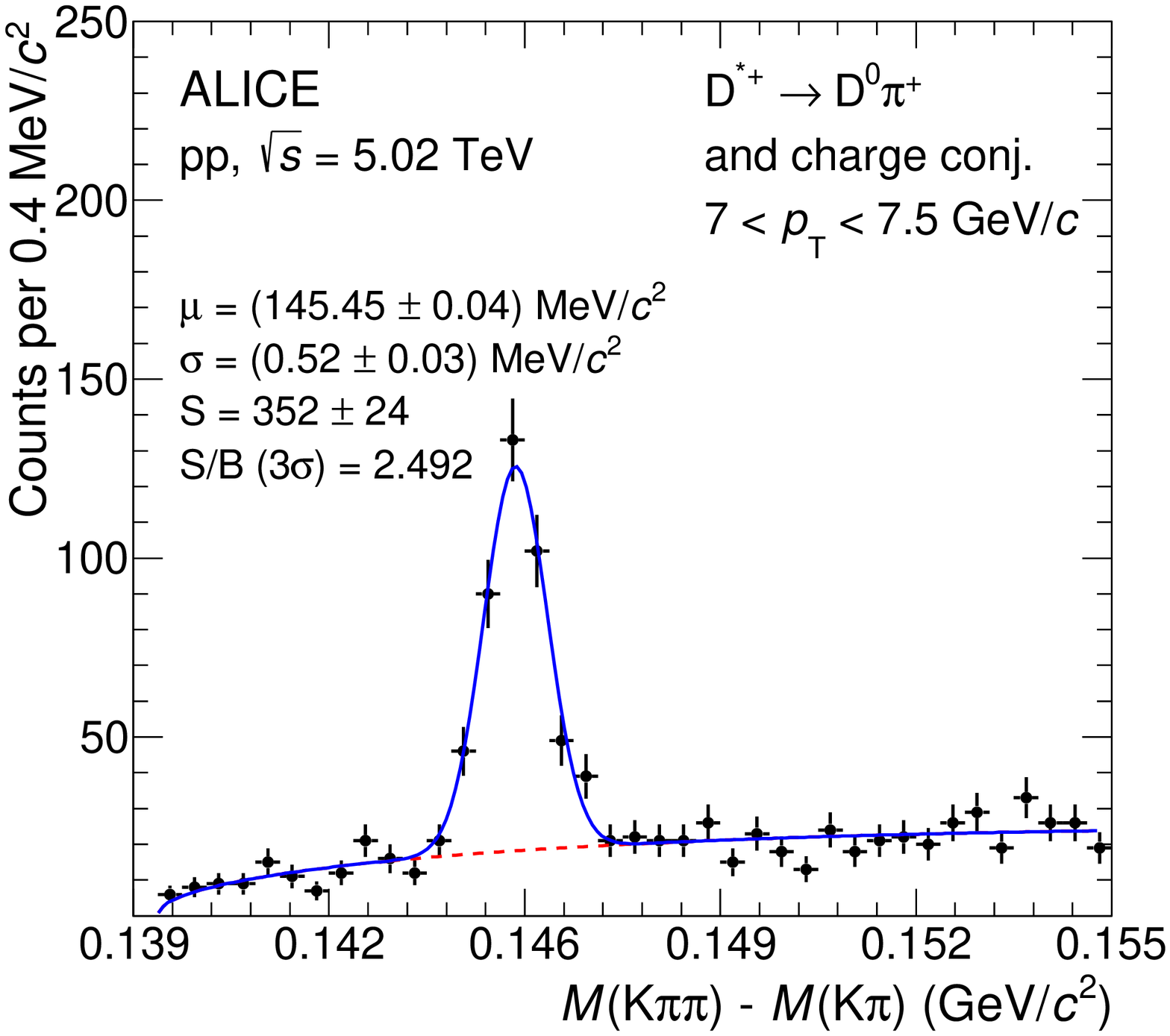}
\includegraphics[width=0.49\textwidth]{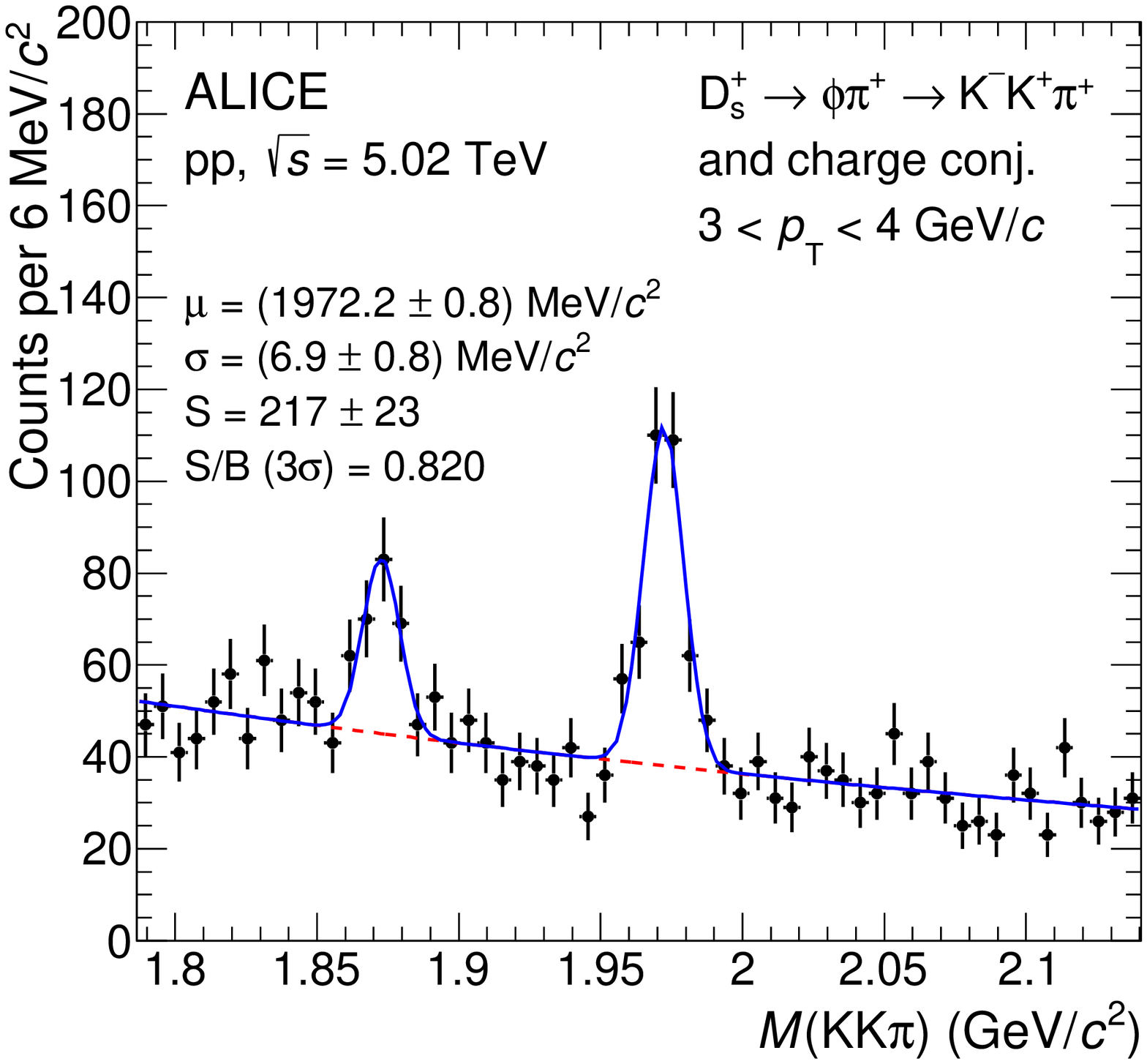}
\caption{Invariant-mass (mass-difference) distributions of $\Dzero$, $\Dplus$, $\Dstar$, and $\Ds$ 
candidates and charge conjugates in $1.5<\pt<2~\gevc$, $16<\pt<24~\gevc$, $7<\pt<7.5~\gevc$, and $3<\pt<4~\gevc$ intervals, respectively. The blue solid lines show the total fit functions as described in the text and the red dashed lines are the combinatorial-background terms. In case of $\Dzero$, the grey dashed line represents the combinatorial background with the contribution of the reflections.  
The values of the mean ($\mu$) and the width ($\sigma$) of the signal peak are reported together with the signal counts ($S$) and the signal over background ratio ($S/B$) in the mass interval ($\mu-3\sigma,\mu+3\sigma$). The reported uncertainties are only the statistical uncertainties from the fit.}
\label{fig:InvMass}
\end{center}
\end{figure}

The D-meson raw yields, including both particles and antiparticles, were obtained from binned maximum likelihood fits to the invariant-mass ($M$) distributions of $\Dzero$, 
$\Dplus$, and $\Ds$ candidates and to the mass difference 
$\Delta M = M (\mathrm{K} \pi \pi) - M(\mathrm{K} \pi)$ distributions of $\Dstar$ candidates, in the transverse-momentum intervals $0.5<\pt<36~\gev/c$ for $\rm D^0$ mesons, $1<\pt<36~\gev/c$ for $\rm D^+$ and $\rm D^{*+}$ mesons, and $2<\pt<24~\gev/c$ for $\rm D^+_s$ mesons. The signal extraction was performed in finer $\pt$ bins with respect the previous measurements at $\sqrt{s}=7~\tev$~\cite{Acharya:2017jgo}, allowing for a more detailed description of the cross-section $\pt$ shape.
The fit function was composed of a Gaussian for the description of the signal and of an exponential term for the background of $\Dzero$, $\Dplus$, and $\Ds$ candidates, and of a threshold function 
for $\Dstar$ candidates~\cite{Acharya:2017jgo}. 
For the $\Dzero$ meson, the contribution of signal candidates present in the invariant-mass distribution with the wrong decay-particle mass assignment (reflections) was included in the fit. It was modelled based on the invariant-mass distributions of the reflected signal in the simulation, which were parametrised as the sum of two Gaussian functions. The contribution of reflections is about $2\%-3\%$ of the raw signal depending on $\pt$.
For the $M({\rm KK\pi})$ distribution, an additional Gaussian was used to describe the signal of the decay ${\rm D^+\rightarrow K^+K^-\pi^+}$, with a branching ratio of ($9.51\pm 0.34)\cdot 10^{-3}$~\cite{Tanabashi:2018oca}, present on the left side of the $\Ds$-meson signal. Figure~\ref{fig:InvMass} shows the invariant mass (mass-difference) distributions together with the result of the fits, in $1.5<\pt<2~\gevc$, $16<\pt<24~\gevc$, $7<\pt<7.5~\gevc$, and $3<\pt<4~\gevc$ intervals for $\Dzero$, $\Dplus$, $\Dstar$, and $\Ds$ candidates, respectively. The statistical significance of the observed signals, $S/\sqrt{(S+B)}$, varies from 4 to 28, depending on the meson species and on the $\pt$ interval. The $S/B$ values obtained applying the selections described above are 0.01--1.85 for $\Dzero$, 0.5--2.2 for $\Dplus$, 0.3--4.2 for $\Dstar$, and 0.3--2.2 for $\Ds$ mesons, depending on $\pt$.

The $\pt$-differential cross section of prompt D mesons in each $\pt$ interval was computed as:
\begin{equation}
  \label{eq:dNdpt}
  \frac{{\rm d}^2 \sigma^{\rm D}}{{\rm d}\pt {\rm d} y}=
  \frac{1}{c_{\Delta y}(\pt)\Delta \pt}\cdot\frac{1}{{\rm BR}}  \cdot\frac{\frac{1}{2} \, \left.f_{\rm prompt}(\pt)\cdot N^{\rm D+\overline D,raw}(\pt)\right|_{|y|<y_{\rm fid}(\pt)}}{ ({\rm Acc}\times\epsilon)_{\rm prompt}(\pt)}    \frac{1}{L_{\rm int}} \,.
\end{equation}
The raw yield values (sum of particles and antiparticles, 
$N^{\rm D+\overline D,raw}$) were divided by a factor of two and multiplied by the prompt fraction $f_{\rm{prompt}}$ to obtain the charged-averaged yields of prompt D mesons. Furthermore, they were divided by the acceptance-times-efficiency of prompt D mesons
$(\rm Acc \times \epsilon)_{\rm{prompt}}$, the BR of the decay channel, 
the width of the $\pt$ interval ($\Delta \pt$), the correction factor for the rapidity coverage $c_{\Delta y}$, and the integrated luminosity $L_{\rm int}=N_{\rm ev} / \sigma_{\rm MB} $, where $N_{\rm ev}$ is the number of analysed events and $\sigma_{\rm MB}=(50.9\pm 0.9)$~mb is the cross section for the MB trigger condition~\cite{ALICE-PUBLIC-2018-014}.

The $(\rm Acc \times \epsilon)$ correction was obtained simulating pp collisions with the PYTHIA 6.4.25 event generator~\cite{Sjostrand:2006za} (Perugia-11 tune~\cite{Skands:2010ak}), and propagating the generated particles through the detector
using GEANT3~\cite{Brun:1994aa}. Each simulated PYTHIA pp event contained a $\rm c\overline c$ or $\rm b\overline b$ pair, and D mesons were forced to decay into the hadronic channels of interest for the analysis. 
The luminous region distribution and the conditions of all the ALICE
detectors in terms of active channels, gain, noise level and alignment, and their evolution with time
during the data taking, were taken into account in the 
simulations.

\begin{figure}[!t]
\begin{center}
\includegraphics[width=0.49\textwidth]{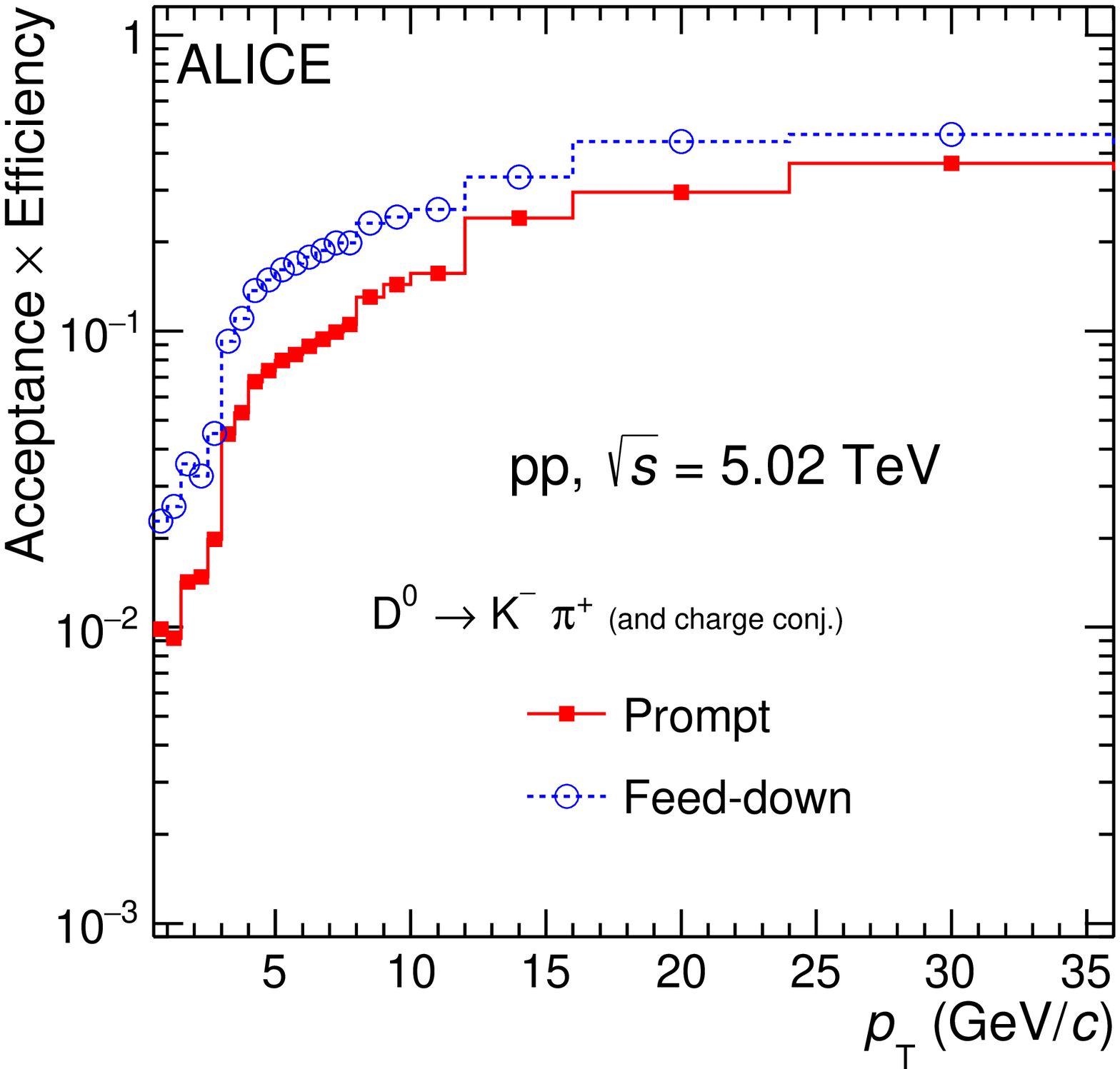}
\includegraphics[width=0.49\textwidth]{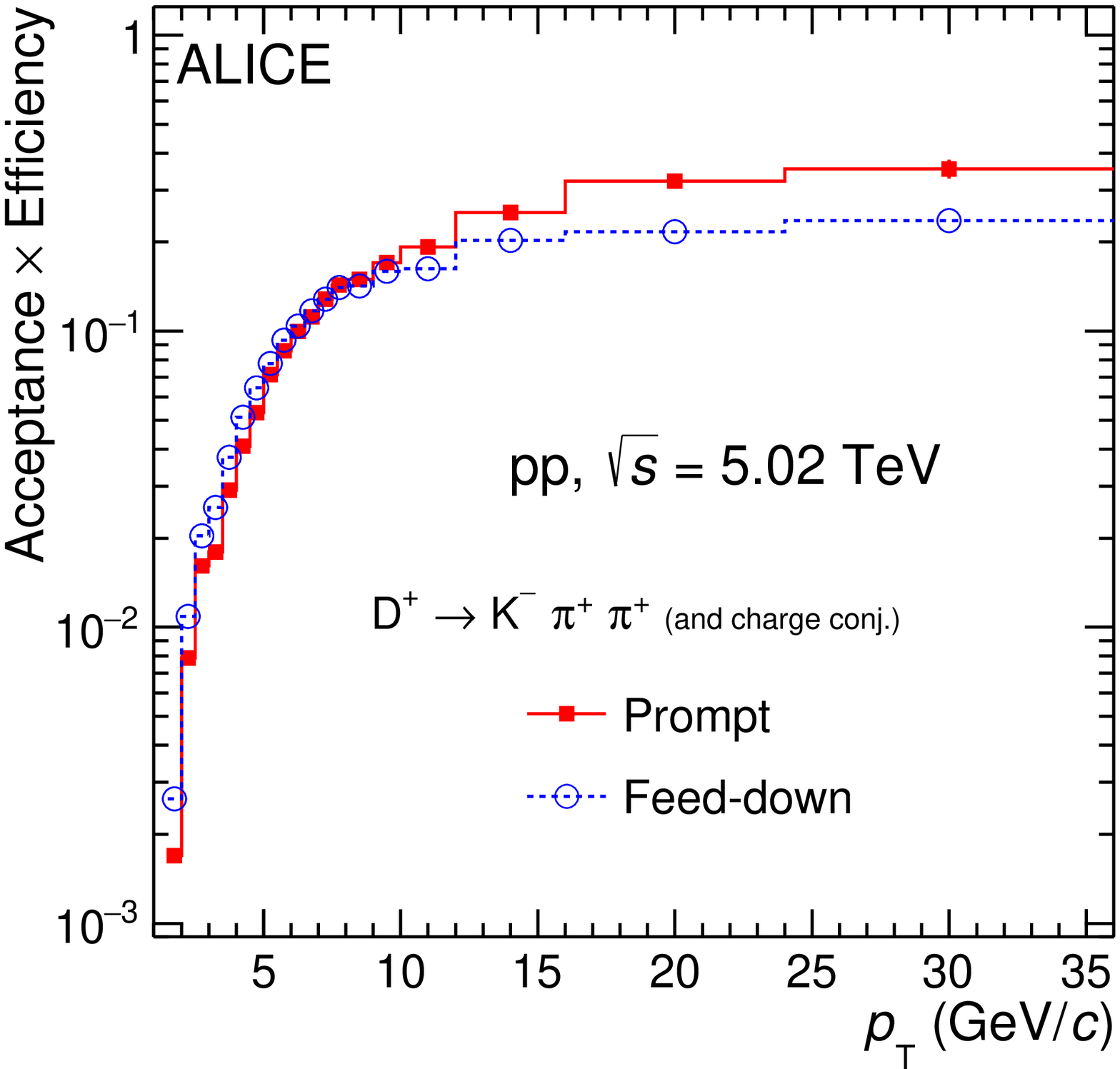}
\includegraphics[width=0.49\textwidth]{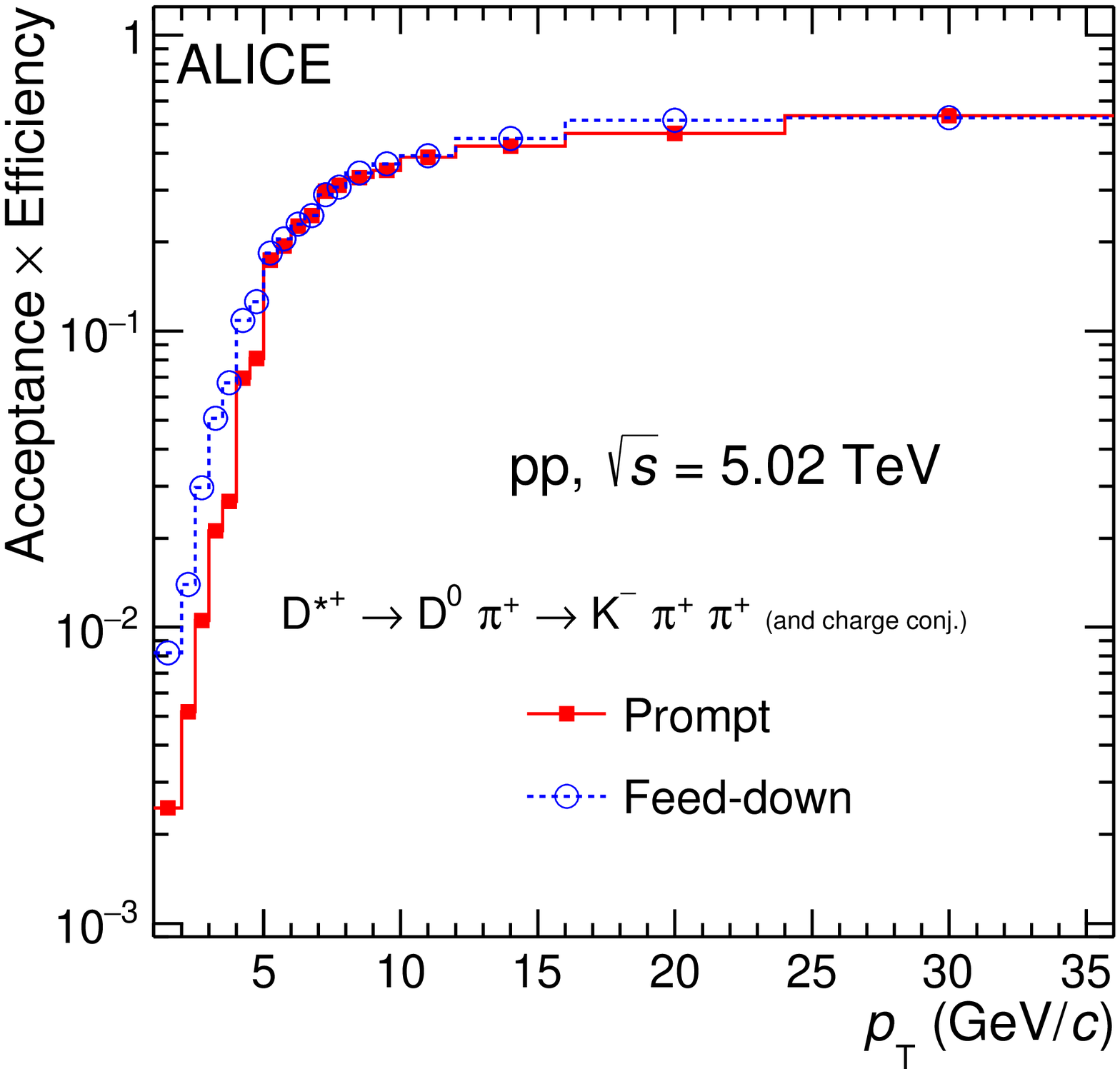}
\includegraphics[width=0.49\textwidth]{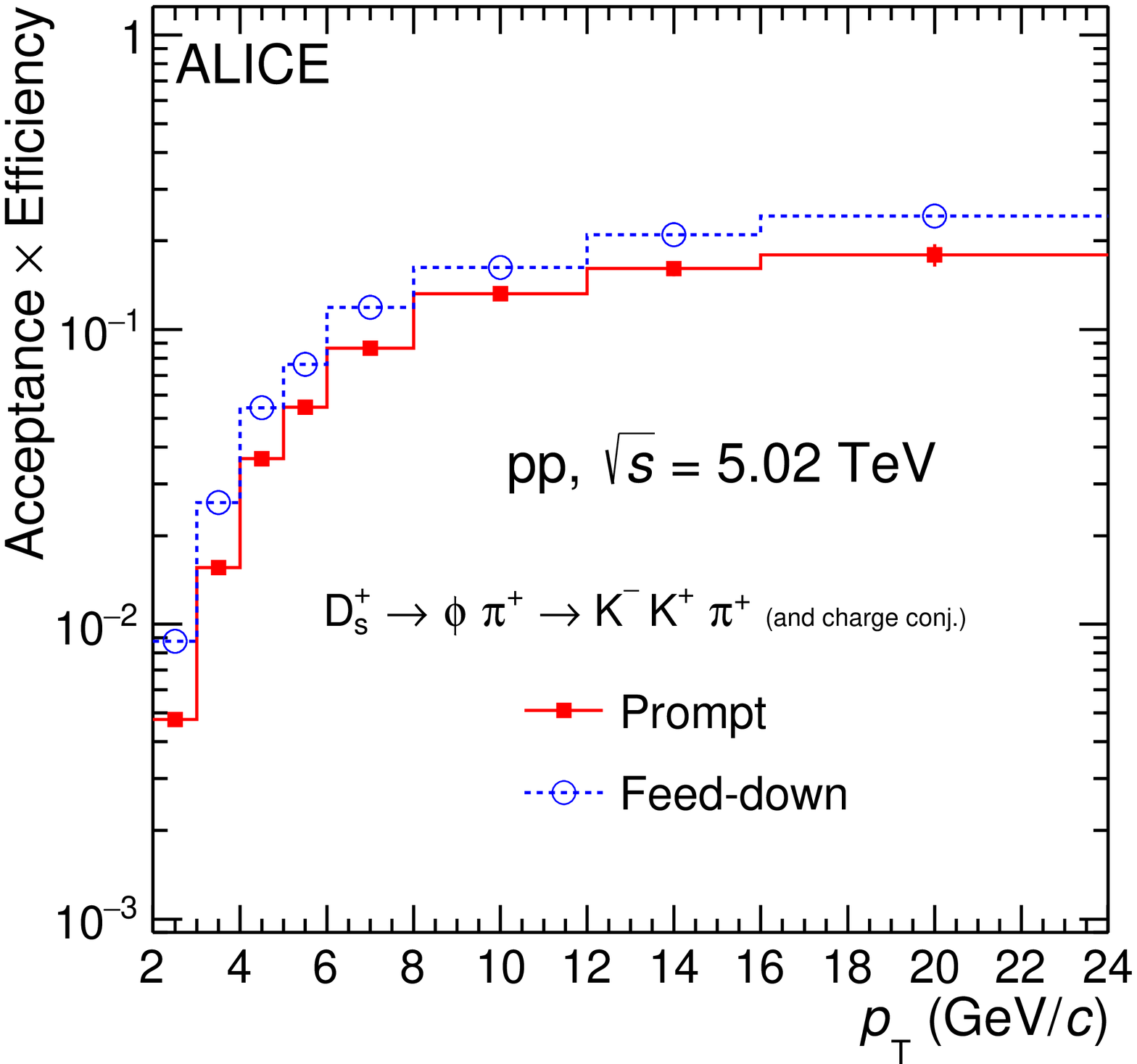}
\caption{Acceptance $\times$ efficiency for $\Dzero$, $\Dplus$, $\Dstar$, and $\Ds$ mesons, as a function of $\pt$.
The efficiencies for prompt (solid lines) and feed-down (dotted lines)
D mesons are shown.}
\label{fig:AccEff}
\end{center}
\end{figure}

Figure~\ref{fig:AccEff} shows the $(\rm Acc \times \epsilon)$ as a function of $\pt$ for prompt and feed-down $\Dzero$, $\Dplus$, 
$\Dstar$, and $\Ds$ mesons within the fiducial acceptance region. 
The average larger displacement from the primary vertex of beauty hadrons due to their long lifetime ($c\tau\approx 500~\mum$~\cite{Tanabashi:2018oca}) results in a more efficient selection of feed-down D mesons compared to prompt D mesons in most of the $\pt$ intervals.

The correction factor for the rapidity acceptance $c_{\Delta y}$ was computed with the PYTHIA 6.4.25 event generator with Perugia-11 tune. It was defined as the ratio between the
generated D-meson yield in $\Delta y = 2\,y_{\rm fid}$,
and that in $|y|<0.5$.
It was checked that calculations of the $c_{\Delta y}$ correction factor based on FONLL pQCD calculations~\cite{Cacciari:2012ny} or
on the assumption of uniform D-meson rapidity distribution in $|y|<y_{\rm fid}$ would give the same result, because both in PYTHIA and in FONLL the D-meson yield is uniform within 1\% in the range $|y|<0.8$.

The $f_{\rm prompt}$ fraction was calculated similarly to previous measurements (see e.g. Refs.~\cite{Acharya:2017jgo,Acharya:2018hre}) using the beauty-hadron production cross sections from  
FONLL calculations~\cite{Cacciari:1998it, Cacciari:2001td}, the 
$\mathrm{beauty~hadron} \rightarrow \mathrm{D} + X$ decay kinematics from the EvtGen package~\cite{Lange:2001uf}, 
and the efficiencies for feed-down D mesons reported in 
Fig.~\ref{fig:AccEff}.
The values of $f_{\rm prompt}$ range between 0.8 and 0.96 depending on 
D-meson species and $\pt$.

\subsection{Analysis without D-meson decay vertex reconstruction}
\begin{figure}[!tb]
\begin{center}
\includegraphics[width=\textwidth]{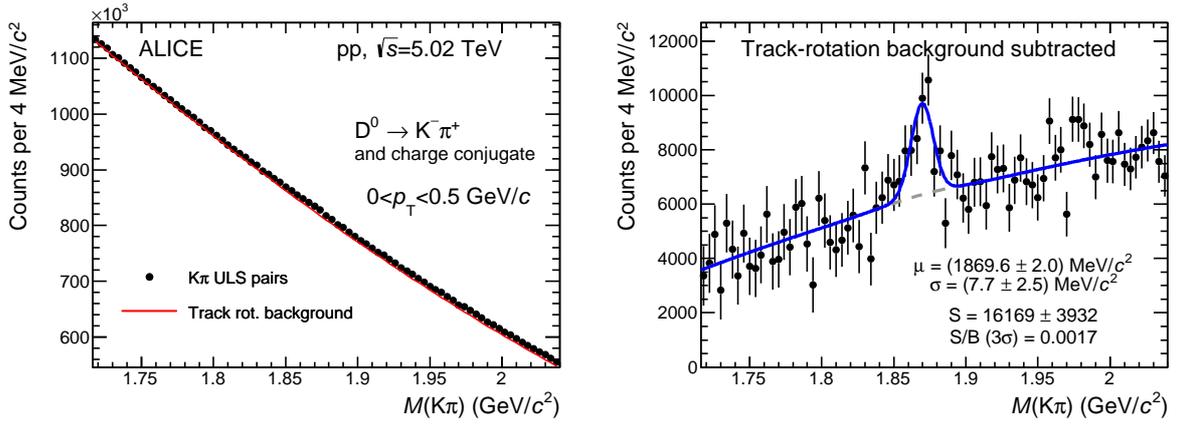}
\caption{Invariant-mass distributions of $\DtoKpi$ candidates (and
charge conjugates) for $0<\pt<0.5~\gev/c$. The left panel displays
the invariant-mass distribution of all opposite-sign K$\pi$ pairs (or unlike sign, ULS in the legend) together with the
background distribution estimated with the 
track-rotation technique. The right panel shows the invariant-mass 
distributions after subtraction of the background from the track-rotation technique. The blue solid line shows the total fit function as described in the text and the grey dashed line is the residual background after the subtraction of the background from the track-rotation technique.}
\label{fig:invmasssplowptpp} 
\end{center}
\end{figure}

A different analysis method, not based on geometrical selections of the displaced decay-vertex topology, was developed for the two-body decay $\DtoKpi$ (and its charge conjugate) in order to 
extend the measurement of the cross section down to $\pt=0$~\cite{Adam:2016ich}. 
Indeed, the poor track impact parameter resolution at very low $\pt$ and the small Lorentz boost limit the effectiveness of the selections based on the displaced decay-vertex topology. Furthermore, geometrical selections based on the displacement of the $\Dzero$-meson decay vertex tend to enhance the contribution of feed-down D mesons, increasing the related systematic uncertainty. 
This alternative analysis technique is mainly based on particle identification and 
on the estimation and subtraction of the combinatorial background.

The $\Dzero$ candidates were formed combining pairs of kaons and pions tracks with opposite charge sign, $|\eta|<0.8$, and $\pt>0.3~\gev/c$.
Track selection and pion and kaon identification were performed with the same strategy used in the analysis with decay-vertex reconstruction described in Section~\ref{subsec:vertexing}.
The resulting $\Dzero$ and $\Dzerobar$ candidates were selected
by applying the same fiducial acceptance selection $|y|<y_{\rm fid}(\pt)$ adopted for the analysis with decay-vertex reconstruction. The invariant-mass distribution of K$\pi$ pairs was obtained in fourteen transverse momentum intervals, in the range $0<\pt<12~\GeV/c$. The background distribution was estimated with the track-rotation technique. For each $\Dzero$ (and $\Dzerobar$) candidate, up to 19 combinatorial-background-like candidates were created by rotating the kaon track by different angles in the range between $\frac{\pi}{10}$ and $\frac{19\pi}{10}$ radians in azimuth.
 The left hand panel of Fig.~\ref{fig:invmasssplowptpp} shows the invariant-mass distribution of opposite-sign K$\pi$ pairs together with that of the background estimated with the track-rotation technique in the interval $0<\pt<0.5~\GeV/c$.


After subtracting the background distribution from the opposite-sign K$\pi$ invariant-mass 
distribution, the $\Dzero$-meson raw signal (sum of particle and antiparticle contributions) was extracted from the resulting distribution via a fit to the background-subtracted 
invariant-mass distribution, as reported in Fig.~\ref{fig:invmasssplowptpp} 
(right panel) for the interval $0<\pt<0.5~\GeV/c$. In the fit function, the signal was modelled with a Gaussian term, while the residual background with second-order polynomial function.
The statistical significance of the signal extracted in $0<\pt<0.5~\gevc$ ($0.5<\pt<1~\gevc$) is $S/\sqrt{S+B}=5.2~(8.0)$.

\begin{figure}[!tb]
\begin{center}
\includegraphics[width=0.48\textwidth]{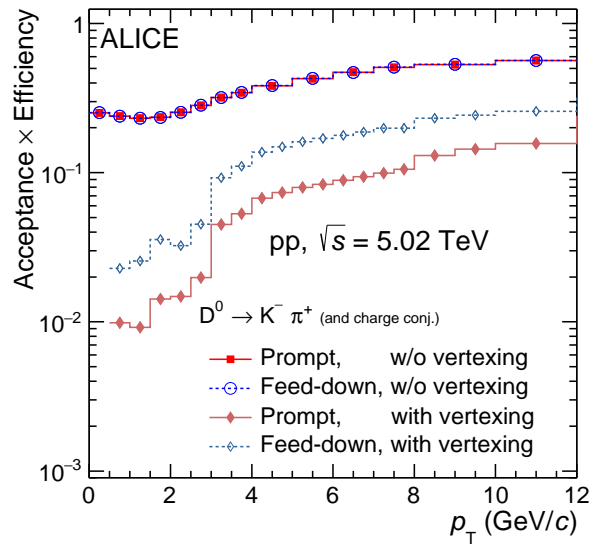}
\caption{Product of acceptance and efficiency of $\DtoKpi$ (and charge conjugates).}
\label{fig:accefflowpt} 
\end{center}
\end{figure}

The $({\rm Acc}\times\epsilon)$ correction factors of prompt and feed-down $\Dzero$ mesons were determined from the same Monte Carlo simulations as those used for the analyses with decay-vertex reconstruction. The $({\rm Acc}\times\epsilon)$ obtained with the two different analyses are compared in Fig.~\ref{fig:accefflowpt}.
For the analysis that does not exploit the selections on the $\Dzero$-meson decay vertex, the efficiency is higher by a factor of about 30 (3) at low (high) $\pt$ and almost independent of $\pt$. The mild increase with the increasing $\pt$ is mainly determined by the geometrical acceptance of the detector. 
Unlike in the analysis with decay-vertex reconstruction, the efficiency is the same for prompt $\Dzero$ and for feed-down $\Dzero$, as expected when no selection is made on the displacement of the $\Dzero$-meson decay vertex from the interaction point.

The prompt fraction to the $\Dzero$-meson raw yield, $f_{\rm prompt}$, was estimated with the same FONLL-based approach used for the analysis with decay-vertex. 
The resulting $f_{\rm prompt}$ values decrease with increasing $\pt$, from a value of about 0.95 for $\pt<4~\gevc$ to about 0.90 in the interval $8<\pt<12~\gev/c$ and are larger compared to the analysis with decay-vertex reconstruction, due to the fact that the feed-down component is not enhanced by the topological selection criteria.

\subsection{Measurement of the fraction of prompt D mesons}
\label{sec:fprompt}

\begin{figure}[!t]
\begin{center}
\includegraphics[width=0.9\textwidth]{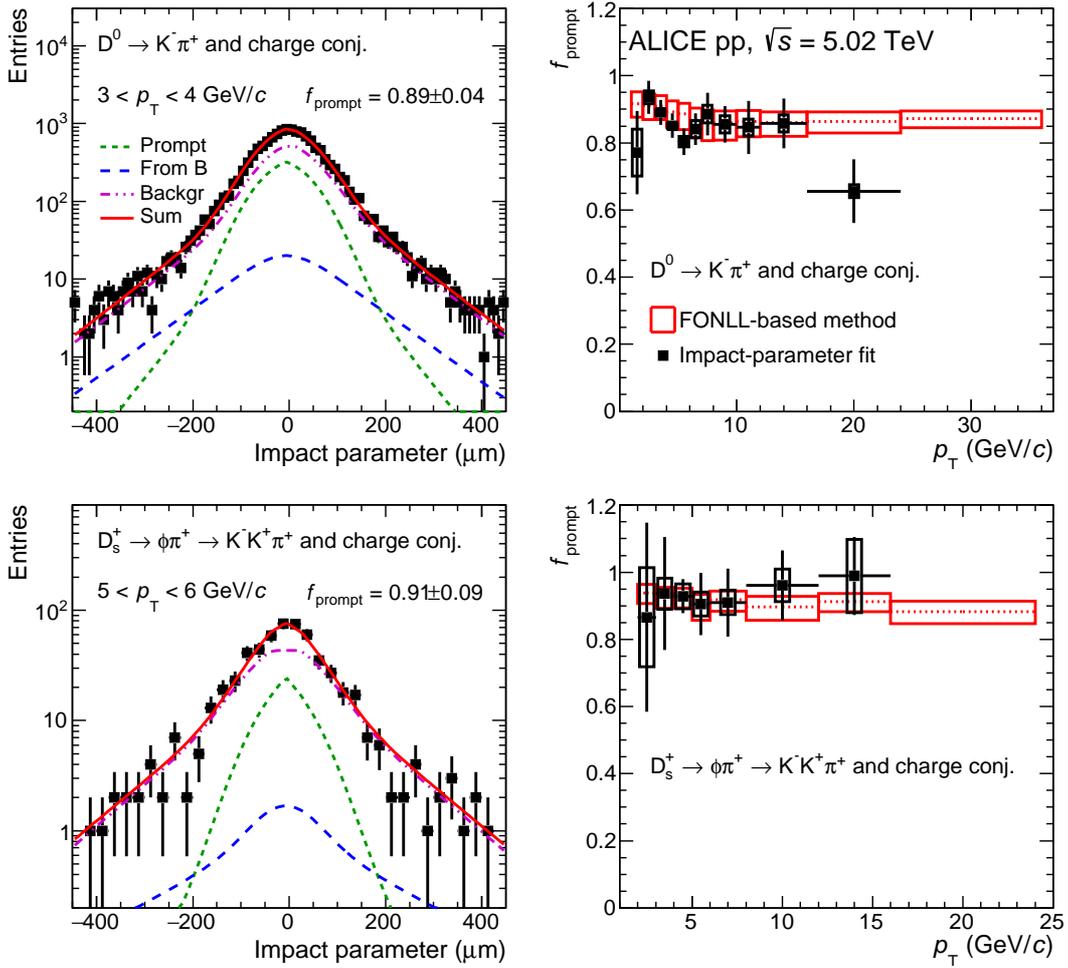}
\caption{Left: examples of fits to the impact-parameter distributions of $\Dzero$ and $\Ds$ candidates. The curves show the fit functions describing the prompt, feed-down, and background contributions, as well as their sum, as described in the text.
Right: fraction of prompt $\Dzero$ and $\Ds$-mesons raw yield as a function of $\pt$ compared to the values obtained with the FONLL-based approach. 
The results from the data-driven method are shown as square markers with the error bars (boxes) representing the statistical (systematic) uncertainty.  
The central values of $\fprompt$ from the FONLL-based approach are shown by the dashed line and their uncertainty by the red boxes.
\label{fig:fprompt} 
}
\end{center}
\end{figure}

In order to cross-check the values obtained with the  FONLL-based method of Section~\ref{subsec:vertexing}, the fractions of prompt $\Dzero$ and $\Ds$ mesons in the raw yields, $\fprompt$, were measured exploiting the different shapes for the distributions of the transverse-plane impact parameter to the primary vertex ($d_0$) of prompt and feed-down D mesons. The prompt fraction was estimated via an unbinned maximum-likelihood fit of the $d_0$ distribution of $\Dzero$ and $\Ds$ candidates with invariant mass $|M-M_{\mathrm{D}}|<2 \sigma$ (where $\sigma$ is the standard deviation of the Gaussian function describing the D-meson signal in the invariant-mass fits), using the fit function 
\begin{equation}
 F(d_0) = S\cdot \left[ (1-f_{\rm prompt}) F^{\rm feed\mbox{-}down} (d_0) + f_{\rm prompt} F^{\rm prompt} (d_0) \right] + B\cdot F^{\rm backgr}(d_0)\,.
  \label{eq:ImpParFit}
\end{equation} 
In this function, $S$ and $B$ are the signal raw yield and background 
in the selected invariant-mass range, fixed to the values obtained from the invariant-mass fit; $F^{\rm prompt} (d_0)$, 
$F^{\rm feed\mbox{-}down} (d_0)$, and $F^{\rm backgr}(d_0)$ 
are the functions describing the impact-parameter distributions of prompt and feed-down D mesons and background, respectively. 
The function $F^{\rm prompt}$ is a detector resolution term modelled with a Gaussian and a symmetric exponential term. 
The function $F^{\rm feed\mbox{-}down}$ is the convolution of a sum of two symmetric exponential functions ($F^{\rm feed\mbox{-}down}_{\rm true}$), which describe the intrinsic impact-parameter distribution of secondary D mesons from beauty-hadron decays, and the detector resolution term ($F^{\rm prompt}$). 
All the parameters of the $F^{\rm prompt}$ and $F^{\rm feed\mbox{-}down}_{\rm true}$ functions were fixed in the data fit to the values obtained by fitting the distributions from Monte Carlo simulations, except for the Gaussian width of the detector-resolution term, which was kept free in order to compensate a possible discrepancy between the impact-parameter resolution in the data and in the simulation.  
The distribution describing the combinatorial background was parameterised with a function composed of a Gaussian and symmetric exponential term ($F^{\rm backgr}$). The parameters were fixed to those obtained by fitting the impact-parameter distribution of background candidates in the side bands of the signal peak in the invariant-mass distributions. 
Figure~\ref{fig:fprompt} (left) shows examples of fits to the impact-parameter distributions of $\Dzero$ and $\Ds$ mesons in the transverse-momentum intervals $3<\pt<4~\gevc$ and $5<\pt<6~\gevc$, respectively. For this study, wider $\pt$ intervals were adopted compared to the analysis, due to the poor quality of the fit when reducing the sample.     
The $\Dzero$ candidates used in the impact-parameter fit were selected with the same criteria described in Section~\ref{subsec:vertexing}. For the $\Ds$ mesons, the impact-parameter selection, used to extract the raw yield from the invariant-mass distribution, was not applied for this study. In this case, the prompt fraction, $\fprompt$, was obtained by integrating the functions obtained from the fit in the restricted impact-parameter range used in the analysis.


The prompt fraction measured with the fits to the impact-parameter distributions of D-meson candidates has three main sources of systematic uncertainty, namely 
(i) the assumption on the shape of the impact-parameter distribution for each contribution (prompt D mesons, feed-down D mesons,  
and combinatorial background); 
(ii) the uncertainty on the signal and background yields extracted from the invariant-mass fits; and
(iii) the consistency of the procedure, evaluated with a Monte Carlo closure test.
These uncertainties were estimated with the procedures described in Ref.~\cite{Adam:2016ich}.
The total systematic uncertainty on $\fprompt$ with the data-driven 
approach ranges, depending on $\pt$, are between 1\% and 9\% for the $\Dzero$ meson, and between 4\% and 17\% for the $\Ds$ meson. 

The prompt fractions in the raw yields of $\Dzero$ and $\Ds$ mesons 
measured with the data-driven method are compared to those calculated with the FONLL-based approach in the right panels of Fig.~\ref{fig:fprompt} and found to be compatible within uncertainties. 
For the interval $24<\pt<36~\gev/c$ ($16<\pt<24~\gev/c$), given the poor precision of 
the impact-parameter fit, it was not possible to determine the data-driven prompt fraction for the $\Dzero$ ($\Ds$) meson.

\section{Systematic uncertainties}
\label{sec:Syst}
Systematic uncertainties on the D-meson cross sections were estimated considering the following sources:
(i) extraction of the raw yield from the invariant-mass distributions; 
(ii) track reconstruction efficiency; 
(iii) D-meson selection efficiency; 
(iv) PID efficiency; 
(v) the shape of the $\pt$ spectrum generated for D mesons in the simulation; 
(vi) subtraction of the feed-down from beauty-hadron decays.
In addition, the uncertainties on the branching ratios and on the integrated luminosity were considered. A summary of the systematic uncertainties is reported in Table~\ref{tab:sumSyst} for different $\pt$ intervals.

The systematic uncertainties on the raw yield extraction were evaluated by repeating the fits several hundred times varying the fit interval and the functional form of the background fit function. The same strategy was performed using a bin-counting method, in which the signal yield was obtained by integrating the invariant-mass distribution after subtracting the background, estimated from a fit to the side-bands only. The systematic uncertainty was defined as the RMS of the distribution of the signal yields obtained from all these variations and ranges between 1\% and 9\% depending on the D-meson species and $\pt$ interval. This includes for the $\Dzero$ mesons a contribution of about 1\% obtained by varying the ratio of the integral of the reflections to the integral of the signal and the shape of the templates used in the invariant-mass fits. For the background estimation of the $\Dzero$-meson analysis without decay-vertex reconstruction with the track-rotation technique, different configurations of the rotation angle were used. In addition, three alternative approaches were tested to estimate the background distribution: like-sign (LS) pairs, event mixing, and side-band fit~\cite{Adam:2016ich}. The raw yield values obtained subtracting these alternative background distributions were found to be consistent with those from the default configuration of the track-rotation method within the uncertainty estimated by varying the fit conditions and therefore no additional systematic uncertainty was assigned.

The systematic uncertainty on the track reconstruction efficiency has two different contributions. The first one is estimated by varying the track-quality selection criteria and the second one is estimated by comparing the probability to match the tracks from the TPC to the ITS hits in data and simulation (matching efficiency). To obtain the matching efficiency, the abundances of primary and secondary particles in data were estimated via template fits to the track impact-parameter distributions, where the relative abundances in the simulation were weighted to match those in data~\cite{ALICE-PUBLIC-2017-005,Acharya:2017jgo}. The estimated uncertainty, a quadratic sum of the two contributions, depends on the D-meson $\pt$ and it ranges from 3\% to 5\% for the two-body decay of $\Dzero$ mesons and from 3.5\% to 7\% for the three-body decays of $\Dplus$, $\Dstar$, and $\Ds$ mesons.

The systematic uncertainty on the D-meson selection efficiency originates from imperfections in the simulation of the D-meson decay kinematics and topology and of the resolutions and alignments of detectors in the simulation. For the analyses with decay-vertex reconstruction, the systematic uncertainty was estimated by repeating the analysis with different sets of selection criteria, resulting in a significant modification of the efficiencies, raw yield, and background values. The systematic uncertainties are largest at low $\pt$ (up to 5\%), where the efficiencies are low and vary steeply with $\pt$, because of the tighter geometrical selections. For the $\Ds$ meson, for which more stringent selection criteria were used, slightly larger uncertainties were estimated, ranging from 5\% at high $\pt$ to 8\% at low $\pt$. In the case of the $\Dzero$-meson analysis without decay-vertex reconstruction, the stability of the corrected yield was tested against variations of the single-track $\pt$ selection and no systematic effect was observed.

To estimate the uncertainty on the PID selection efficiency, the analysis was repeated without PID selection for the three non-strange D-meson species and $\Ds$ mesons with $\pt>6~\gevc$. The resulting cross sections were found to be compatible with those obtained with the PID selection and therefore no systematic uncertainty was assigned. For $\Ds$ mesons with $\pt<6~\gevc$ and the $\Dzero$-meson analysis without decay-vertex reconstruction, an analysis without applying PID selections could not be performed due to the insufficient statistical significance of the signal. The systematic uncertainty for low-$\pt$ $\Ds$ mesons was therefore estimated by comparing the pion and kaon PID selection efficiencies in the data and in the simulation and combining the observed differences using the $\Ds$-meson decay kinematics~\cite{Acharya:2018hre}. A 3\% systematic uncertainty was assigned for $4<\pt<6~\gevc$, and 2.5\% for $\pt<4~\gevc$. For the $\Dzero$-meson analysis without decay-vertex reconstruction, compatible cross sections were obtained when using more stringent PID criteria. Based on this result and on the fact that the PID selections are the same as used in the analysis with decay-vertex reconstruction, no uncertainty due to PID was assigned.

The systematic uncertainty due to the generated D-meson $\pt$ shape was estimated by using FONLL as an alternative generator with respect to PYTHIA to simulate the D-meson $\pt$ distribution~\cite{ALICE:2011aa}, and was found to be 0\%--5\% for $\pt<3~\gevc$ and negligible at higher $\pt$. The $\pt$ shape of both considered distributions were found to be compatible with the measured one within uncertainties. Finally, the systematic uncertainty on the subtraction of feed-down from beauty-hadron decays (i.e.\ the calculation of the $f_{\rm prompt}$ fraction) was estimated by varying the FONLL parameters (b-quark mass, factorisation, and renormalisation scales) as prescribed in Ref.~\cite{Cacciari:2012ny}. It ranges between $^{+1.0}_{-1.2}$\% and $^{+4.4}_{-6.3}$\% depending on the D-meson species and $\pt$ interval.

The contributions of these different sources of uncertainties were summed in quadrature to obtain the total systematic uncertainty in each $\pt$ interval, which varies from 6.5\%--10.0\%, 6.5\%--10.5\%, 5.4\%--11.3\%, and 8.7\%--12.1\% for the $\Dzero$, $\Dplus$, $\Dstar$, and $\Ds$ mesons, respectively. The systematic uncertainty on PID, tracking, and selection efficiencies are mainly correlated among the different $\pt$ intervals, while the raw-yield extraction uncertainty is mostly uncorrelated. The $\pt$-differential cross sections have an additional global normalisation uncertainty due to the uncertainties on the integrated luminosity~\cite{ALICE-PUBLIC-2018-014} and on the branching ratios of the considered D-meson decays~\cite{Tanabashi:2018oca}.

\begin{table}[t!]
    \centering
    \begin{tabular}{l|ccc|cc|cc|cc}
         & \multicolumn{3}{c|}{$\Dzero$} & \multicolumn{2}{c|}{$\Dplus$} & \multicolumn{2}{c|}{$\Dstar$} & \multicolumn{2}{c}{$\Ds$} \\
         $\pt$ ($\gevc$) & 0-0.5 & 2-2.5 & 10-12 & 2-2.5 & 10-12 & 2-2.5 & 10-12 & 2-3 & 8-12 \\
         \hline
         Signal yield & 9\% & 3\% & 2\%  & 3\% & 3\% & 3\% & 1\% & 7\% & 3\% \\
         Tracking efficiency & 3\% & 4\% & 5\%  & 4.5\% & 7\% & 4\% & 5\% & 4.5\% & 7\% \\
         Selection efficiency & 0 & 5\% & 3\%  & 4\% & 3\% & 5\% & 1\% & 8\% & 5\% \\
         PID efficiency & 0 & 0 & 0  & 0 & 0 & 0 & 0 & 2.5\% & 0 \\
         $\pt$ shape in MC & 0 & 0 & 0  & 1\% & 0 & 1\% & 0 & 1\% & 0 \\
         Feed-down & $^{+1.1}_{-1.3}$\% & $^{+3.6}_{-4.3}$\% & $^{+3.8}_{-5.3}$\% & $^{+2.4}_{-2.8}$\% & $^{+2.3}_{-3.1}$\% & $^{+3.0}_{-3.5}$\% & $^{+1.8}_{-2.5}$\% & $^{+2.8}_{-3.3}$\% & $^{+3.4}_{-4.5}$\% \\
         Branching ratio & \multicolumn{3}{c|}{1.0\%} &  \multicolumn{2}{c|}{3.1\%} & \multicolumn{2}{c|}{1.3\%} & \multicolumn{2}{c}{3.5\%} \\
         \hline
         Luminosity & \multicolumn{9}{c}{2.1\%} \\
         \hline 
    \end{tabular}
    \caption{Summary of relative systematic uncertainties on $\Dzero$, $\Dplus$, $\Dstar$, and $\Ds$ measurements in different $\pt$ intervals.}
    \label{tab:sumSyst}
\end{table}

\section{Results}
\label{sec:results}
\subsection{Transverse momentum-differential cross sections}
\begin{figure}[!th]
\begin{center}
\includegraphics[width=0.65\textwidth]{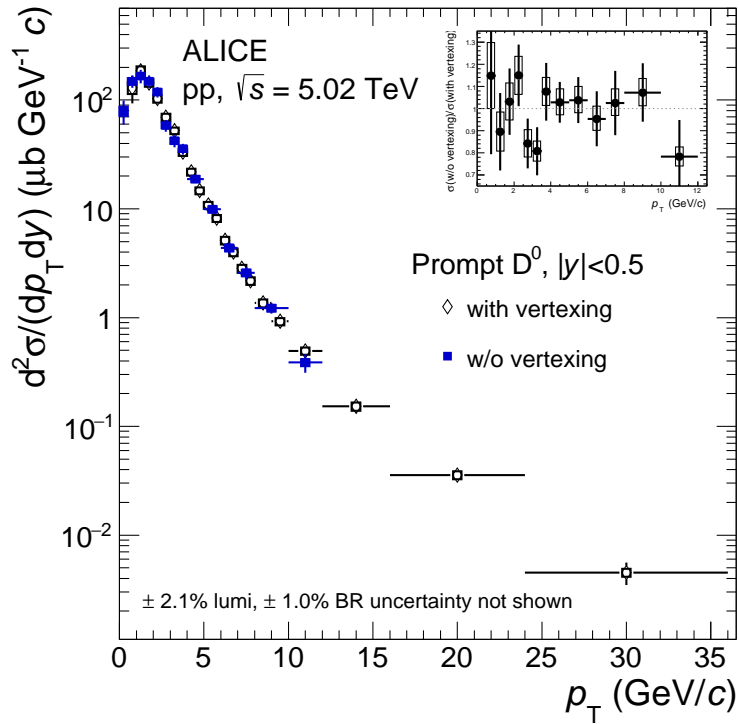}
\caption{ \label{fig:CrossSecD0PP}
Prompt $\Dzero$-meson $\pt$-differential production cross section in $|y|<0.5$ in pp collisions at $\sqrts =5.02~\tev$ measured with and without decay-vertex reconstruction.  The inset shows the ratio of the measurements in their common $\pt$ range. 
The vertical error bars and the empty boxes represent the statistical and systematic uncertainties, respectively.}
\end{center}
\end{figure}

The $\pt$-differential production cross section for prompt $\Dzero$ mesons in $|y|<0.5$ in pp collisions at $\sqrts=5.02~\tev$ was obtained from the analyses 
with and without decay-vertex reconstruction.
The two results are compared in Fig.~\ref{fig:CrossSecD0PP} with the inset
showing their ratio in the common $\pt$ range.
In all the figures in this section, the vertical error bars represent the statistical uncertainties and the systematic uncertainties are depicted as boxes around the data points. 
In each $\pt$ interval the symbols are positioned horizontally at the center 
of the bin and the horizontal bars represents the width of the $\pt$ interval.
The two results for prompt $\Dzero$-meson cross section are found to be 
consistent within statistical uncertainties, which are independent between 
the two measurements because of their very different signal-to-background 
ratios and efficiencies.
The most precise measurement of the prompt $\Dzero$-meson production cross section is 
obtained using the results of the analysis without decay-vertex reconstruction 
in the interval $0<\pt<1~\gev/c$ and those of the analysis with decay-vertex reconstruction for $\pt>1~\gev/c$.

\begin{figure}[!t]
\begin{center}
\includegraphics[width=0.8\textwidth]{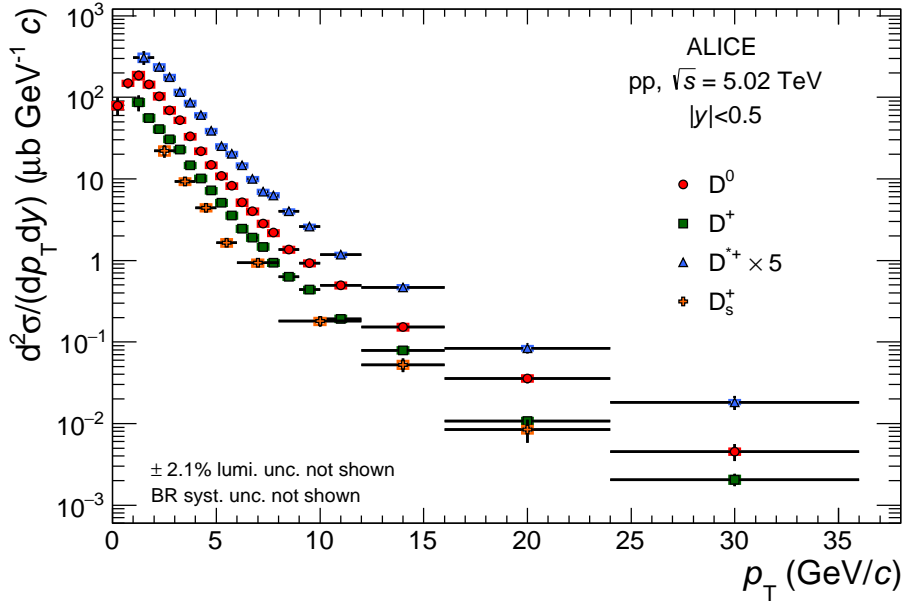}
\caption{ \label{fig:CrossSecFourMesons}
$\pt$-differential production cross section of prompt 
$\Dzero$, $\Dplus$, $\Dstar$, and $\Ds$ mesons in pp collisions
at $\sqrts=5.02~\tev$. Statistical uncertainties (bars) and systematic 
uncertainties (boxes) are shown. For the $\Dzero$ meson, the results in $0<\pt<1~\gev/c$ are obtained from the analysis without decay-vertex reconstruction, while those in \mbox{$1<\pt<36~\gev/c$} are taken from the analysis with decay-vertex reconstruction. The $\Dstar$-meson cross section is scaled by a factor of 5 for better visibility.
}
 \end{center}
    \end{figure}

The $\pt$-differential cross sections for prompt $\Dzero$, $\Dplus$, $\Dstar$, 
and $\Ds$-meson production in $|y|<0.5$ are depicted in Fig.~\ref{fig:CrossSecFourMesons}.
The prompt $\Dzero$-meson $\pt$-differential cross section is compatible with 
the one measured by the CMS collaboration at the same centre-of-mass energy 
in $|y|<1$ and $2<\pt<100~\GeV/c$~\cite{Sirunyan:2017xss}.

\begin{figure}[!th]
 \begin{center}
      \includegraphics[width=0.48\textwidth]{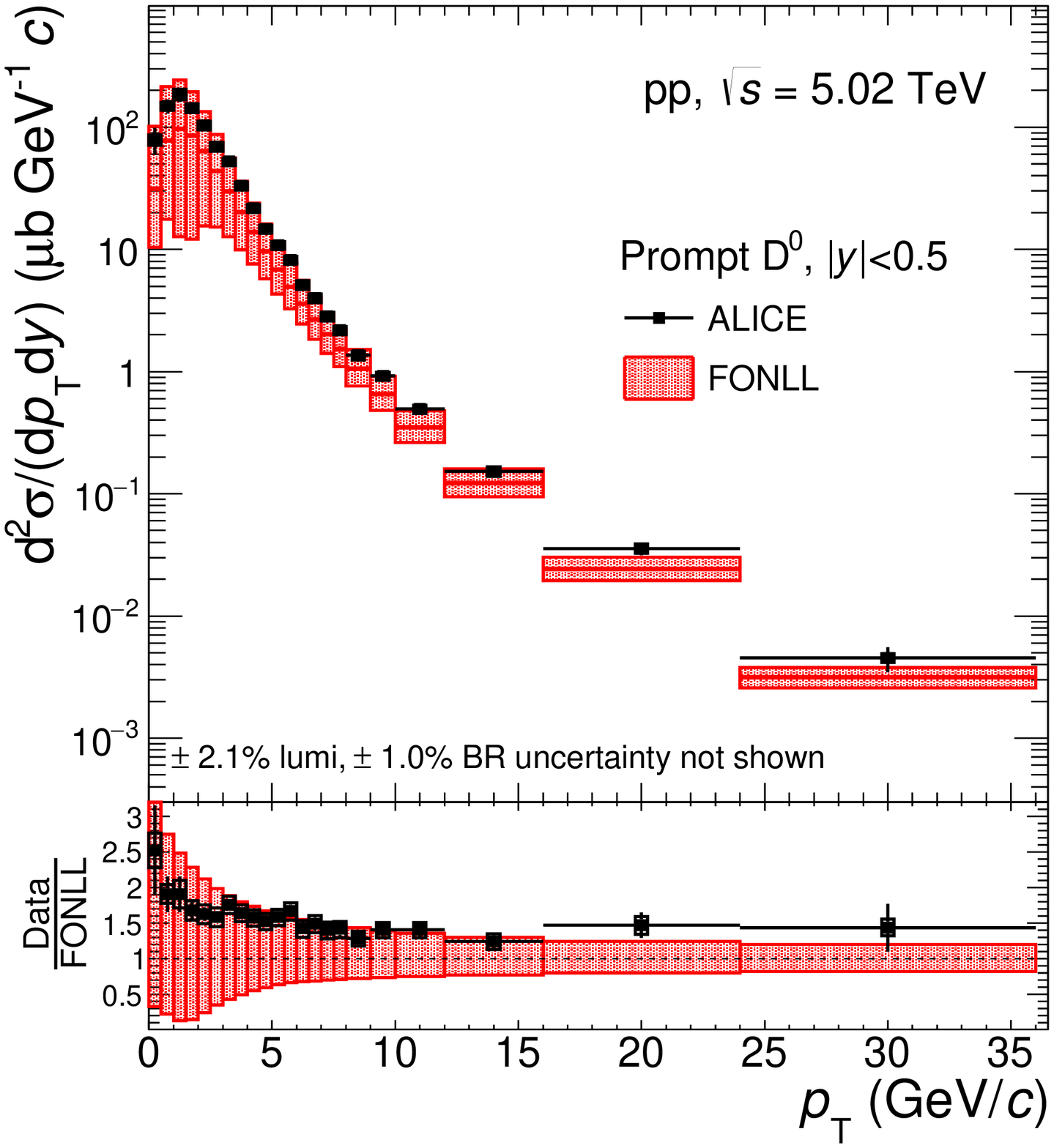}
        \includegraphics[width=0.48\textwidth]{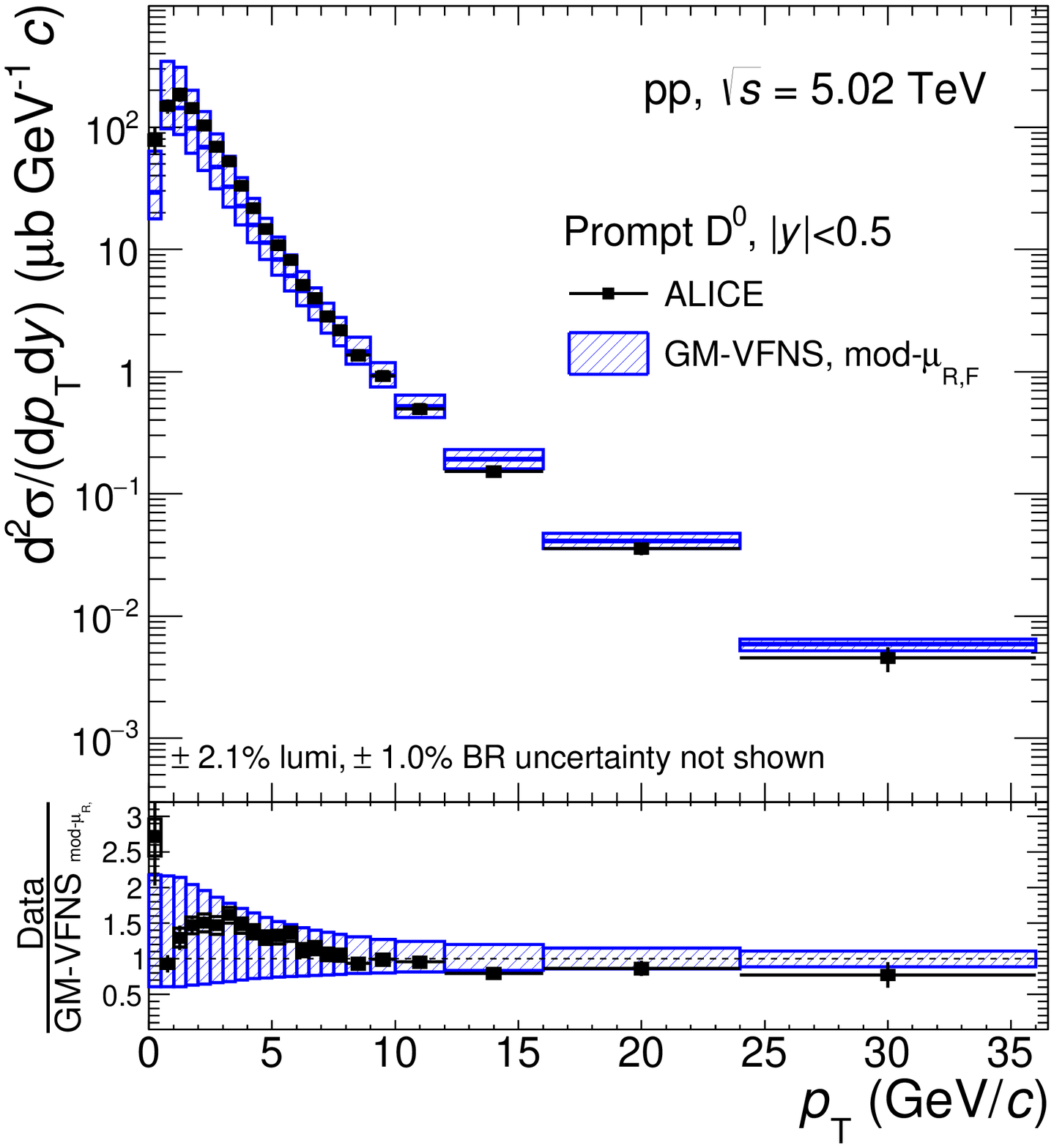}
         \includegraphics[width=0.48\textwidth]{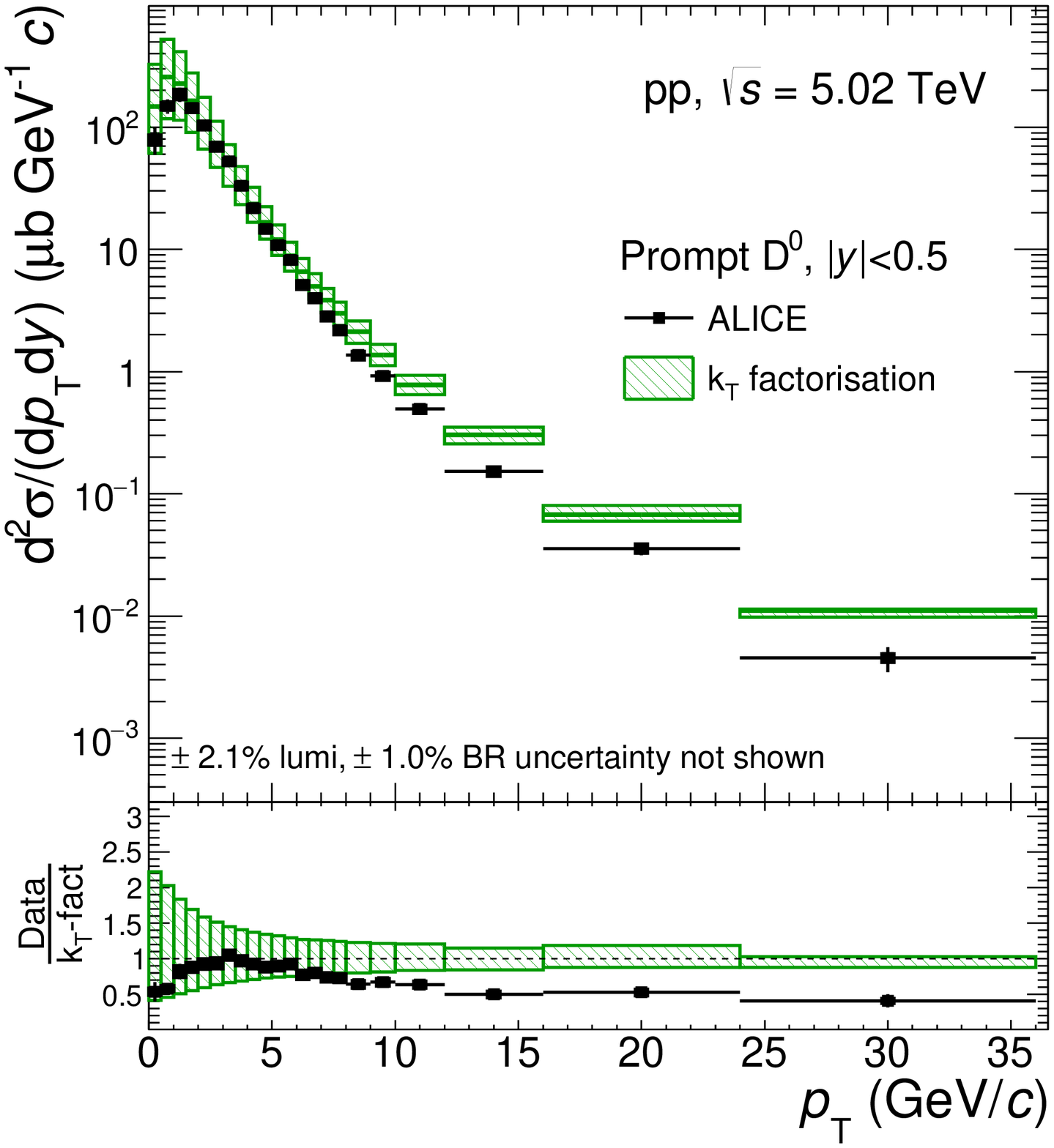}
        \includegraphics[width=0.48\textwidth]{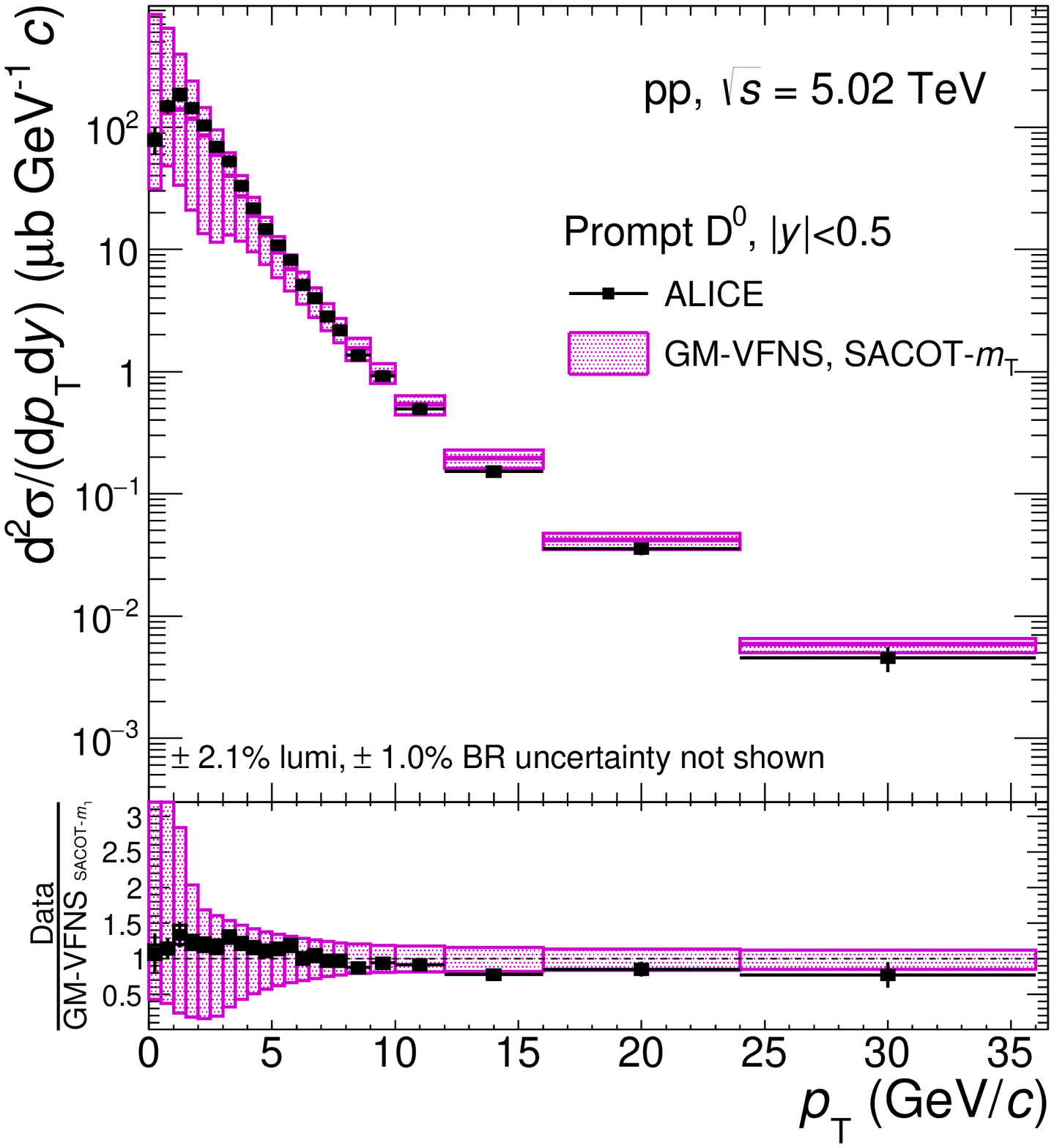}
         \caption{ \label{fig:dsdptppDzero}   
        $\pt$-differential production cross sections for prompt $\Dzero$ meson compared to pQCD calculations: FONLL~\cite{Cacciari:1998it,Cacciari:2012ny}, GM-VFNS(mod-$\mu_{\rm R,F}$)~\cite{Benzke:2017yjn,Kramer:2017gct}, GM-VFNS(SACOT-$m_{\rm T}$)~\cite{Helenius:2018uul}, and $k_{\rm T}$-factorisation~\cite{Maciula:2018iuh}. The ratios of the data to the theoretical predictions are shown in the lower part of each panel.}
 \end{center}
 \end{figure}
  
\begin{figure}[!th]
 \begin{center}
      \includegraphics[width=0.48\textwidth]{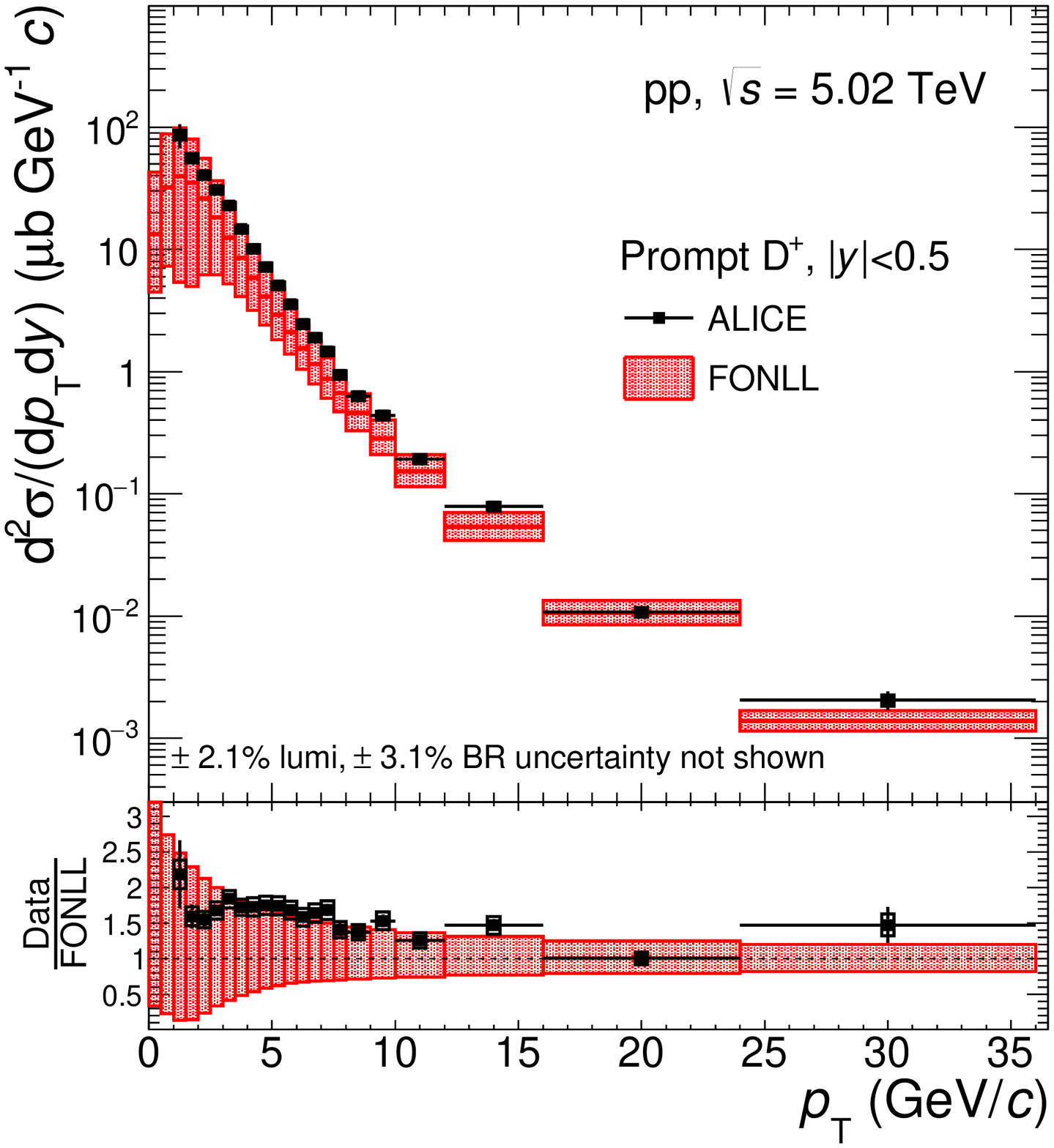}
       \includegraphics[width=0.48\textwidth]{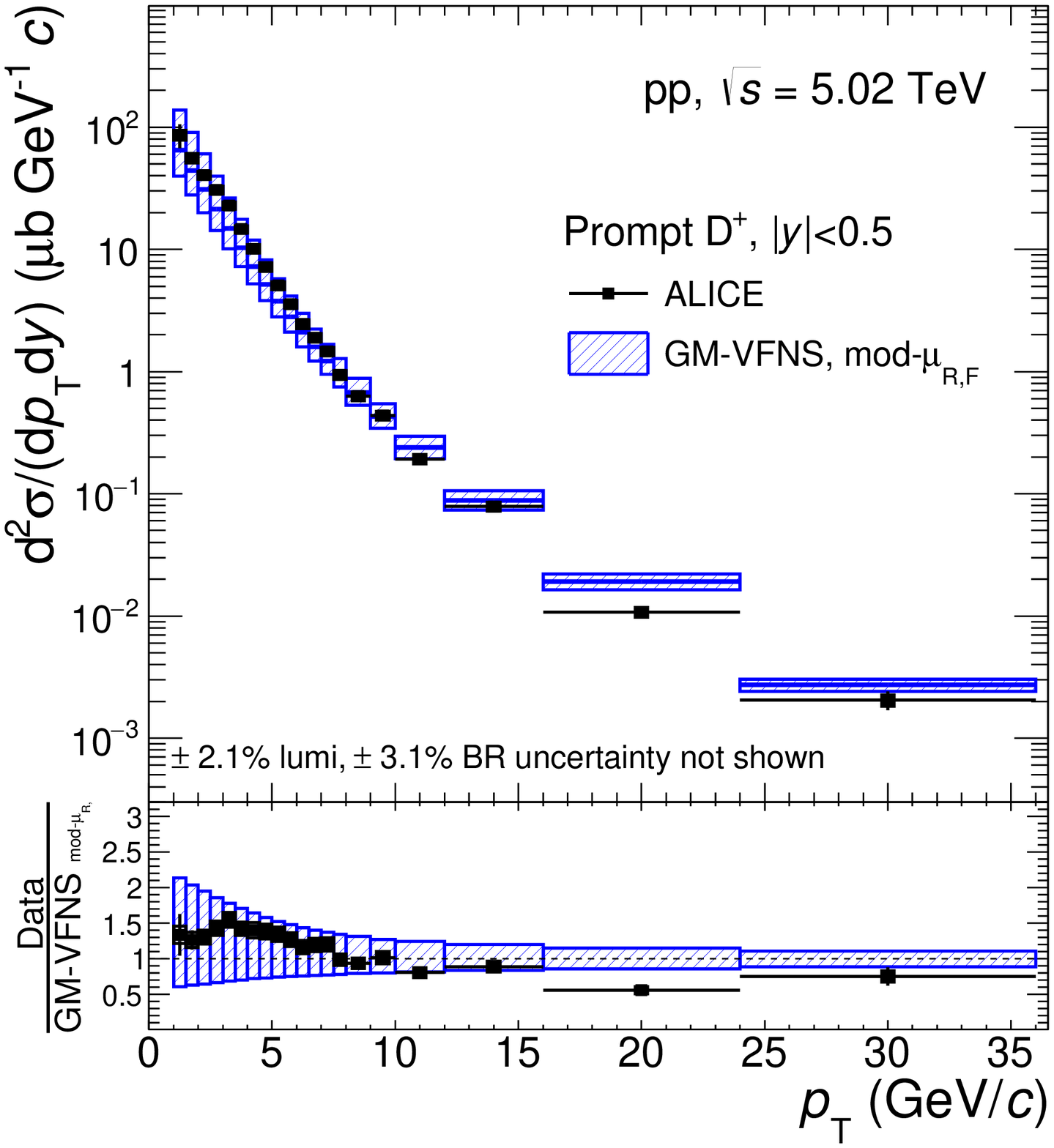} 
         \includegraphics[width=0.48\textwidth]{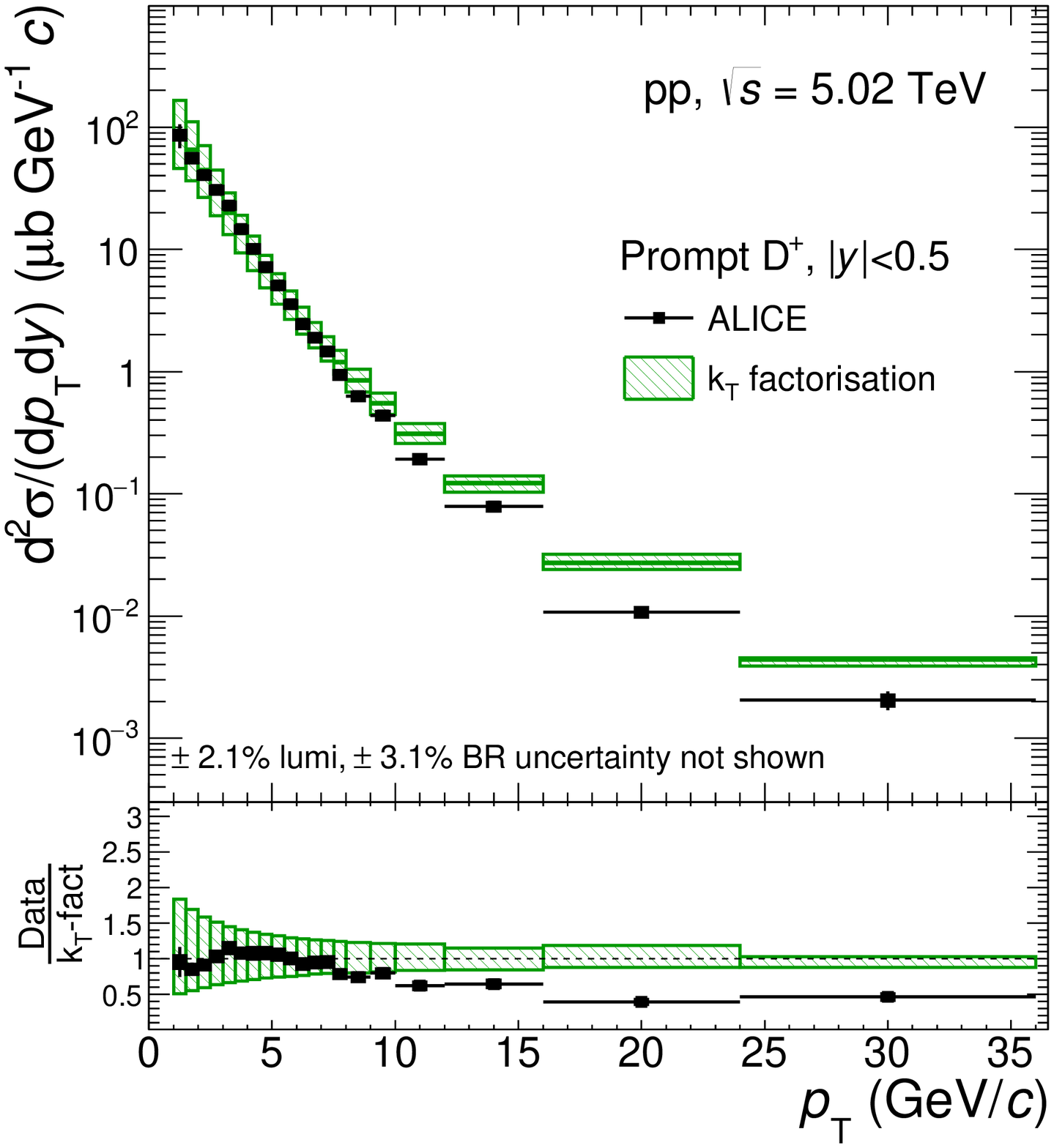} 
         \includegraphics[width=0.48\textwidth]{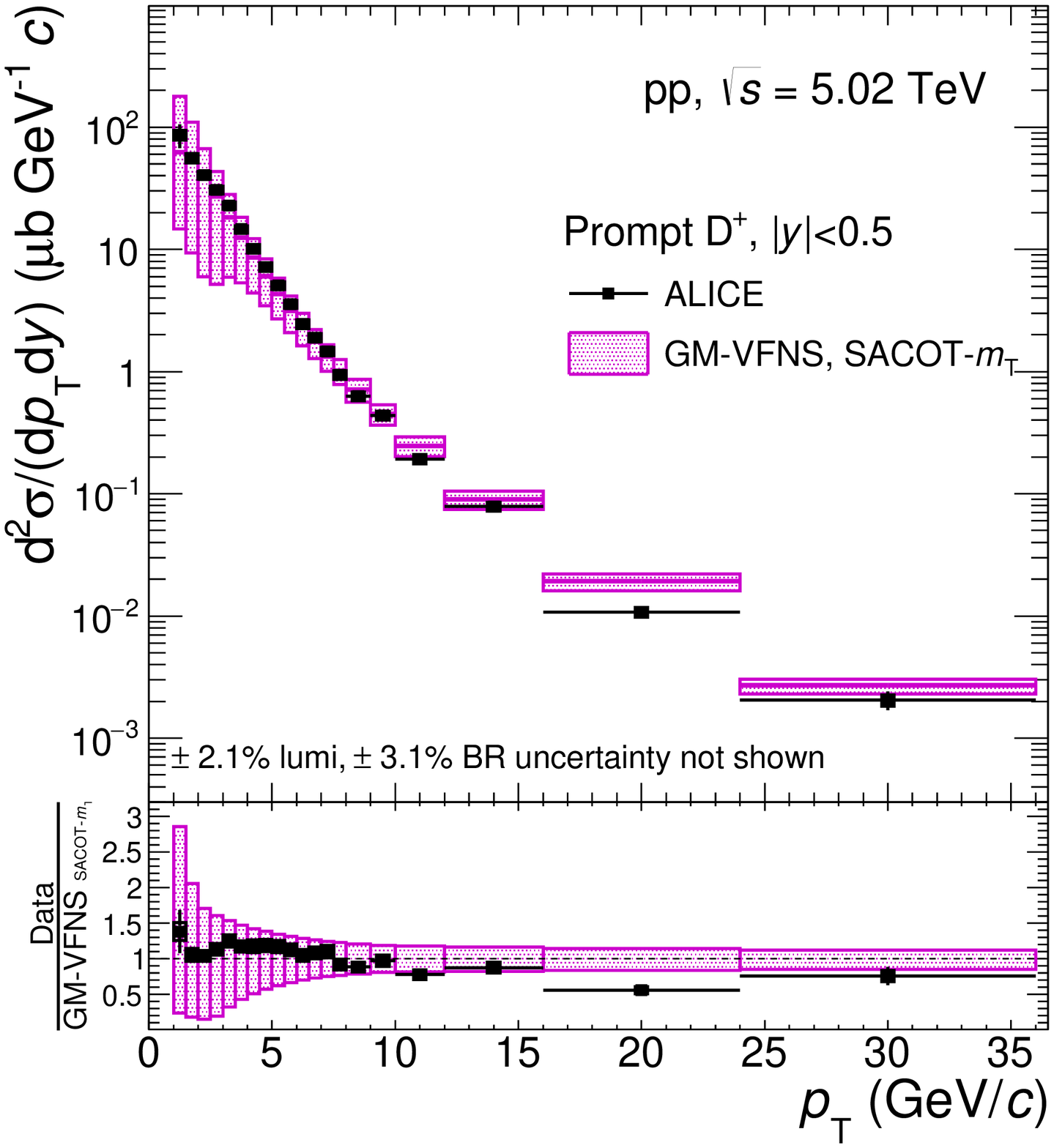}
            \caption{ \label{fig:dsdptppDplus}   
        $\pt$-differential production cross sections for prompt $\Dplus$ meson compared to pQCD calculations: FONLL~\cite{Cacciari:1998it,Cacciari:2012ny}, GM-VFNS(mod-$\mu_{\rm R,F}$)~\cite{Benzke:2017yjn,Kramer:2017gct}, GM-VFNS(SACOT-$m_{\rm T}$)~\cite{Helenius:2018uul}, and $k_{\rm T}$-factorisation~\cite{Maciula:2018iuh}. The ratios of the data to the theoretical predictions are shown in the lower part of each panel.}
 \end{center}
 \end{figure}

 \begin{figure}[!th]
 \begin{center}
      \includegraphics[width=0.48\textwidth]{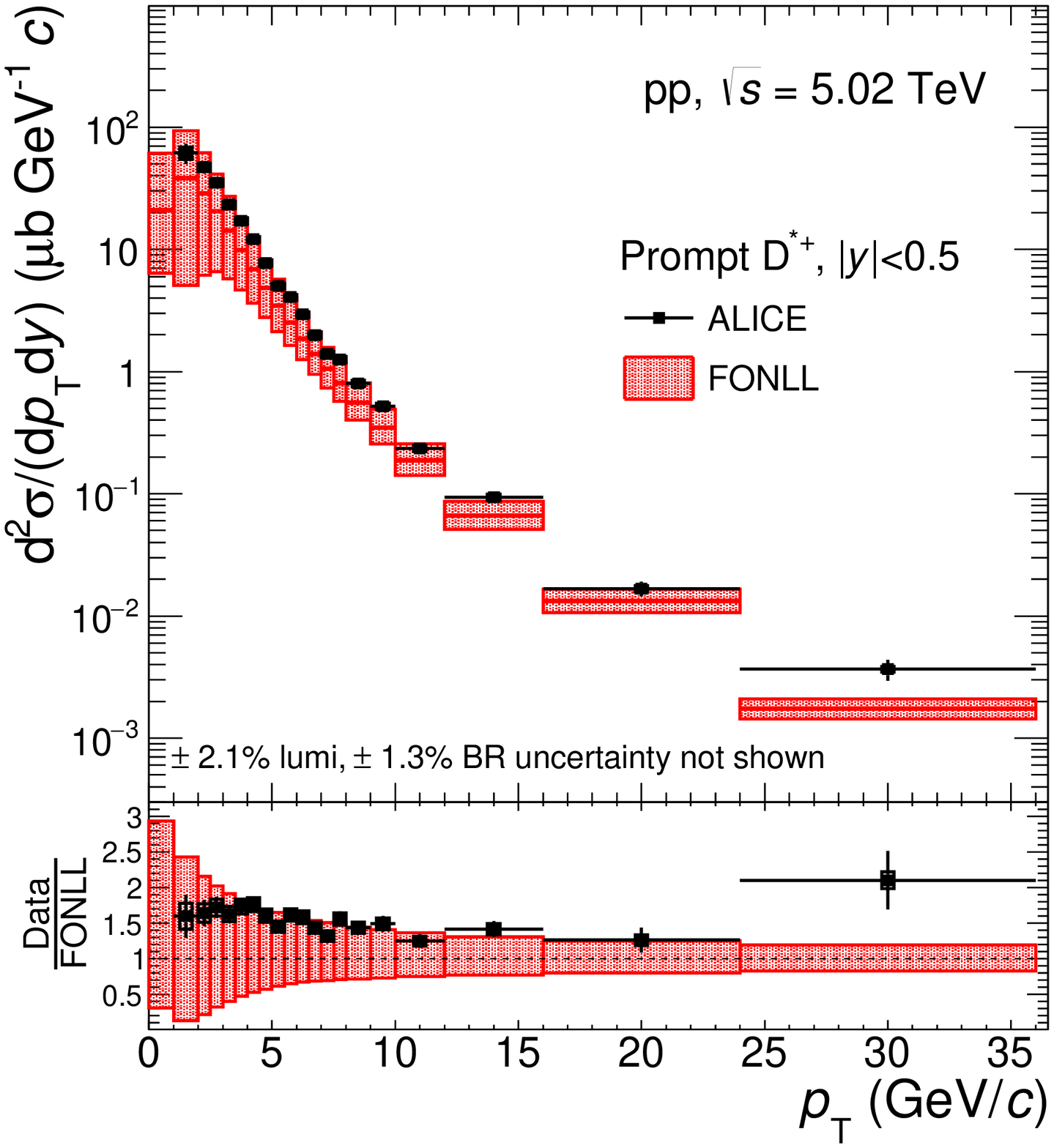}
      \includegraphics[width=0.48\textwidth]{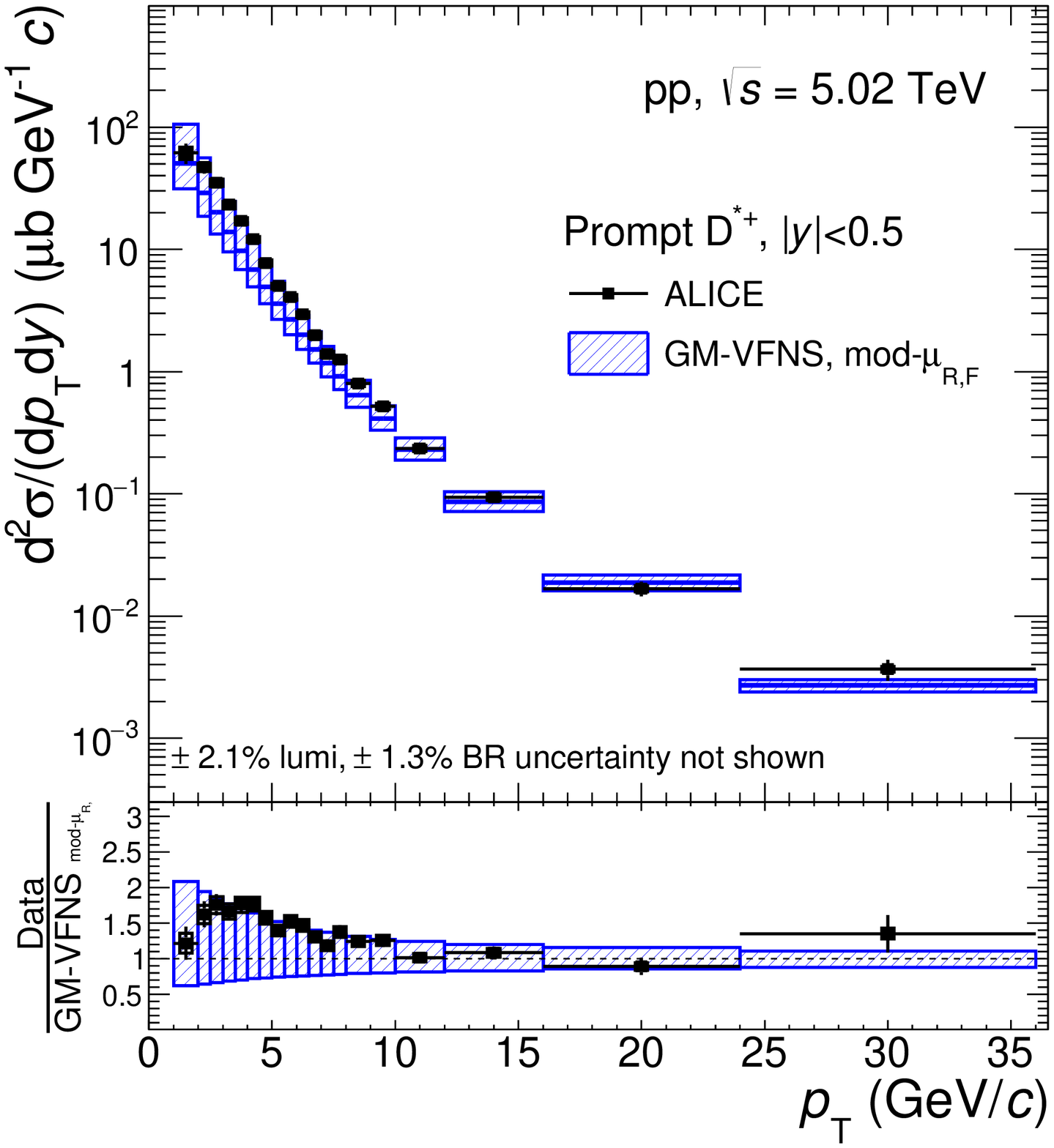}
       \includegraphics[width=0.48\textwidth]{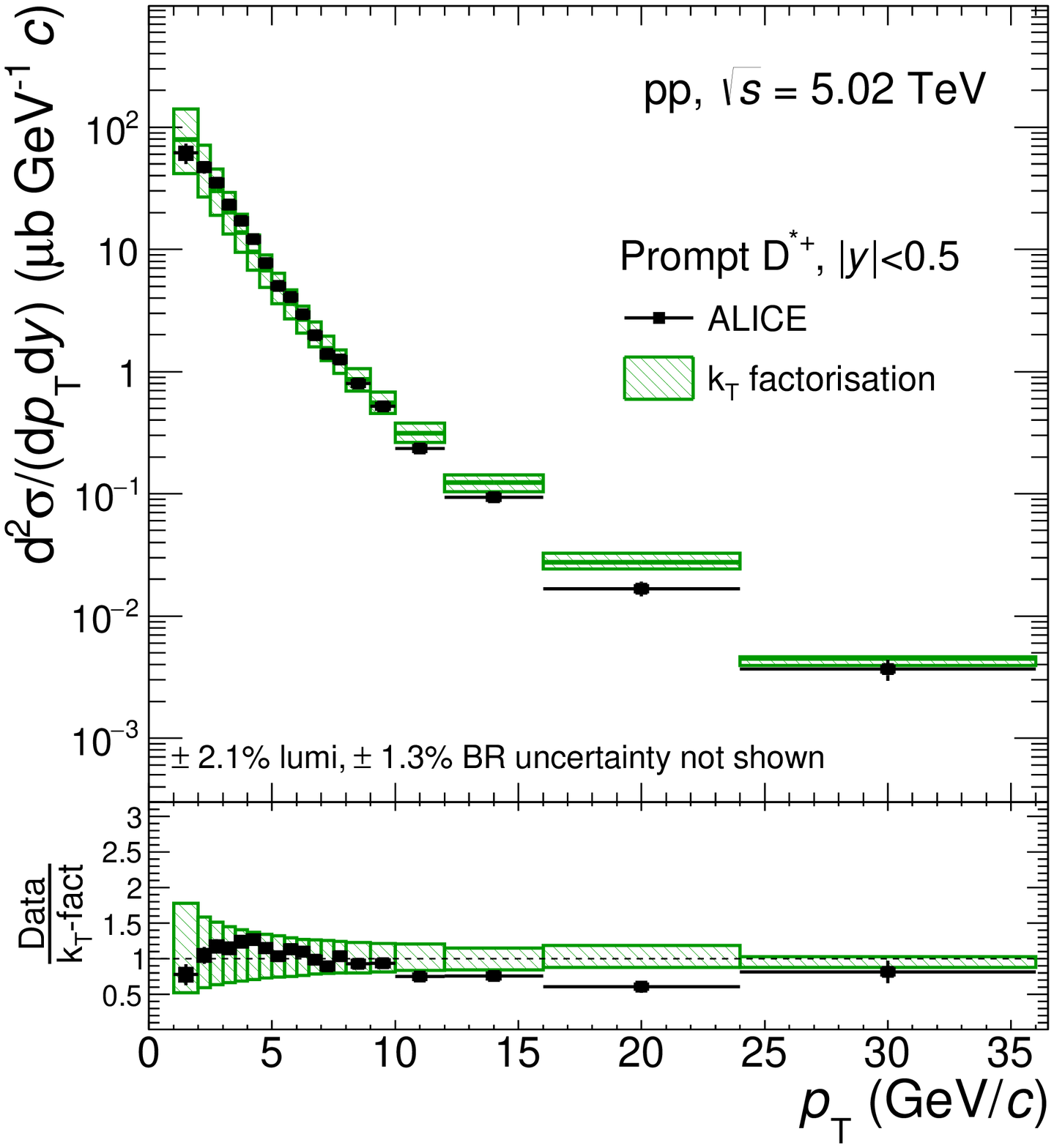}
       \includegraphics[width=0.48\textwidth]{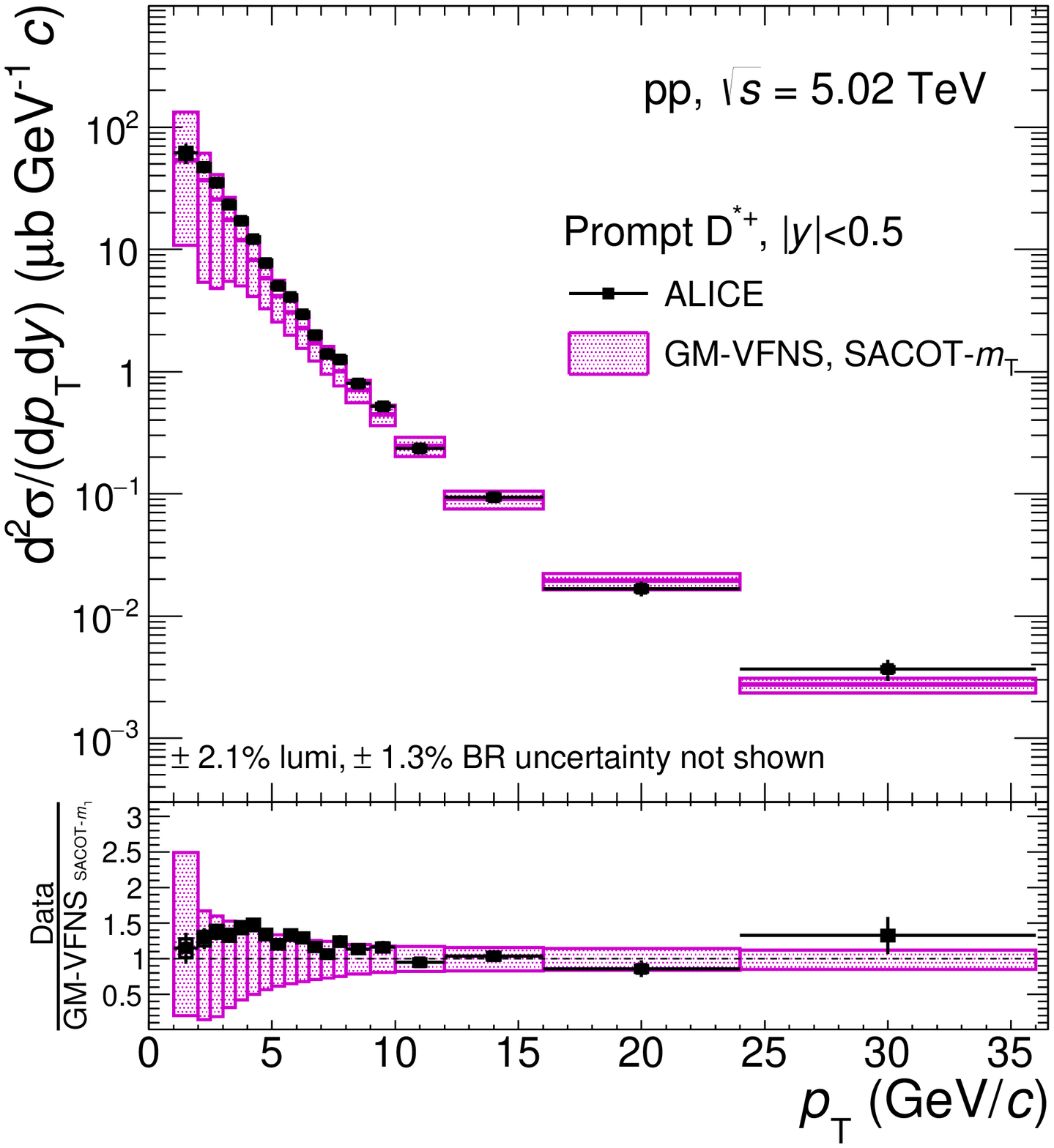}
 \caption{ \label{fig:dsdptppDstar}   
        $\pt$-differential production cross sections for prompt $\Dstar$ meson compared to pQCD calculations: FONLL~\cite{Cacciari:1998it,Cacciari:2012ny}, GM-VFNS(mod-$\mu_{\rm R,F}$)~\cite{Benzke:2017yjn,Kramer:2017gct}, GM-VFNS(SACOT-$m_{\rm T}$)~\cite{Helenius:2018uul}, and $k_{\rm T}$-factorisation~\cite{Maciula:2018iuh}. The ratios of the data to the theoretical predictions are shown in the lower part of each panel.}
 \end{center}
 \end{figure}
 
 \begin{figure}[!th]
     \begin{center}
\includegraphics[width=0.48\textwidth]{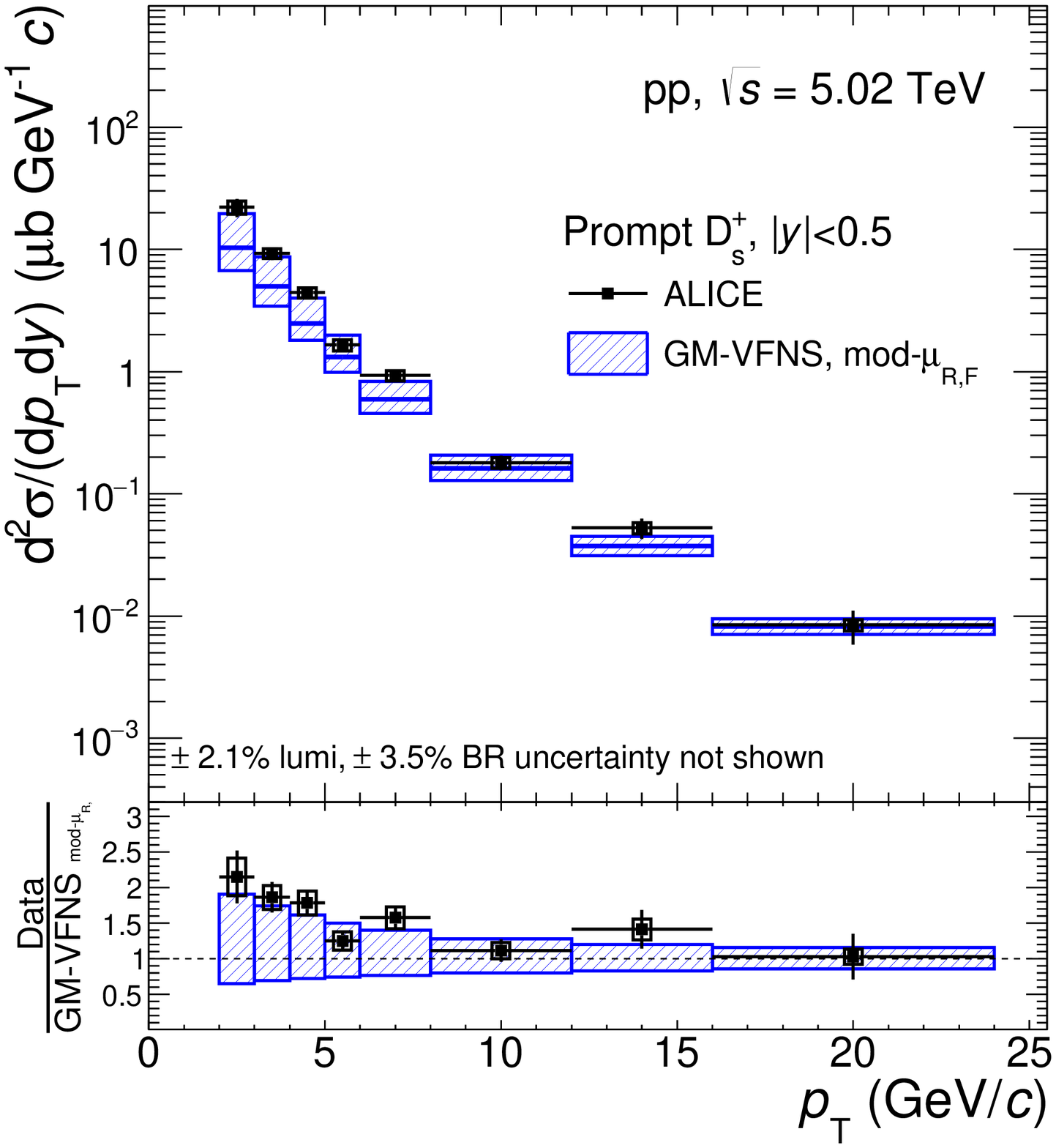}
\includegraphics[width=0.48\textwidth]{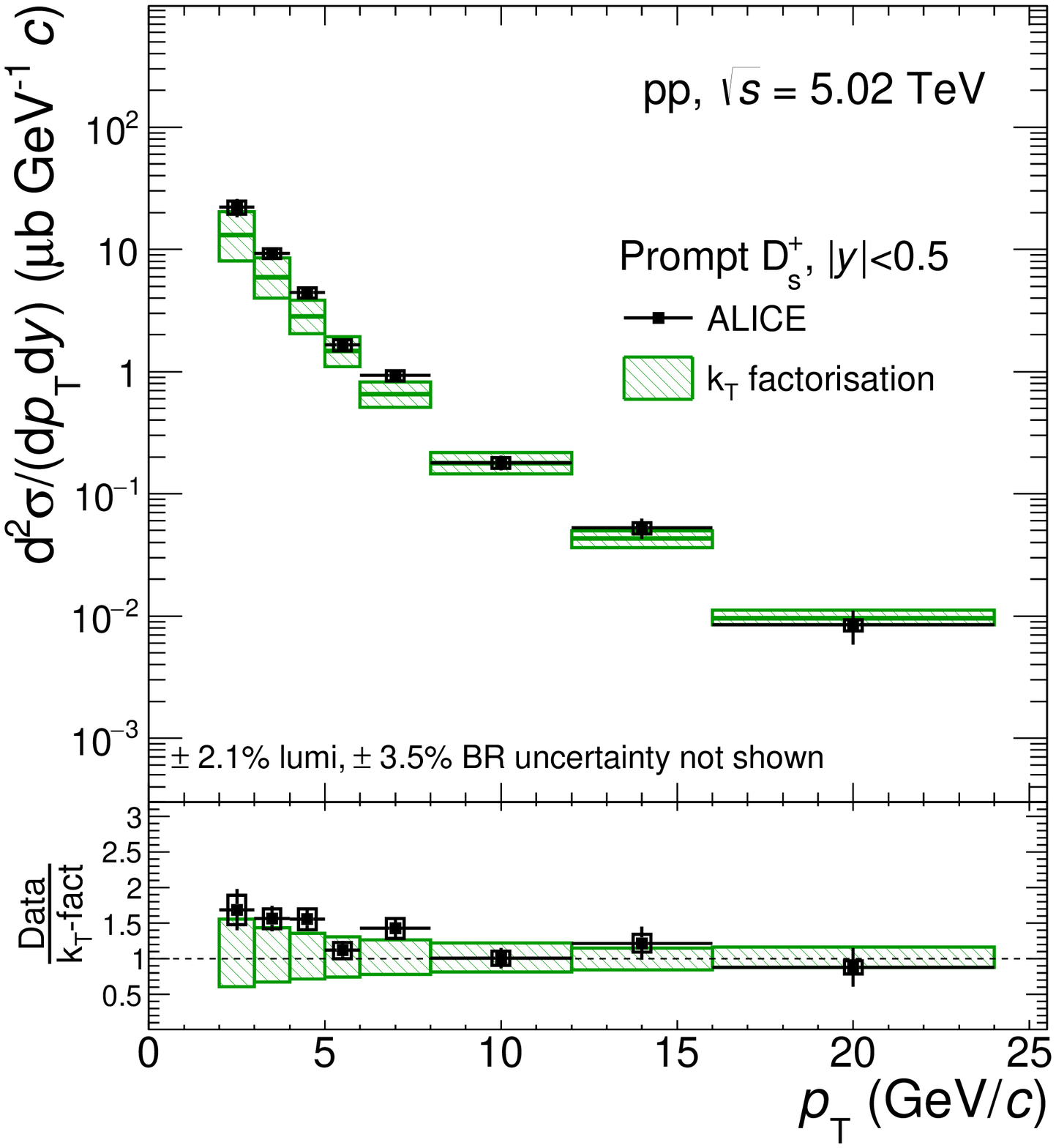} 
\includegraphics[width=0.48\textwidth]{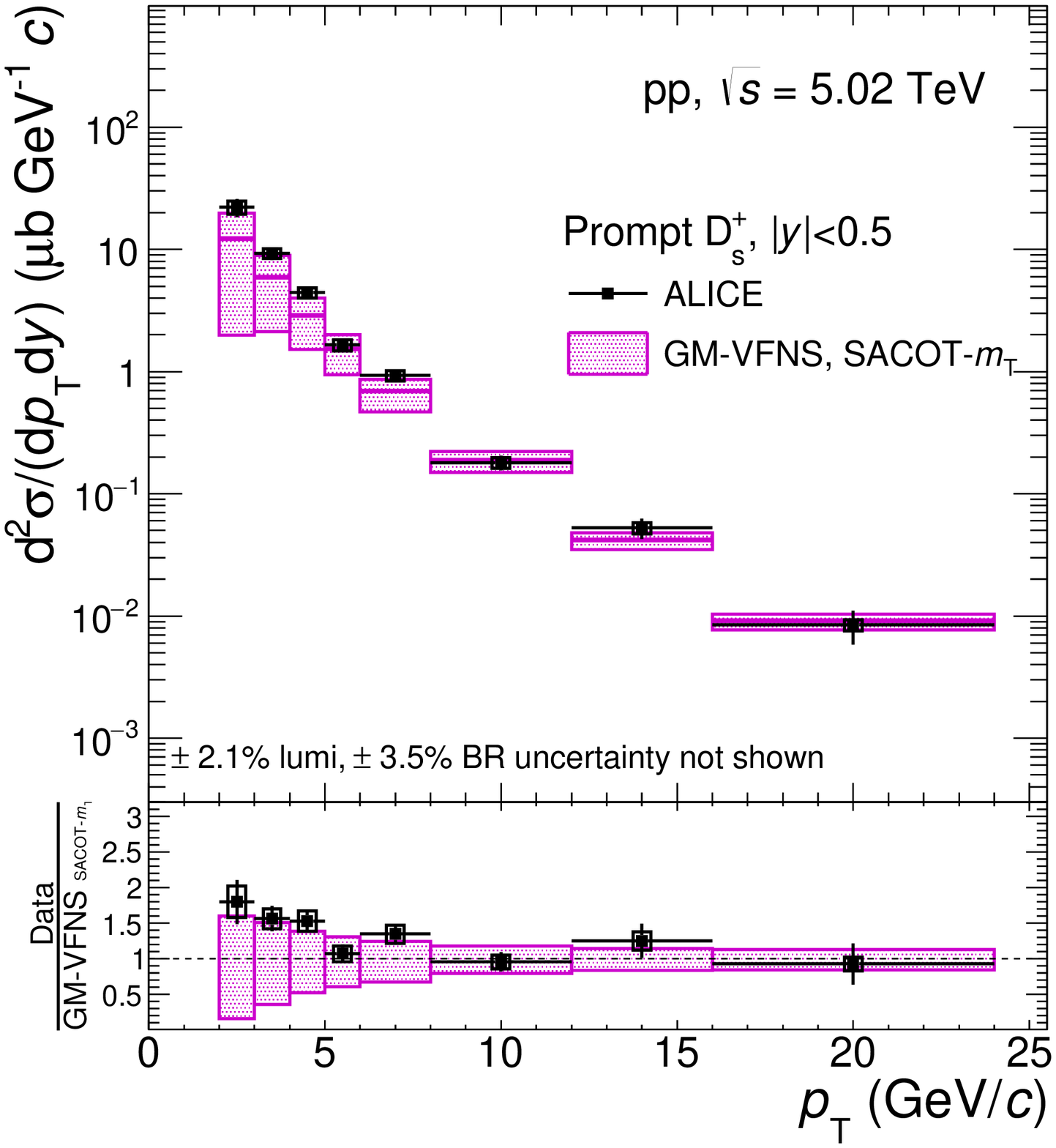}    
       \caption{ \label{fig:dsdptppDs}   
        $\pt$-differential production cross sections for prompt $\Ds$ meson compared to GM-VFNS(mod-$\mu_{\rm R,F}$)~\cite{Benzke:2017yjn,Kramer:2017gct}, GM-VFNS(SACOT-$m_{\rm T}$)~\cite{Helenius:2018uul}, and $k_{\rm T}$-factorisation~\cite{Maciula:2018iuh} pQCD calculations. The ratios of the data to the theoretical predictions are shown in the lower part of each panel.}
 \end{center}
 \end{figure}
   
In Figs.~\ref{fig:dsdptppDzero},~\ref{fig:dsdptppDplus},~\ref{fig:dsdptppDstar} and~\ref{fig:dsdptppDs} the measured prompt $\Dzero$, $\Dplus$, $\Dstar$, $\Ds$-meson $\pt$-differential cross sections are compared with results of pQCD calculations performed with different schemes: FONLL~\cite{Cacciari:1998it,Cacciari:2012ny} (not available for the $\Ds$ meson), two calculations using the GM-VNFS framework with different prescriptions to regulate the divergences at small transverse momentum, dubbed as GM-VFNS(mod-$\mu_{\rm R,F}$)~\cite{Benzke:2017yjn,Kramer:2017gct} and GM-VFNS(SACOT-$m_{\rm T}$)~\cite{Helenius:2018uul}, and a calculation based on $k_{\rm T}$-factorisation~\cite{Maciula:2018iuh}. 
The GM-VFNS(mod-$\mu_{\rm R,F}$) calculations were performed with a different choice of the factorisation and renormalisation scales ${\mu_{\rm F}}$ and ${\mu_{\rm R}}$ with respect to the GM-VFNS predictions of Ref.~\cite{Kniehl:2012ti} that were compared in Ref.~\cite{Acharya:2017jgo} to the cross sections measured at $\sqrt{s}=7~\tev$. With this modification of QCD scale, the calculations could be extended to lower $\pt$.
In GM-VFNS(SACOT-$m_{\rm T}$), the divergences of the heavy-quark PDFs and light-parton fragmentation functions at low $\pt$ are regulated by the heavy-quark mass, thus allowing the calculation of the D-meson cross section down to $\pt=0$.
Note also that the authors of the $k_{\rm T}$-factorisation calculations changed 
the treatment of the running strong coupling constant $\alpha_{\rm S}$ and the 
gluon distributions~\cite{Maciula:2018iuh}, with respect to the predictions
shown in Ref.~\cite{Acharya:2017jgo}.
In GM-VFNS(mod-$\mu_{\rm R,F}$) the value of charm mass is set to 1.3~$\gev/c^{2}$, while in FONLL, GM-VFNS(SACOT-$m_{\rm T}$) and $k_{\rm T}$-factorisation predictions the mass is set to 1.5~$\gev/c^{2}$. 
The four frameworks utilise different sets of PDFs (CTEQ6.6~\cite{Pumplin:2002vw}, CTEQ14~\cite{Dulat:2015mca}, NNPDF3.1~\cite{Ball:2017nwa} and MMHT2014~\cite{Harland-Lang:2014zoa} for FONLL, GM-VFNS(mod-$\mu_{\rm R,F}$), GM-VFNS(SACOT-$m_{\rm T}$) and $k_{\rm T}$-factorisation, respectively) and different fragmentation functions.
The theoretical uncertainties are estimated by varying the factorisation and
renormalisation scales in FONLL, GM-VFNS(SACOT-$m_{\rm T}$) and $k_{\rm T}$-factorisation, while only the renormalisation scale  ${\mu_{\rm R}}$ is varied in
GM-VFNS(mod-$\mu_{\rm R,F}$).
In FONLL and $k_{\rm T}$-factorisation calculations the charm-quark mass is also varied. 
The uncertainties on the PDFs are included in the GM-VFNS(SACOT-$m_{\rm T}$) and FONLL predictions.
The theoretical calculations are performed in the same $\pt$ intervals as the measurements, except for the first bin of the $\Dzero$ prediction with GM-VFNS(mod-$\mu_{\rm R,F}$) that starts from 0.1~$\gev/c$.
The results of these calculations are shown as filled boxes spanning the theoretical uncertainties
and a solid line representing the values obtained with the central values of the pQCD parameters.

The measured cross sections of non-strange D mesons are described within uncertainties by FONLL and the two GM-VFNS calculations.
The data lie systematically on the upper edge of the uncertainty band of the FONLL predictions. 
For the two calculations in the GM-VFNS framework, the central values of the predictions tend to underestimate the data at low and intermediate $\pt$ and to overestimate them at high $\pt$. 
The $k_{\rm T}$-factorisation predictions describe the data at low and intermediate $\pt$, but overshoots them for $\pt > 7~\gev/c$. 
The $\Ds$-meson production tends to be underestimated by the three pQCD calculations in the measured $\pt$ range. 
 
\begin{figure}[!th]
\begin{center}
 \includegraphics[width=0.7\textwidth]{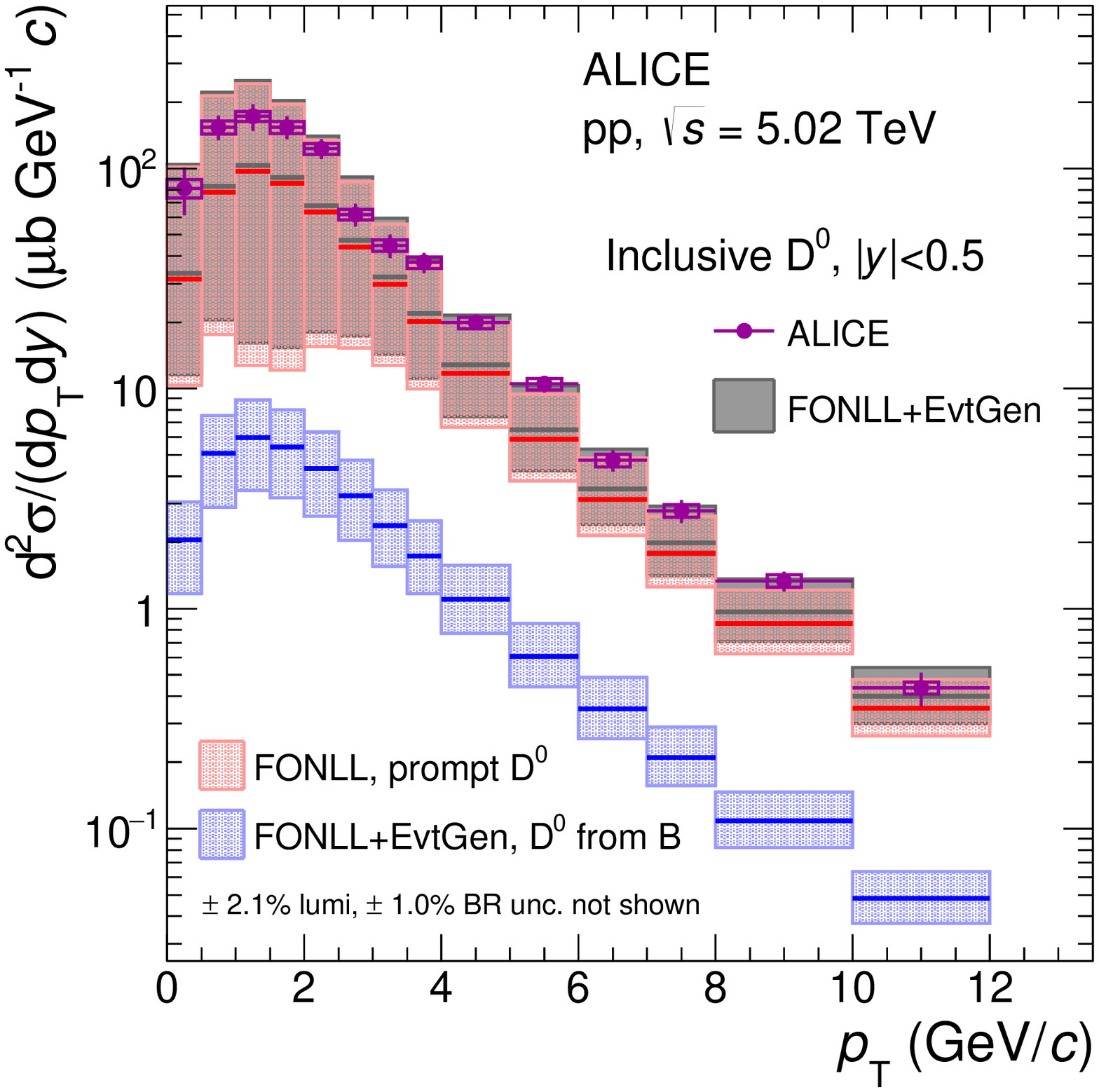}
\caption{ \label{fig:CrossSecD0Inclusive}
Inclusive $\Dzero$ mesons (including also $\Dzero$ mesons from beauty-hadron decays) in $|y|<0.5$ in pp collisions at $\sqrts =5.02~\tev$, from the analysis without decay-vertex reconstruction, compared to FONLL pQCD calculations~\cite{Cacciari:1998it,Cacciari:2012ny} with the $\mathrm{B} \rightarrow \mathrm{D} + X$ decay kinematics from the EvtGen package~\cite{Lange:2001uf} (grey boxes). The contributions of prompt $\Dzero$ from FONLL (red) and $\Dzero$ from B-meson decays from FONLL+EvtGen (blue) are also shown separately. The vertical error bars and the empty boxes represent the statistical and systematic uncertainties, respectively.}
\end{center}
\end{figure}

The analysis without decay-vertex reconstruction provides also a direct 
measurement of the inclusive $\Dzero$-meson cross section because no selections are applied on the decay topology, which alter the fraction of prompt and feed-down D mesons.  
The inclusive $\Dzero$-meson cross section is shown in Fig.~\ref{fig:CrossSecD0Inclusive} and compared with results from FONLL calculations~\cite{Cacciari:1998it,Cacciari:2012ny} with the $\mathrm{B} \rightarrow \mathrm{D} + X$ decay kinematics from the EvtGen package~\cite{Lange:2001uf}. 
The contributions of prompt $\Dzero$-meson poduction from FONLL and $\Dzero$ mesons from B-meson decays from FONLL+EvtGen are also shown separately.
The measured cross sections are described by the calculation within the theoretical uncertainties, with the central value of the prediction lying below the data in all the $\pt$ intervals, similarly to what observed for prompt D mesons.

The mean $\pt$ of prompt $\Dzero$ mesons, $\meanpt$, was evaluated for $\pt>$0 with a fit of the prompt $\Dzero$-meson cross section, that is measured down to $\pt$ = 0, using a power-law function, as was done in Ref.~\cite{Acharya:2017jgo}. 
The result is:
\begin{equation}
\meanpt^{\rm prompt\,D^0}_{\rm pp,\,5.02\,TeV} = 2.06 \pm 0.03\,({\rm stat.})\,\pm 0.03\,({\rm syst.})~\gev/c\,,
\end{equation}
which is slightly smaller than the one computed for pp collisions at $\sqrts=7~\tev$~\cite{Acharya:2017jgo}: 
\begin{equation}
\meanpt^{\rm prompt\,D^0}_{\rm pp,\,7\,TeV} = 2.19 \pm 0.06\,({\rm stat.})\,\pm 0.04\,({\rm syst.})~\gev/c\,.
\end{equation}

The systematic uncertainty on the $\meanpt$ was estimated as described in 
Refs.~\cite{Adam:2016ich,Acharya:2017jgo}.  The contributions due to the correlated and uncorrelated systematic uncertainties on the measured $\pt$-differential cross section were taken into account separately and the contribution due to the choice of the fit function has been estimated by comparing results obtained using different functions and using a method based on direct calculations of  $\meanpt$ from the data points.

\subsection{D-meson cross-section ratios}
\begin{figure}[!th]
\begin{center}
\includegraphics[width=\textwidth]{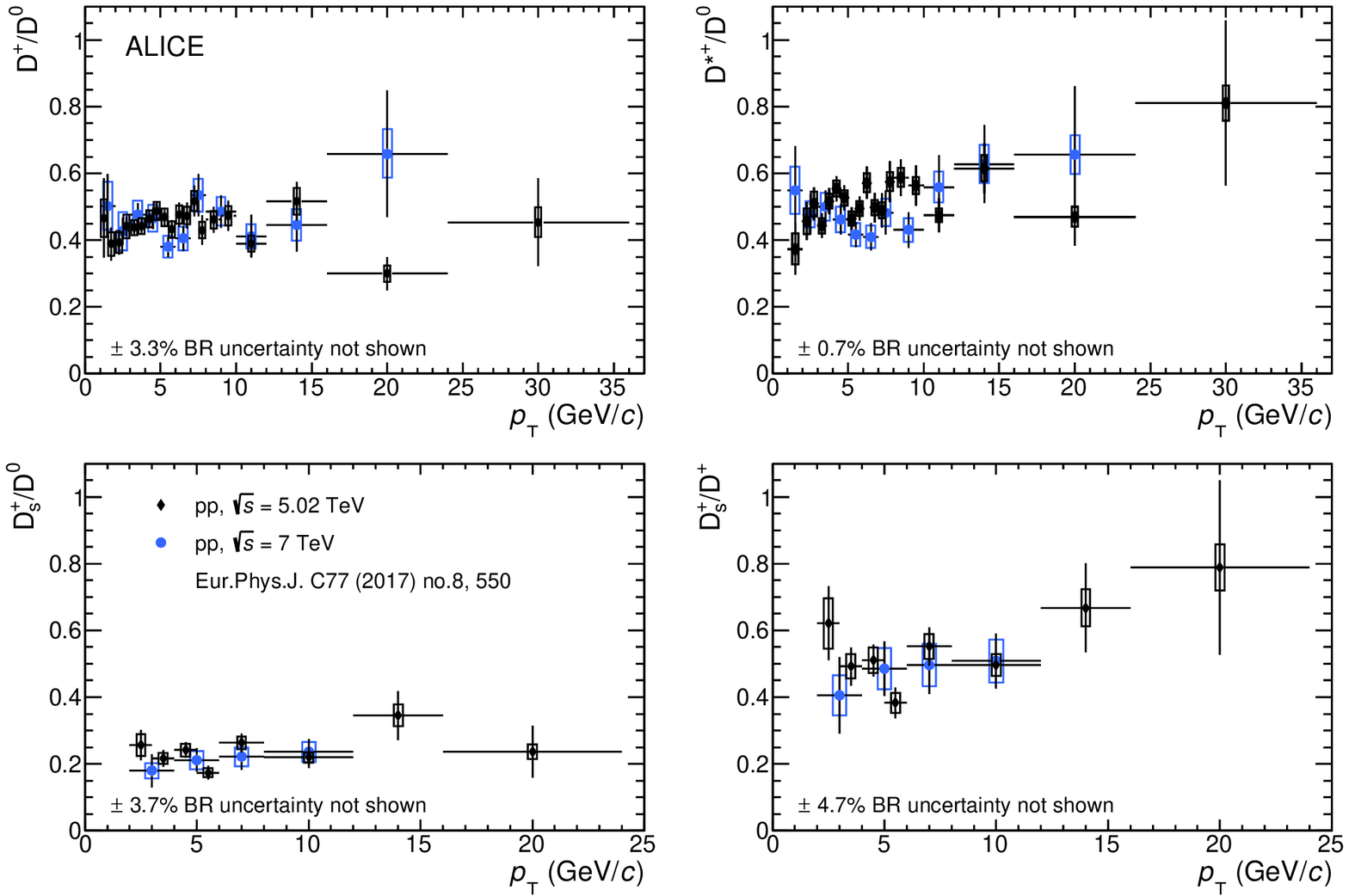}
\caption{Ratios of D-meson production cross sections as a function of $\pt$ in pp collisions at $\sqrts=5.02$~TeV and $\sqrts=7$~TeV~\cite{Acharya:2017jgo}.}
\label{fig:DratiosVsPt}
\end{center}
\end{figure}

The ratios of the $\pt$-differential cross sections of prompt $\Dzero$, $\Dplus$, $\Dstar$, and $\Ds$ mesons in pp collisions at $\sqrts=5.02$~TeV are reported in Fig.~\ref{fig:DratiosVsPt}.
In the evaluation of the systematic uncertainties on these ratios, 
the sources of correlated and uncorrelated systematic effects were treated 
separately. 
In particular, the contributions of the yield extraction and cut efficiency 
were considered as uncorrelated, while those of the feed-down from 
beauty-hadron decays and the tracking efficiency were treated as fully 
correlated among the different D-meson species.
The measured D-meson cross-section ratios do not show a significant $\pt$ dependence within 
the experimental uncertainties, thus suggesting no discernible difference 
between the fragmentation functions of charm quarks to pseudoscalar 
($\Dzero$, $\Dplus$, and $\Ds$) and vector ($\Dstar$) mesons and to strange and 
non-strange mesons. The results are compatible within uncertainties with the ratios measured in pp collisions at $\sqrts=7$~TeV~\cite{Acharya:2017jgo}\footnote{The cross section for $\Dzero$ and $\Dplus$ mesons in pp collisions at $\sqrts=7$~TeV were updated with respect to Ref.~\cite{Acharya:2017jgo} to account for the change of the world-average BR of $\DtoKpi$ and $\DtoKpipi$ from ($3.93\% \pm 0.04$) to ($3.89\% \pm 0.04$), and from ($9.46\% \pm 0.24$) to ($8.98\% \pm 0.28$), respectively.}.


\begin{figure}[!th]
\begin{center}
\includegraphics[width=0.48\textwidth]{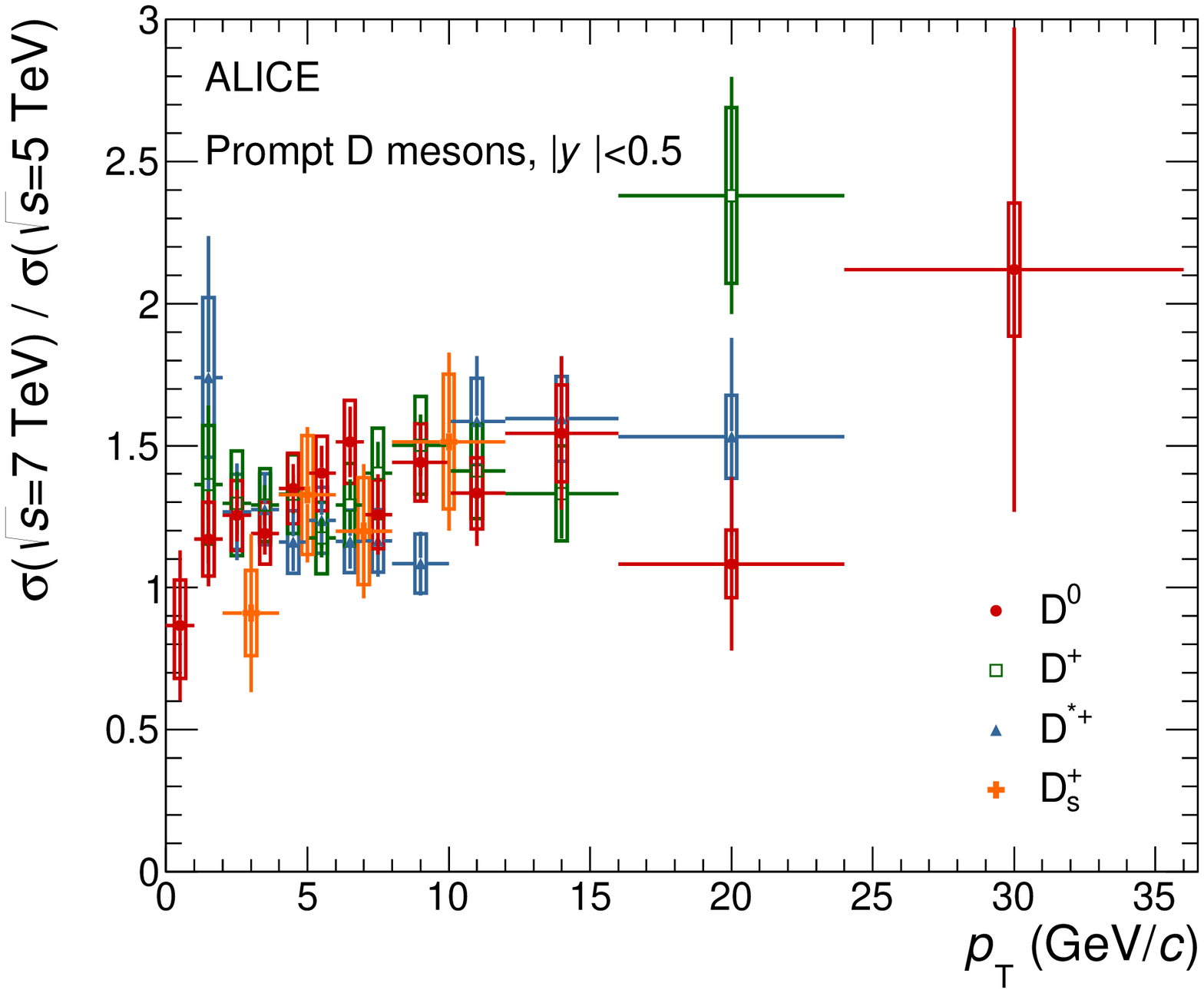}
\includegraphics[width=0.48\textwidth]{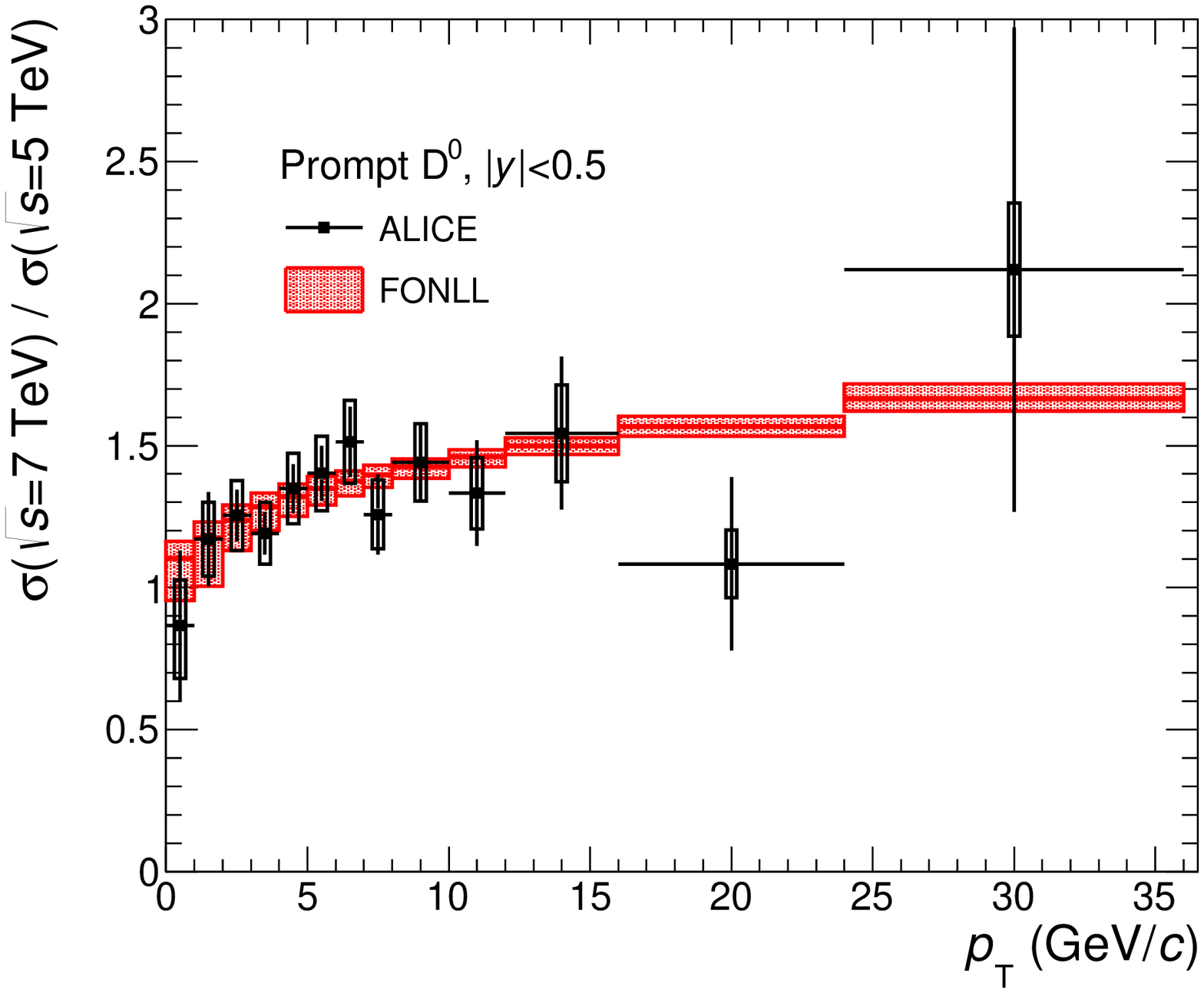}
\caption{Ratios of $\Dzero$, $\Dplus$, $\Dstar$ and $\Ds$-meson production cross sections in pp collisions at $\sqrts=7$~TeV~\cite{Acharya:2017jgo} and $\sqrts=5.02$~TeV as a function of $\pt$ (left panel).  $\Dzero$ ratio compared to FONLL pQCD calculations~\cite{Cacciari:1998it,Cacciari:2012ny} (right panel).}
\label{fig:DEnergyratiosVsPt}
\end{center}
\end{figure}

To study the evolution of prompt D-meson production with the centre-of-mass energy of the collision, the ratios of the production cross sections in pp collisions at $\sqrts=7$~TeV~\cite{Acharya:2017jgo} and $\sqrts=5.02$~TeV were computed for $\Dzero$, $\Dplus$, $\Dstar$ and $\Ds$ mesons.
The systematic uncertainties on the measured ratios were obtained treating the contribution originating from the subtraction of the feed-down from 
beauty-hadron decays as correlated, while all the other systematic uncertainties on the cross sections were propagated as uncorrelated between the measurements at the two different energies, except for the uncertainty on the BR, which cancels out in the ratio.
The results for $\Dzero$, $\Dplus$, $\Dstar$ and $\Ds$ are compared in Fig.~\ref{fig:DEnergyratiosVsPt}, on the left panel. The ratios for the different D-meson species are compatible within uncertainties. 
 In the right panel, the $\Dzero$-meson results are compared to FONLL calculations, which describe consistently the increasing trend as a function of $\pt$ observed in the data. 
In the FONLL predictions, the uncertainties originating from scale variations and from PDFs cancel out to a large extent in the ratio~\cite{Cacciari:2015fta}, thus making the magnitude of the theoretical uncertainties comparable with those of the data.

\begin{figure}[!tb]
\begin{center}
\includegraphics[width=\textwidth]{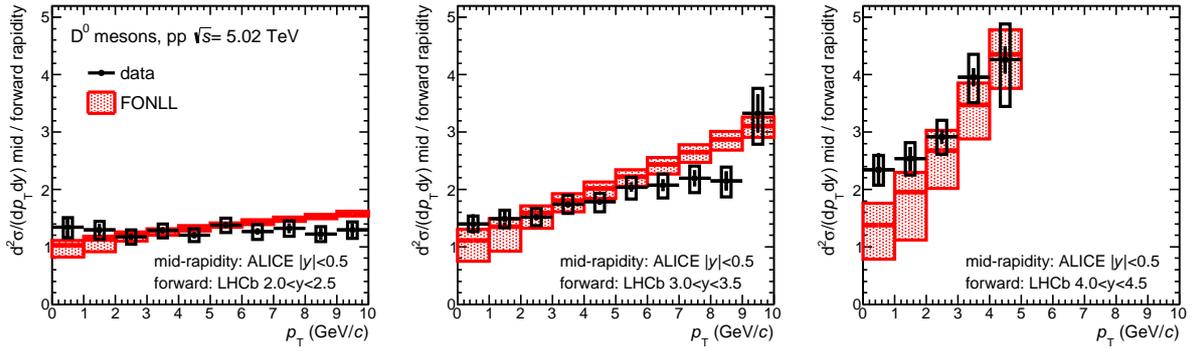}
\caption{Ratios of $\Dzero$-meson production cross section per unit of rapidity at mid-rapidity ($|y|<0.5$) to those measured by the LHCb 
Collaboration~\cite{Aaij:2016jht} in three 
rapidity ranges, $2<y<2.5$ (left panel), 
$3<y<3.5$ (middle panel), and $4<y<4.5$ (right panel), as a function 
of $\pt$. 
The error bars and boxes represent the statistical and systematic uncertainty, respectively.
Predictions from FONLL calculations are compared to the data points.}
\label{fig:CentForwRatios}
\end{center}
\end{figure}

The rapidity dependence of $\Dzero$-meson production in pp collisions at $\sqrts=5~\tev$ can be studied from the ratios between our measurements at midrapidity and the LHCb results in different $y$ intervals at forward rapidity~\cite{Aaij:2016jht}. The precise measurement of the $\Dzero$-meson cross section down to $\pt=0$ presented in this paper, when analysed together with other results at different centre-of-mass energies and rapidities, can provide sensitivity to the gluon PDF at small values of Bjorken-$x$ ($10^{-4}$--$10^{-5}$)~\cite{Cacciari:2015fta}.
In Fig.~\ref{fig:CentForwRatios} the ratios of the $\Dzero$-meson production cross sections per unit of rapidity measured with ALICE at mid-rapidity 
($|y|<0.5$) and by the LHCb collaboration in three rapidity intervals at forward rapidity $2<y<2.5$ (left panel), 
$3<y<3.5$ (middle panel), $4<y<4.5$ (right panel)~\cite{Aaij:2016jht} are shown as a function of $\pt$.
The error bars and boxes represent the uncertainty obtained from the propagation 
of the statistical and systematic uncertainties, respectively, from the $\pt$-differential 
cross sections. 
The systematic uncertainties, including the one on the luminosity determination, were treated as uncorrelated between the ALICE and LHCb results, except for the uncertainty on the BR, which cancels out in the ratio.
The central values and the uncertainties of the FONLL calculations are evaluated as described in Ref.~\cite{Acharya:2017jgo}.
The measured ratios are described by FONLL calculations, shown as red boxes in Fig.~\ref{fig:CentForwRatios}.
Nevertheless the comparison seems to hint at a different slope in data with respect to FONLL, since at low (high) $\pt$ the data tend to stay above (below) the FONLL central values, in all rapidity intervals. 


\subsection{Transverse momentum-integrated cross sections and ratios}
The visible production cross sections of prompt D mesons were evaluated by integrating the $\pt$-differential cross sections over the narrower $\pt$ intervals of the $\Dplus$, $\Dstar$, and $\Ds$-meson measurements, in the measured $\pt$ range. The results are reported in Table~\ref{tab:ptintegcs}. 
The systematic uncertainty was evaluated by propagating all the uncertainties as correlated among $\pt$ intervals, except for the yield extraction uncertainty which is treated as uncorrelated owing to the bin-by-bin variation, significant especially at low $\pt$, of S/B and background invariant-mass shape.

\begin{table}
\centering
\begin{tabular}{c|c|l} 
 & Kinematic range & Visible cross section ($\mub$) \\
\hline
\rule{0pt}{12pt} 
$\Dzero$  & $0<\pt<36~\GeV/c$ & $447 \pm 20 ({\rm stat}) \pm 30 ({\rm syst}) \pm 9 ({\rm lumi}) \pm 5 ({\rm BR})$ \\[1ex]
\hline
\rule{0pt}{12pt} 
$\Dplus$  & $1<\pt<36~\GeV/c$ & $144 \pm 10 ({\rm stat}) \pm 10 ({\rm syst}) \pm 3 ({\rm lumi}) \pm 4 ({\rm BR})$\\[1ex]
\hline
\rule{0pt}{12pt} 
$\Dstar$  & $1<\pt<36~\GeV/c$ & $143 \pm 12 ({\rm stat}) \pm 11 ({\rm syst}) \pm 3 ({\rm lumi}) \pm 2 ({\rm BR})$\\[1ex]
\hline
\rule{0pt}{12pt} 
$\Ds$     & $2<\pt<24~\GeV/c$ & $\phantom{0}40 \pm \phantom{0}4 ({\rm stat}) \pm \phantom{0}4 ({\rm syst}) \pm 1 ({\rm lumi}) \pm 1 ({\rm BR})$\\[1ex]
\hline
\end{tabular}
\caption{Visible production cross sections of prompt D mesons in $|y| < 0.5$ in pp collisions at $\sqrts = 5.02~\TeV$.}
\label{tab:ptintegcs}
\end{table}

The ratios of the $\pt$-integrated yields of the different D-meson species were 
computed from the cross sections integrated over the common $\pt$ range.
The systematic uncertainties on the ratios were computed treating the BR, yield
extraction and cut efficiency uncertainties as uncorrelated among the different species and the other sources as correlated.
The results are reported in Table~\ref{tab:ptintegrat}.

The measured ratios are compatible within uncertainties with the results at
$\sqrts=2.76~\TeV$ and $\sqrts=7~\TeV$~\cite{Abelev:2012vra,Acharya:2017jgo} and with the measurements of
the LHCb collaboration at forward rapidity ($2.0<y<4.5$) at three 
different collision energies $\sqrts=5.02$, $7$, and 
$13~\TeV$~\cite{Aaij:2013mga,Aaij:2016jht,Aaij:2015bpa}.

\begin{table}
\centering
\begin{tabular}{c|c|l} 
 & Kinematic range & Production cross section ratio \\
\hline
\rule{0pt}{12pt} 
$\sigma(\Dplus) / \sigma(\Dzero)$ & $1<\pt<36~\GeV/c$ & $0.43 \pm 0.04 ({\rm stat}) \pm 0.03 ({\rm syst}) \pm 0.01 ({\rm BR})$\\[1ex]
$\sigma(\Dstar) / \sigma(\Dzero)$ & $1<\pt<36~\GeV/c$ & $0.43 \pm 0.04 ({\rm stat}) \pm 0.03 ({\rm syst}) \pm 0.003 ({\rm BR})$\\[1ex]
$\sigma(\Ds) / \sigma(\Dzero)$    & $2<\pt<24~\GeV/c$ & $0.24 \pm 0.02 ({\rm stat}) \pm 0.02 ({\rm syst}) \pm 0.01 ({\rm BR})$\\[1ex]
$\sigma(\Ds) / \sigma(\Dplus)$    & $2<\pt<24~\GeV/c$ & $0.56 \pm 0.06 ({\rm stat}) \pm 0.05 ({\rm syst}) \pm 0.03 ({\rm BR})$\\[1ex]
\hline
\end{tabular}
\caption{Ratios of the measured $\pt$-integrated cross sections of prompt D mesons in $|y| < 0.5$ in pp collisions at $\sqrts = 5.02~\TeV$.}
\label{tab:ptintegrat}
\end{table}


The production cross sections per unit of rapidity, ${\rm d}\sigma/{\rm d}y$, 
at mid-rapidity were computed for each D-meson species by extrapolating the 
visible cross section to the full $\pt$ range.
The extrapolation factor for a given D-meson species was computed using the FONLL central parameters to evaluate the ratio 
between the total production cross section in $|y|<0.5$ and that 
in the experimentally covered phase space.
It was verified that the extrapolation factors computed with FONLL were compatible with those resulting from  GM-VFNS calculations. 
The systematic uncertainty on the extrapolation factor was estimated as proposed in Ref.~\cite{Cacciari:2012ny}, 
considering sources due to (i) the CTEQ6.6 PDFs uncertainties~\cite{Pumplin:2002vw},  (ii) the variation of the charm-quark 
mass and (iii) the renormalisation and factorisation scales in the FONLL 
calculation. 
For $\Dzero$ mesons, for which the measurement extends down to $\pt=0$, the extrapolation factor accounts only for the very small contribution of D mesons with
$\pt>36~\GeV/c$ and therefore its value is very close to unity with 
negligible uncertainty.
The FONLL predictions are not available for $\Ds$ mesons, hence in this case
the central value of the extrapolation factor was computed as described in Ref.~\cite{Acharya:2017jgo}, combining the prediction based on the $\pt$-differential cross section of charm 
quarks from FONLL, the fractions $f({\rm c\rightarrow \Ds})$ and 
$f({\rm c\rightarrow D_{\rm s}^{*+}})$ from ALEPH~\cite{Barate:1999bg}, and the 
fragmentation functions from Ref.~\cite{Braaten:1994bz}, which have one parameter, 
$r$, that was set to 0.1 as done in FONLL~\cite{Cacciari:2003zu}.
An additional contribution to the systematic uncertainty was assigned
based on the envelope of the results obtained using the 
FONLL $\pt$-differential cross sections of non-strange D mesons to compute the $\Ds$-meson extrapolation factor.
The computed extrapolation factors and the prompt D-meson 
production cross sections per unit of rapidity ${\rm d}\sigma/{\rm d}y$ in $|y| < 0.5$, are 
presented in Table~\ref{tab:dsdy}.

\begin{table}
\centering
\begin{tabular}{c|c|l} 
 & Extr. factor to $\pt>0$ & ${\rm d}\sigma/{\rm d}y \mid_{|y|<0.5}$ ($\mub$) \\
\hline
\rule{0pt}{12pt} 
$\Dzero$  & $1.0000 ^{+0.0003}_{-0.0000}$ & $447 \pm 20 ({\rm stat}) \pm 30 ({\rm syst}) \pm 9 ({\rm lumi}) \pm 5 ({\rm BR})$\\[1ex]
\hline
\rule{0pt}{12pt} 
$\Dplus$  & $1.28 ^{+0.35}_{-0.09}$ & $184 \pm 13 ({\rm stat}) \pm 13 ({\rm syst}) \pm 4 ({\rm lumi}) \pm 6 ({\rm BR})^{+50}_{-13} ({\rm extrap})$ \\[1ex]
\hline
\rule{0pt}{12pt} 
$\Dstar$  & $1.24 ^{+0.34}_{-0.08}$ & $178 \pm 15 ({\rm stat}) \pm 14 ({\rm syst}) \pm 4 ({\rm lumi}) \pm 2 ({\rm BR})^{+48}_{-12} ({\rm extrap})$\\[1ex]
\hline
\rule{0pt}{12pt} 
$\Ds$     & $2.35^{+0.78}_{-0.66}$ &  $\phantom{0}95 \pm \phantom{0}9 ({\rm stat}) \pm 10 ({\rm syst}) \pm 2 ({\rm lumi}) \pm 3 ({\rm BR})^{+31}_{-26} ({\rm extrap})$\\[1ex]
\hline
\end{tabular}
\caption{Production cross sections of prompt D mesons in $|y| < 0.5$ and full $\pt$ range in pp collisions at $\sqrts = 5.02~\TeV$.}
\label{tab:dsdy}
\end{table}

In Ref.~\cite{Acharya:2017jgo}, the ${\rm c\overline{c}}$ production cross section per unit of rapidity at mid-rapidity ($|y|<0.5$) and the total charm production cross sections in pp collisions at $\sqrt{s}=7~\TeV$ were reported.
They were computed from the prompt $\Dzero$-meson production cross section, which was divided by the fraction of charm quarks hadronising into $\Dzero$ mesons, $f(\rm c\rightarrow\Dzero)=0.542 \pm 0.024$, derived in Ref.~\cite{Gladilin:2014tba} by averaging the measurements in e$^+$e$^-$ collisions at LEP.
However, recent measurements of the $\rm \Lambda^{+}_{c}$ baryon production cross section in pp collisions at $\sqrt{s}=7~\TeV$ and in p--Pb collisions
at $\sqrt{s}=5.02~\TeV$~\cite{Acharya:2017kfy} show a significant enhancement of the $\rm \Lambda^{+}_{c}$/$\Dzero$ ratio for $\pt>$1 GeV/$c$ as compared to
the values measured in e$^+$e$^-$ and ep collisions at lower centre-of-mass energies.
This suggests that the fragmentation fractions of charm quarks into charmed baryons in pp collisions at LHC energies might differ significantly from the 
LEP results reported in Ref.~\cite{Gladilin:2014tba} and that measurements of charmed-baryon production cross sections in pp collisions at $\sqrt{s}=5.02~\TeV$ 
are needed for an accurate calculation of the charm production cross section.


\section{Summary}
\label{sec:summary}
We have reported the measurement of the inclusive $\pt$-differential
production cross sections of prompt $\Dzero$, $\Dplus$, $\Dstar$, and $\Ds$ 
mesons at mid-rapidity ($|y|<0.5$) in pp collisions at a centre-of-mass energy of
$\sqrt{s}=5.02~\tev$, obtained with the data collected at the end of 2017 with the ALICE detector. 
The measurement was performed in the transverse-momentum range $0<\pt<36~\GeV/c$
for $\Dzero$, $1<\pt<36~\GeV/c$ for $\Dplus$ and $\Dstar$, 
and $2<\pt<24~\GeV/c$ for $\Ds$ mesons. It is measured in finer $\pt$ bins with respect to the previous measurements at $\sqrt{s}=7~\tev$~\cite{Acharya:2017jgo}, providing a more detailed description of the cross-section $\pt$ shape. The results were compared and found compatible with different pQCD calculations performed with different schemes: FONLL~\cite{Cacciari:1998it,Cacciari:2012ny}, two calculations using the GM-VNFS framework with different prescriptions ~\cite{Benzke:2017yjn,Kramer:2017gct, Helenius:2018uul}, and a calculation based on $k_{\rm T}$-factorisation~\cite{Maciula:2018iuh}.  
The ratios of $\Dzero$-meson production cross sections measured with ALICE and LHCb 
in different rapidity intervals were compatible with FONLL calculations, indicating a slightly smaller slope in data with respect to theoretical predictions. 
The ratios of the cross sections of $\Dzero$, $\Dplus$, and
$\Dstar$ mesons at $\sqrts=7$~TeV~\cite{Acharya:2017jgo} and $\sqrts=5.02$~TeV are consistent with FONLL pQCD calculations. The ratios of the $\pt$-differential cross sections of $\Dzero$, $\Dplus$, 
$\Dstar$, and $\Ds$ mesons were found to be compatible within uncertainties with the D-meson cross-section ratios measured in pp collisions at $\sqrts=7$~TeV~\cite{Acharya:2017jgo}.
The new measurement will allow for a more accurate determination of the nuclear modification factor $R_{\rm pA}$ in p--Pb collisions and $\Raa$ in Pb--Pb collisions at $\sqrt{s_{\rm NN}}=5.02$~TeV, due to the larger statistics available and since it is performed at the same centre-of-mass energy of the other collision systems.


%
%
\newenvironment{acknowledgement}{\relax}{\relax}
\begin{acknowledgement}
\section*{Acknowledgements}

The ALICE Collaboration would like to thank all its engineers and technicians for their invaluable contributions to the construction of the experiment and the CERN accelerator teams for the outstanding performance of the LHC complex.
The ALICE Collaboration gratefully acknowledges the resources and support provided by all Grid centres and the Worldwide LHC Computing Grid (WLCG) collaboration.
The ALICE Collaboration acknowledges the following funding agencies for their support in building and running the ALICE detector:
A. I. Alikhanyan National Science Laboratory (Yerevan Physics Institute) Foundation (ANSL), State Committee of Science and World Federation of Scientists (WFS), Armenia;
Austrian Academy of Sciences and Nationalstiftung f\"{u}r Forschung, Technologie und Entwicklung, Austria;
Ministry of Communications and High Technologies, National Nuclear Research Center, Azerbaijan;
Conselho Nacional de Desenvolvimento Cient\'{\i}fico e Tecnol\'{o}gico (CNPq), Universidade Federal do Rio Grande do Sul (UFRGS), Financiadora de Estudos e Projetos (Finep) and Funda\c{c}\~{a}o de Amparo \`{a} Pesquisa do Estado de S\~{a}o Paulo (FAPESP), Brazil;
Ministry of Science \& Technology of China (MSTC), National Natural Science Foundation of China (NSFC) and Ministry of Education of China (MOEC) , China;
Croatian Science Foundation and Ministry of Science and Education, Croatia;
Centro de Aplicaciones Tecnol\'{o}gicas y Desarrollo Nuclear (CEADEN), Cubaenerg\'{\i}a, Cuba;
Ministry of Education, Youth and Sports of the Czech Republic, Czech Republic;
The Danish Council for Independent Research | Natural Sciences, the Carlsberg Foundation and Danish National Research Foundation (DNRF), Denmark;
Helsinki Institute of Physics (HIP), Finland;
Commissariat \`{a} l'Energie Atomique (CEA) and Institut National de Physique Nucl\'{e}aire et de Physique des Particules (IN2P3) and Centre National de la Recherche Scientifique (CNRS), France;
Bundesministerium f\"{u}r Bildung, Wissenschaft, Forschung und Technologie (BMBF) and GSI Helmholtzzentrum f\"{u}r Schwerionenforschung GmbH, Germany;
General Secretariat for Research and Technology, Ministry of Education, Research and Religions, Greece;
National Research, Development and Innovation Office, Hungary;
Department of Atomic Energy Government of India (DAE), Department of Science and Technology, Government of India (DST), University Grants Commission, Government of India (UGC) and Council of Scientific and Industrial Research (CSIR), India;
Indonesian Institute of Science, Indonesia;
Centro Fermi - Museo Storico della Fisica e Centro Studi e Ricerche Enrico Fermi and Istituto Nazionale di Fisica Nucleare (INFN), Italy;
Institute for Innovative Science and Technology , Nagasaki Institute of Applied Science (IIST), Japan Society for the Promotion of Science (JSPS) KAKENHI and Japanese Ministry of Education, Culture, Sports, Science and Technology (MEXT), Japan;
Consejo Nacional de Ciencia (CONACYT) y Tecnolog\'{i}a, through Fondo de Cooperaci\'{o}n Internacional en Ciencia y Tecnolog\'{i}a (FONCICYT) and Direcci\'{o}n General de Asuntos del Personal Academico (DGAPA), Mexico;
Nederlandse Organisatie voor Wetenschappelijk Onderzoek (NWO), Netherlands;
The Research Council of Norway, Norway;
Commission on Science and Technology for Sustainable Development in the South (COMSATS), Pakistan;
Pontificia Universidad Cat\'{o}lica del Per\'{u}, Peru;
Ministry of Science and Higher Education and National Science Centre, Poland;
Korea Institute of Science and Technology Information and National Research Foundation of Korea (NRF), Republic of Korea;
Ministry of Education and Scientific Research, Institute of Atomic Physics and Ministry of Research and Innovation and Institute of Atomic Physics, Romania;
Joint Institute for Nuclear Research (JINR), Ministry of Education and Science of the Russian Federation, National Research Centre Kurchatov Institute, Russian Science Foundation and Russian Foundation for Basic Research, Russia;
Ministry of Education, Science, Research and Sport of the Slovak Republic, Slovakia;
National Research Foundation of South Africa, South Africa;
Swedish Research Council (VR) and Knut \& Alice Wallenberg Foundation (KAW), Sweden;
European Organization for Nuclear Research, Switzerland;
National Science and Technology Development Agency (NSDTA), Suranaree University of Technology (SUT) and Office of the Higher Education Commission under NRU project of Thailand, Thailand;
Turkish Atomic Energy Agency (TAEK), Turkey;
National Academy of  Sciences of Ukraine, Ukraine;
Science and Technology Facilities Council (STFC), United Kingdom;
National Science Foundation of the United States of America (NSF) and United States Department of Energy, Office of Nuclear Physics (DOE NP), United States of America.   
\end{acknowledgement}

\bibliographystyle{utphys}
\bibliography{Dpp5TeV2017_paper}

\providecommand{\href}[2]{#2}\begingroup\raggedright\begin{thebibliography}{10}

\bibitem{Collins:1989gx}
J.~C. Collins, D.~E. Soper, and G.~F. Sterman, ``{Factorization of Hard
  Processes in QCD}'', \href{http://dx.doi.org/10.1142/9789814503266_0001}{{\em
  Adv. Ser. Direct. High Energy Phys.} {\bfseries 5} (1989) },
\href{http://arxiv.org/abs/hep-ph/0409313}{{\ttfamily arXiv:hep-ph/0409313
  [hep-ph]}}.

\bibitem{Catani:1990eg}
S.~Catani, M.~Ciafaloni, and F.~Hautmann, ``{High-energy factorization and
  small x heavy flavor production}'',
\href{http://dx.doi.org/10.1016/0550-3213(91)90055-3}{{\em Nucl. Phys.}
  {\bfseries B366} (1991) }.

\bibitem{Kniehl:2004fy}
B.~A. Kniehl, G.~Kramer, I.~Schienbein, and H.~Spiesberger, ``{Inclusive D*+-
  production in p anti-p collisions with massive charm quarks}'',
  \href{http://dx.doi.org/10.1103/PhysRevD.71.014018}{{\em Phys. Rev.}
  {\bfseries D71} (2005) },
\href{http://arxiv.org/abs/hep-ph/0410289}{{\ttfamily arXiv:hep-ph/0410289
  [hep-ph]}}.

\bibitem{Kniehl:2005mk}
B.~A. Kniehl, G.~Kramer, I.~Schienbein, and H.~Spiesberger, ``{Collinear
  subtractions in hadroproduction of heavy quarks}'',
  \href{http://dx.doi.org/10.1140/epjc/s2005-02200-7}{{\em Eur. Phys. J.}
  {\bfseries C41} (2005) },
\href{http://arxiv.org/abs/hep-ph/0502194}{{\ttfamily arXiv:hep-ph/0502194
  [hep-ph]}}.

\bibitem{Kniehl:2012ti}
B.~A. Kniehl, G.~Kramer, I.~Schienbein, and H.~Spiesberger, ``{Inclusive
  Charmed-Meson Production at the CERN LHC}'',
  \href{http://dx.doi.org/10.1140/epjc/s10052-012-2082-2}{{\em Eur. Phys. J.}
  {\bfseries C72} (2012) },
\href{http://arxiv.org/abs/1202.0439}{{\ttfamily arXiv:1202.0439 [hep-ph]}}.

\bibitem{Helenius:2018uul}
I.~Helenius and H.~Paukkunen, ``{Revisiting the D-meson hadroproduction in
  general-mass variable flavour number scheme}'',
  \href{http://dx.doi.org/10.1007/JHEP05(2018)196}{{\em JHEP} {\bfseries 05}
  (2018) },
\href{http://arxiv.org/abs/1804.03557}{{\ttfamily arXiv:1804.03557 [hep-ph]}}.

\bibitem{Cacciari:1998it}
M.~Cacciari, M.~Greco, and P.~Nason, ``{The $p_{\rm T}$ spectrum in heavy
  flavor hadroproduction}'',
  \href{http://dx.doi.org/10.1088/1126-6708/1998/05/007}{{\em JHEP} {\bfseries
  05} (1998) },
\href{http://arxiv.org/abs/hep-ph/9803400}{{\ttfamily arXiv:hep-ph/9803400
  [hep-ph]}}.

\bibitem{Cacciari:2012ny}
M.~Cacciari, S.~Frixione, N.~Houdeau, M.~L. Mangano, P.~Nason, and G.~Ridolfi,
  ``{Theoretical predictions for charm and bottom production at the LHC}'',
  \href{http://dx.doi.org/10.1007/JHEP10(2012)137}{{\em JHEP} {\bfseries 10}
  (2012) },
\href{http://arxiv.org/abs/1205.6344}{{\ttfamily arXiv:1205.6344 [hep-ph]}}.

\bibitem{Luszczak:2008je}
M.~Luszczak, R.~Maciula, and A.~Szczurek, ``{Nonphotonic electrons at RHIC
  within k(t)-factorization approach and with experimental semileptonic decay
  functions}'', \href{http://dx.doi.org/10.1103/PhysRevD.79.034009}{{\em Phys.
  Rev.} {\bfseries D79} (2009) },
\href{http://arxiv.org/abs/0807.5044}{{\ttfamily arXiv:0807.5044 [hep-ph]}}.

\bibitem{Maciula:2013wg}
R.~Maciula and A.~Szczurek, ``{Open charm production at the LHC -
  $k_{t}$-factorization approach}'',
  \href{http://dx.doi.org/10.1103/PhysRevD.87.094022}{{\em Phys. Rev.}
  {\bfseries D87} no.~9, (2013) },
\href{http://arxiv.org/abs/1301.3033}{{\ttfamily arXiv:1301.3033 [hep-ph]}}.

\bibitem{Andronic:2015wma}
A.~Andronic {\em et~al.}, ``{Heavy-flavour and quarkonium production in the LHC
  era: from proton-proton to heavy-ion collisions}'',
  \href{http://dx.doi.org/10.1140/epjc/s10052-015-3819-5}{{\em Eur. Phys. J.}
  {\bfseries C76} no.~3, (2016) },
\href{http://arxiv.org/abs/1506.03981}{{\ttfamily arXiv:1506.03981 [nucl-ex]}}.

\bibitem{Adamczyk:2012af}
{\bfseries STAR} Collaboration, L.~Adamczyk {\em et~al.}, ``{Measurements of
  $D^0$ and $D^{*+}$ Production in pp Collisions at $\sqrt s$ = 200 GeV}'',
  \href{http://dx.doi.org/10.1103/PhysRevD.86.072013}{{\em Phys. Rev.}
  {\bfseries D86} (2012) },
\href{http://arxiv.org/abs/1204.4244}{{\ttfamily arXiv:1204.4244 [nucl-ex]}}.

\bibitem{Acosta:2003ax}
{\bfseries CDF} Collaboration, D.~Acosta {\em et~al.}, ``{Measurement of prompt
  charm meson production cross sections in $p\bar{p}$ collisions at $\sqrt{s} =
  1.96$ TeV}'', \href{http://dx.doi.org/10.1103/PhysRevLett.91.241804}{{\em
  Phys. Rev. Lett.} {\bfseries 91} (2003) },
\href{http://arxiv.org/abs/hep-ex/0307080}{{\ttfamily arXiv:hep-ex/0307080
  [hep-ex]}}.

\bibitem{Aad:2015zix}
{\bfseries ATLAS} Collaboration, G.~Aad {\em et~al.}, ``{Measurement of
  $D^{*\pm}$, $D^\pm$ and $D_s^\pm$ meson production cross sections in $pp$
  collisions at $\sqrt{s}=7$ TeV with the ATLAS detector}'',
  \href{http://dx.doi.org/10.1016/j.nuclphysb.2016.04.032}{{\em Nucl. Phys.}
  {\bfseries B907} (2016) },
\href{http://arxiv.org/abs/1512.02913}{{\ttfamily arXiv:1512.02913 [hep-ex]}}.

\bibitem{ALICE:2011aa}
{\bfseries ALICE} Collaboration, B.~Abelev {\em et~al.}, ``{Measurement of
  charm production at central rapidity in proton-proton collisions at $\sqrt{s}
  = 7$ TeV}'', \href{http://dx.doi.org/10.1007/JHEP01(2012)128}{{\em JHEP}
  {\bfseries 01} (2012) },
\href{http://arxiv.org/abs/1111.1553}{{\ttfamily arXiv:1111.1553 [hep-ex]}}.

\bibitem{Abelev:2012vra}
{\bfseries ALICE} Collaboration, B.~Abelev {\em et~al.}, ``{Measurement of
  charm production at central rapidity in proton-proton collisions at
  $\sqrt{s}=2.76$ TeV}'', \href{http://dx.doi.org/10.1007/JHEP07(2012)191}{{\em
  JHEP} {\bfseries 07} (2012) },
\href{http://arxiv.org/abs/1205.4007}{{\ttfamily arXiv:1205.4007 [hep-ex]}}.

\bibitem{Abelev:2012xe}
{\bfseries ALICE} Collaboration, B.~Abelev {\em et~al.}, ``{Measurement of
  electrons from semileptonic heavy-flavour hadron decays in pp collisions at
  $\sqrt{s} = 7 TeV$}'',
  \href{http://dx.doi.org/10.1103/PhysRevD.86.112007}{{\em Phys. Rev.}
  {\bfseries D86} (2012) },
\href{http://arxiv.org/abs/1205.5423}{{\ttfamily arXiv:1205.5423 [hep-ex]}}.

\bibitem{Abelev:2012tca}
{\bfseries ALICE} Collaboration, B.~Abelev {\em et~al.}, ``{$D_{s}^+$ meson
  production at central rapidity in proton--proton collisions at $\sqrt{s}=7$
  TeV}'', \href{http://dx.doi.org/10.1016/j.physletb.2012.10.049}{{\em Phys.
  Lett.} {\bfseries B718} (2012) },
\href{http://arxiv.org/abs/1208.1948}{{\ttfamily arXiv:1208.1948 [hep-ex]}}.

\bibitem{Adam:2016ich}
{\bfseries ALICE} Collaboration, J.~Adam {\em et~al.}, ``{D-meson production in
  p-Pb collisions at $\sqrt{s_{\rm NN}}=5.02$ TeV and in pp collisions at
  $\sqrt{s}=7$ TeV}'', \href{http://dx.doi.org/10.1103/PhysRevC.94.054908}{{\em
  Phys. Rev.} {\bfseries C94} no.~5, (2016) },
\href{http://arxiv.org/abs/1605.07569}{{\ttfamily arXiv:1605.07569 [nucl-ex]}}.

\bibitem{Sirunyan:2017xss}
{\bfseries CMS} Collaboration, A.~M. Sirunyan {\em et~al.}, ``{Nuclear
  modification factor of D$^0$ mesons in PbPb collisions at
  $\sqrt{s_\mathrm{NN}} = 5.02$ TeV}'',
  \href{http://dx.doi.org/10.1016/j.physletb.2018.05.074}{{\em Phys. Lett.}
  {\bfseries B782} (2018) },
\href{http://arxiv.org/abs/1708.04962}{{\ttfamily arXiv:1708.04962 [nucl-ex]}}.

\bibitem{Aaij:2013mga}
{\bfseries LHCb} Collaboration, R.~Aaij {\em et~al.}, ``{Prompt charm
  production in pp collisions at $\sqrt s$ = 7 TeV}'',
  \href{http://dx.doi.org/10.1016/j.nuclphysb.2013.02.010}{{\em Nucl. Phys.}
  {\bfseries B871} (2013) },
\href{http://arxiv.org/abs/1302.2864}{{\ttfamily arXiv:1302.2864 [hep-ex]}}.

\bibitem{Aaij:2016jht}
{\bfseries LHCb} Collaboration, R.~Aaij {\em et~al.}, ``{Measurements of prompt
  charm production cross-sections in $pp$ collisions at $\sqrt{s}=5 $ TeV}'',
  \href{http://dx.doi.org/10.1007/JHEP06(2017)147}{{\em JHEP} {\bfseries 06}
  (2017) },
\href{http://arxiv.org/abs/1610.02230}{{\ttfamily arXiv:1610.02230 [hep-ex]}}.

\bibitem{Aaij:2015bpa}
{\bfseries LHCb} Collaboration, R.~Aaij {\em et~al.}, ``{Measurements of prompt
  charm production cross-sections in $pp$ collisions at $ \sqrt{s}=13 $ TeV}'',
  \href{http://dx.doi.org/10.1007/JHEP03(2016)159}{{\em JHEP} {\bfseries 03}
  (2016) }, \href{http://arxiv.org/abs/1510.01707}{{\ttfamily arXiv:1510.01707
  [hep-ex]}}.
[Erratum: {\it JHEP} {\textbf {09}} (2016) 013].

\bibitem{Cacciari:2015fta}
M.~Cacciari, M.~L. Mangano, and P.~Nason, ``{Gluon PDF constraints from the
  ratio of forward heavy-quark production at the LHC at $\sqrt{s}=7$ and 13
  TeV}'', \href{http://dx.doi.org/10.1140/epjc/s10052-015-3814-x}{{\em Eur.
  Phys. J.} {\bfseries C75} no.~12, (2015) },
\href{http://arxiv.org/abs/1507.06197}{{\ttfamily arXiv:1507.06197 [hep-ph]}}.

\bibitem{Prino:2016cni}
F.~Prino and R.~Rapp, ``{Open Heavy Flavor in QCD Matter and in Nuclear
  Collisions}'', \href{http://dx.doi.org/10.1088/0954-3899/43/9/093002}{{\em J.
  Phys.} {\bfseries G43} no.~9, (2016) },
\href{http://arxiv.org/abs/1603.00529}{{\ttfamily arXiv:1603.00529 [nucl-ex]}}.

\bibitem{Aarts:2016hap}
G.~Aarts {\em et~al.}, ``{Heavy-flavor production and medium properties in
  high-energy nuclear collisions - What next?}'',
  \href{http://dx.doi.org/10.1140/epja/i2017-12282-9}{{\em Eur. Phys. J.}
  {\bfseries A53} no.~5, (2017) },
\href{http://arxiv.org/abs/1612.08032}{{\ttfamily arXiv:1612.08032 [nucl-th]}}.

\bibitem{Acharya:2017jgo}
{\bfseries ALICE} Collaboration, S.~Acharya {\em et~al.}, ``{Measurement of
  D-meson production at mid-rapidity in pp collisions at ${\sqrt{s}=7}$ TeV}'',
  \href{http://dx.doi.org/10.1140/epjc/s10052-017-5090-4}{{\em Eur. Phys. J.}
  {\bfseries C77} no.~8, (2017) },
\href{http://arxiv.org/abs/1702.00766}{{\ttfamily arXiv:1702.00766 [hep-ex]}}.

\bibitem{Aamodt:2008zz}
{\bfseries ALICE} Collaboration, K.~Aamodt {\em et~al.}, ``{The ALICE
  experiment at the CERN LHC}'',
\href{http://dx.doi.org/10.1088/1748-0221/3/08/S08002}{{\em JINST} {\bfseries
  3} (2008) }.

\bibitem{Abelev:2014ffa}
{\bfseries ALICE} Collaboration, B.~B. Abelev {\em et~al.}, ``{Performance of
  the ALICE Experiment at the CERN LHC}'',
  \href{http://dx.doi.org/10.1142/S0217751X14300440}{{\em Int. J. Mod. Phys.}
  {\bfseries A29} (2014) },
\href{http://arxiv.org/abs/1402.4476}{{\ttfamily arXiv:1402.4476 [nucl-ex]}}.

\bibitem{Tanabashi:2018oca}
{\bfseries Particle Data Group} Collaboration, M.~Tanabashi {\em et~al.},
  ``{Review of Particle Physics}'',
\href{http://dx.doi.org/10.1103/PhysRevD.98.030001}{{\em Phys. Rev.} {\bfseries
  D98} no.~3, (2018) }.

\bibitem{Acharya:2018hre}
{\bfseries ALICE} Collaboration, S.~Acharya {\em et~al.}, ``{Measurement of
  D$^0$, D$^+$, D$^{*+}$ and D$^+_{\rm s}$ production in Pb-Pb collisions at
  $\mathbf{\sqrt{s_{\rm NN}}}= 5.02$ TeV}'', {\em Submitted to: JHEP} (2018) ,
\href{http://arxiv.org/abs/1804.09083}{{\ttfamily arXiv:1804.09083 [nucl-ex]}}.

\bibitem{ALICE-PUBLIC-2018-014}
{\bfseries ALICE Collaboration} Collaboration, ``{ALICE 2017 luminosity
  determination for pp collisions at $\sqrt{s}$ = 5 TeV}'',.
  \url{http://cds.cern.ch/record/2648933}.

\bibitem{Sjostrand:2006za}
T.~Sjostrand, S.~Mrenna, and P.~Z. Skands, ``{PYTHIA 6.4 Physics and Manual}'',
  \href{http://dx.doi.org/10.1088/1126-6708/2006/05/026}{{\em JHEP} {\bfseries
  05} (2006) },
\href{http://arxiv.org/abs/hep-ph/0603175}{{\ttfamily arXiv:hep-ph/0603175
  [hep-ph]}}.

\bibitem{Skands:2010ak}
P.~Z. Skands, ``{Tuning Monte Carlo Generators: The Perugia Tunes}'',
  \href{http://dx.doi.org/10.1103/PhysRevD.82.074018}{{\em Phys. Rev.}
  {\bfseries D82} (2010) },
\href{http://arxiv.org/abs/1005.3457}{{\ttfamily arXiv:1005.3457 [hep-ph]}}.

\bibitem{Brun:1994aa}
R.~Brun, F.~Carminati, and S.~Giani, ``{CERN Program Library Long Write-up,
  W5013 GEANT Detector Description and Simulation Tool}'', Tech. Rep.
  CERN-W-5013, 1994.

\bibitem{Cacciari:2001td}
M.~Cacciari, S.~Frixione, and P.~Nason, ``{The p(T) spectrum in heavy flavor
  photoproduction}'',
  \href{http://dx.doi.org/10.1088/1126-6708/2001/03/006}{{\em JHEP} {\bfseries
  03} (2001) },
\href{http://arxiv.org/abs/hep-ph/0102134}{{\ttfamily arXiv:hep-ph/0102134
  [hep-ph]}}.

\bibitem{Lange:2001uf}
D.~J. Lange, ``{The EvtGen particle decay simulation package}'',
\href{http://dx.doi.org/10.1016/S0168-9002(01)00089-4}{{\em Nucl. Instrum.
  Meth.} {\bfseries A462} (2001) }.

\bibitem{ALICE-PUBLIC-2017-005}
{\bfseries ALICE} Collaboration, ``{The ALICE definition of primary
  particles}'', Jun, 2017.
\newblock \url{https://cds.cern.ch/record/2270008}. ALICE-PUBLIC-2017-005.

\bibitem{Benzke:2017yjn}
M.~Benzke, M.~V. Garzelli, B.~Kniehl, G.~Kramer, S.~Moch, and G.~Sigl,
  ``{Prompt neutrinos from atmospheric charm in the general-mass
  variable-flavor-number scheme}'',
  \href{http://dx.doi.org/10.1007/JHEP12(2017)021}{{\em JHEP} {\bfseries 12}
  (2017) },
\href{http://arxiv.org/abs/1705.10386}{{\ttfamily arXiv:1705.10386 [hep-ph]}}.

\bibitem{Kramer:2017gct}
G.~Kramer and H.~Spiesberger, ``{Study of heavy meson production in p–Pb
  collisions at $\sqrt{s}$= 5.02 TeV in the general-mass
  variable-flavour-number scheme}'',
  \href{http://dx.doi.org/10.1016/j.nuclphysb.2017.10.016}{{\em Nucl. Phys.}
  {\bfseries B925} (2017) },
\href{http://arxiv.org/abs/1703.04754}{{\ttfamily arXiv:1703.04754 [hep-ph]}}.

\bibitem{Maciula:2018iuh}
R.~Maciula and A.~Szczurek, ``{Production of $\Lambda_c$ baryons at the LHC
  within the $k_T$-factorization approach and independent parton fragmentation
  picture}'', \href{http://dx.doi.org/10.1103/PhysRevD.98.014016}{{\em Phys.
  Rev.} {\bfseries D98} no.~1, (2018) },
\href{http://arxiv.org/abs/1803.05807}{{\ttfamily arXiv:1803.05807 [hep-ph]}}.

\bibitem{Pumplin:2002vw}
J.~Pumplin, D.~R. Stump, J.~Huston, H.~L. Lai, P.~M. Nadolsky, and W.~K. Tung,
  ``{New generation of parton distributions with uncertainties from global QCD
  analysis}'', \href{http://dx.doi.org/10.1088/1126-6708/2002/07/012}{{\em
  JHEP} {\bfseries 07} (2002) },
\href{http://arxiv.org/abs/hep-ph/0201195}{{\ttfamily arXiv:hep-ph/0201195
  [hep-ph]}}.

\bibitem{Dulat:2015mca}
S.~Dulat, T.-J. Hou, J.~Gao, M.~Guzzi, J.~Huston, P.~Nadolsky, J.~Pumplin,
  C.~Schmidt, D.~Stump, and C.~P. Yuan, ``{New parton distribution functions
  from a global analysis of quantum chromodynamics}'',
  \href{http://dx.doi.org/10.1103/PhysRevD.93.033006}{{\em Phys. Rev.}
  {\bfseries D93} no.~3, (2016) },
\href{http://arxiv.org/abs/1506.07443}{{\ttfamily arXiv:1506.07443 [hep-ph]}}.

\bibitem{Ball:2017nwa}
{\bfseries NNPDF} Collaboration, R.~D. Ball {\em et~al.}, ``{Parton
  distributions from high-precision collider data}'',
  \href{http://dx.doi.org/10.1140/epjc/s10052-017-5199-5}{{\em Eur. Phys. J.}
  {\bfseries C77} no.~10, (2017) },
\href{http://arxiv.org/abs/1706.00428}{{\ttfamily arXiv:1706.00428 [hep-ph]}}.

\bibitem{Harland-Lang:2014zoa}
L.~A. Harland-Lang, A.~D. Martin, P.~Motylinski, and R.~S. Thorne, ``{Parton
  distributions in the LHC era: MMHT 2014 PDFs}'',
  \href{http://dx.doi.org/10.1140/epjc/s10052-015-3397-6}{{\em Eur. Phys. J.}
  {\bfseries C75} no.~5, (2015) },
\href{http://arxiv.org/abs/1412.3989}{{\ttfamily arXiv:1412.3989 [hep-ph]}}.

\bibitem{Barate:1999bg}
{\bfseries ALEPH} Collaboration, R.~Barate {\em et~al.}, ``{Study of charm
  production in Z decays}'',
  \href{http://dx.doi.org/10.1007/s100520000421}{{\em Eur. Phys. J.} {\bfseries
  C16} (2000) },
\href{http://arxiv.org/abs/hep-ex/9909032}{{\ttfamily arXiv:hep-ex/9909032
  [hep-ex]}}.

\bibitem{Braaten:1994bz}
E.~Braaten, K.-M. Cheung, S.~Fleming, and T.~C. Yuan, ``{Perturbative QCD
  fragmentation functions as a model for heavy quark fragmentation}'',
  \href{http://dx.doi.org/10.1103/PhysRevD.51.4819}{{\em Phys. Rev.} {\bfseries
  D51} (1995) },
\href{http://arxiv.org/abs/hep-ph/9409316}{{\ttfamily arXiv:hep-ph/9409316
  [hep-ph]}}.

\bibitem{Cacciari:2003zu}
M.~Cacciari and P.~Nason, ``{Charm cross-sections for the Tevatron Run II}'',
  \href{http://dx.doi.org/10.1088/1126-6708/2003/09/006}{{\em JHEP} {\bfseries
  09} (2003) },
\href{http://arxiv.org/abs/hep-ph/0306212}{{\ttfamily arXiv:hep-ph/0306212
  [hep-ph]}}.

\bibitem{Gladilin:2014tba}
L.~Gladilin, ``{Fragmentation fractions of $c$ and $b$ quarks into charmed
  hadrons at LEP}'',
  \href{http://dx.doi.org/10.1140/epjc/s10052-014-3250-3}{{\em Eur. Phys. J.}
  {\bfseries C75} no.~1, (2015) },
\href{http://arxiv.org/abs/1404.3888}{{\ttfamily arXiv:1404.3888 [hep-ex]}}.

\bibitem{Acharya:2017kfy}
{\bfseries ALICE} Collaboration, S.~Acharya {\em et~al.}, ``{$\Lambda_{\rm
  c}^+$ production in pp collisions at $\sqrt{s} = 7$ TeV and in p-Pb
  collisions at $\sqrt{s_{\rm NN}} = 5.02$ TeV}'',
  \href{http://dx.doi.org/10.1007/JHEP04(2018)108}{{\em JHEP} {\bfseries 04}
  (2018) },
\href{http://arxiv.org/abs/1712.09581}{{\ttfamily arXiv:1712.09581 [nucl-ex]}}.

\end{thebibliography}\endgroup

\newpage
\appendix

\section{The ALICE Collaboration}
\label{app:collab} 

\begingroup
\small
\begin{flushleft}
S.~Acharya\Irefn{org140}\And 
D.~Adamov\'{a}\Irefn{org93}\And 
S.P.~Adhya\Irefn{org140}\And 
A.~Adler\Irefn{org74}\And 
J.~Adolfsson\Irefn{org80}\And 
M.M.~Aggarwal\Irefn{org98}\And 
G.~Aglieri Rinella\Irefn{org34}\And 
M.~Agnello\Irefn{org31}\And 
Z.~Ahammed\Irefn{org140}\And 
S.~Ahmad\Irefn{org17}
% ...
\And
\renewcommand\labelenumi{\textsuperscript{\theenumi}~}

\section*{Affiliation notes}
\renewcommand\theenumi{\roman{enumi}}
\begin{Authlist}
\item \Adef{org*}Deceased
\item \Adef{orgI}Dipartimento DET del Politecnico di Torino, Turin, Italy
\item \Adef{orgII}M.V. Lomonosov Moscow State University, D.V. Skobeltsyn Institute of Nuclear, Physics, Moscow, Russia
\item \Adef{orgIII}Department of Applied Physics, Aligarh Muslim University, Aligarh, India
\item \Adef{orgIV}Institute of Theoretical Physics, University of Wroclaw, Poland
\end{Authlist}

\section*{Collaboration Institutes}
\renewcommand\theenumi{\arabic{enumi}~}
\begin{Authlist}
\item \Idef{org1}A.I. Alikhanyan National Science Laboratory (Yerevan Physics Institute) Foundation, Yerevan, Armenia
\item \Idef{org2}Bogolyubov Institute for Theoretical Physics, National Academy of Sciences of Ukraine, Kiev, Ukraine
\item \Idef{org3}Bose Institute, Department of Physics  and Centre for Astroparticle Physics and Space Science (CAPSS), Kolkata, India
\item \Idef{org4}Budker Institute for Nuclear Physics, Novosibirsk, Russia
\item \Idef{org5}California Polytechnic State University, San Luis Obispo, California, United States
\item \Idef{org6}Central China Normal University, Wuhan, China
\item \Idef{org7}Centre de Calcul de l'IN2P3, Villeurbanne, Lyon, France
\item \Idef{org8}Centro de Aplicaciones Tecnol\'{o}gicas y Desarrollo Nuclear (CEADEN), Havana, Cuba
\item \Idef{org9}Centro de Investigaci\'{o}n y de Estudios Avanzados (CINVESTAV), Mexico City and M\'{e}rida, Mexico
\item \Idef{org10}Centro Fermi - Museo Storico della Fisica e Centro Studi e Ricerche ``Enrico Fermi', Rome, Italy
\item \Idef{org11}Chicago State University, Chicago, Illinois, United States
\item \Idef{org12}China Institute of Atomic Energy, Beijing, China
\item \Idef{org13}Chonbuk National University, Jeonju, Republic of Korea
\item \Idef{org14}Comenius University Bratislava, Faculty of Mathematics, Physics and Informatics, Bratislava, Slovakia
\item \Idef{org15}COMSATS University Islamabad, Islamabad, Pakistan
\item \Idef{org16}Creighton University, Omaha, Nebraska, United States
\item \Idef{org17}Department of Physics, Aligarh Muslim University, Aligarh, India
\item \Idef{org18}Department of Physics, Pusan National University, Pusan, Republic of Korea
\item \Idef{org19}Department of Physics, Sejong University, Seoul, Republic of Korea
\item \Idef{org20}Department of Physics, University of California, Berkeley, California, United States
\item \Idef{org21}Department of Physics, University of Oslo, Oslo, Norway
\item \Idef{org22}Department of Physics and Technology, University of Bergen, Bergen, Norway
\item \Idef{org23}Dipartimento di Fisica dell'Universit\`{a} 'La Sapienza' and Sezione INFN, Rome, Italy
\item \Idef{org24}Dipartimento di Fisica dell'Universit\`{a} and Sezione INFN, Cagliari, Italy
\item \Idef{org25}Dipartimento di Fisica dell'Universit\`{a} and Sezione INFN, Trieste, Italy
\item \Idef{org26}Dipartimento di Fisica dell'Universit\`{a} and Sezione INFN, Turin, Italy
\item \Idef{org27}Dipartimento di Fisica e Astronomia dell'Universit\`{a} and Sezione INFN, Bologna, Italy
\item \Idef{org28}Dipartimento di Fisica e Astronomia dell'Universit\`{a} and Sezione INFN, Catania, Italy
\item \Idef{org29}Dipartimento di Fisica e Astronomia dell'Universit\`{a} and Sezione INFN, Padova, Italy
\item \Idef{org30}Dipartimento di Fisica `E.R.~Caianiello' dell'Universit\`{a} and Gruppo Collegato INFN, Salerno, Italy
\item \Idef{org31}Dipartimento DISAT del Politecnico and Sezione INFN, Turin, Italy
\item \Idef{org32}Dipartimento di Scienze e Innovazione Tecnologica dell'Universit\`{a} del Piemonte Orientale and INFN Sezione di Torino, Alessandria, Italy
\item \Idef{org33}Dipartimento Interateneo di Fisica `M.~Merlin' and Sezione INFN, Bari, Italy
\item \Idef{org34}European Organization for Nuclear Research (CERN), Geneva, Switzerland
\item \Idef{org35}Faculty of Electrical Engineering, Mechanical Engineering and Naval Architecture, University of Split, Split, Croatia
\item \Idef{org36}Faculty of Engineering and Science, Western Norway University of Applied Sciences, Bergen, Norway
\item \Idef{org37}Faculty of Nuclear Sciences and Physical Engineering, Czech Technical University in Prague, Prague, Czech Republic
\item \Idef{org38}Faculty of Science, P.J.~\v{S}af\'{a}rik University, Ko\v{s}ice, Slovakia
\item \Idef{org39}Frankfurt Institute for Advanced Studies, Johann Wolfgang Goethe-Universit\"{a}t Frankfurt, Frankfurt, Germany
\item \Idef{org40}Gangneung-Wonju National University, Gangneung, Republic of Korea
\item \Idef{org41}Gauhati University, Department of Physics, Guwahati, India
\item \Idef{org42}Helmholtz-Institut f\"{u}r Strahlen- und Kernphysik, Rheinische Friedrich-Wilhelms-Universit\"{a}t Bonn, Bonn, Germany
\item \Idef{org43}Helsinki Institute of Physics (HIP), Helsinki, Finland
\item \Idef{org44}High Energy Physics Group,  Universidad Aut\'{o}noma de Puebla, Puebla, Mexico
\item \Idef{org45}Hiroshima University, Hiroshima, Japan
\item \Idef{org46}Hochschule Worms, Zentrum  f\"{u}r Technologietransfer und Telekommunikation (ZTT), Worms, Germany
\item \Idef{org47}Horia Hulubei National Institute of Physics and Nuclear Engineering, Bucharest, Romania
\item \Idef{org48}Indian Institute of Technology Bombay (IIT), Mumbai, India
\item \Idef{org49}Indian Institute of Technology Indore, Indore, India
\item \Idef{org50}Indonesian Institute of Sciences, Jakarta, Indonesia
\item \Idef{org51}INFN, Laboratori Nazionali di Frascati, Frascati, Italy
\item \Idef{org52}INFN, Sezione di Bari, Bari, Italy
\item \Idef{org53}INFN, Sezione di Bologna, Bologna, Italy
\item \Idef{org54}INFN, Sezione di Cagliari, Cagliari, Italy
\item \Idef{org55}INFN, Sezione di Catania, Catania, Italy
\item \Idef{org56}INFN, Sezione di Padova, Padova, Italy
\item \Idef{org57}INFN, Sezione di Roma, Rome, Italy
\item \Idef{org58}INFN, Sezione di Torino, Turin, Italy
\item \Idef{org59}INFN, Sezione di Trieste, Trieste, Italy
\item \Idef{org60}Inha University, Incheon, Republic of Korea
\item \Idef{org61}Institut de Physique Nucl\'{e}aire d'Orsay (IPNO), Institut National de Physique Nucl\'{e}aire et de Physique des Particules (IN2P3/CNRS), Universit\'{e} de Paris-Sud, Universit\'{e} Paris-Saclay, Orsay, France
\item \Idef{org62}Institute for Nuclear Research, Academy of Sciences, Moscow, Russia
\item \Idef{org63}Institute for Subatomic Physics, Utrecht University/Nikhef, Utrecht, Netherlands
\item \Idef{org64}Institute for Theoretical and Experimental Physics, Moscow, Russia
\item \Idef{org65}Institute of Experimental Physics, Slovak Academy of Sciences, Ko\v{s}ice, Slovakia
\item \Idef{org66}Institute of Physics, Homi Bhabha National Institute, Bhubaneswar, India
\item \Idef{org67}Institute of Physics of the Czech Academy of Sciences, Prague, Czech Republic
\item \Idef{org68}Institute of Space Science (ISS), Bucharest, Romania
\item \Idef{org69}Institut f\"{u}r Kernphysik, Johann Wolfgang Goethe-Universit\"{a}t Frankfurt, Frankfurt, Germany
\item \Idef{org70}Instituto de Ciencias Nucleares, Universidad Nacional Aut\'{o}noma de M\'{e}xico, Mexico City, Mexico
\item \Idef{org71}Instituto de F\'{i}sica, Universidade Federal do Rio Grande do Sul (UFRGS), Porto Alegre, Brazil
\item \Idef{org72}Instituto de F\'{\i}sica, Universidad Nacional Aut\'{o}noma de M\'{e}xico, Mexico City, Mexico
\item \Idef{org73}iThemba LABS, National Research Foundation, Somerset West, South Africa
\item \Idef{org74}Johann-Wolfgang-Goethe Universit\"{a}t Frankfurt Institut f\"{u}r Informatik, Fachbereich Informatik und Mathematik, Frankfurt, Germany
\item \Idef{org75}Joint Institute for Nuclear Research (JINR), Dubna, Russia
\item \Idef{org76}Korea Institute of Science and Technology Information, Daejeon, Republic of Korea
\item \Idef{org77}KTO Karatay University, Konya, Turkey
\item \Idef{org78}Laboratoire de Physique Subatomique et de Cosmologie, Universit\'{e} Grenoble-Alpes, CNRS-IN2P3, Grenoble, France
\item \Idef{org79}Lawrence Berkeley National Laboratory, Berkeley, California, United States
\item \Idef{org80}Lund University Department of Physics, Division of Particle Physics, Lund, Sweden
\item \Idef{org81}Nagasaki Institute of Applied Science, Nagasaki, Japan
\item \Idef{org82}Nara Women{'}s University (NWU), Nara, Japan
\item \Idef{org83}National and Kapodistrian University of Athens, School of Science, Department of Physics , Athens, Greece
\item \Idef{org84}National Centre for Nuclear Research, Warsaw, Poland
\item \Idef{org85}National Institute of Science Education and Research, Homi Bhabha National Institute, Jatni, India
\item \Idef{org86}National Nuclear Research Center, Baku, Azerbaijan
\item \Idef{org87}National Research Centre Kurchatov Institute, Moscow, Russia
\item \Idef{org88}Niels Bohr Institute, University of Copenhagen, Copenhagen, Denmark
\item \Idef{org89}Nikhef, National institute for subatomic physics, Amsterdam, Netherlands
\item \Idef{org90}NRC Kurchatov Institute IHEP, Protvino, Russia
\item \Idef{org91}NRNU Moscow Engineering Physics Institute, Moscow, Russia
\item \Idef{org92}Nuclear Physics Group, STFC Daresbury Laboratory, Daresbury, United Kingdom
\item \Idef{org93}Nuclear Physics Institute of the Czech Academy of Sciences, \v{R}e\v{z} u Prahy, Czech Republic
\item \Idef{org94}Oak Ridge National Laboratory, Oak Ridge, Tennessee, United States
\item \Idef{org95}Ohio State University, Columbus, Ohio, United States
\item \Idef{org96}Petersburg Nuclear Physics Institute, Gatchina, Russia
\item \Idef{org97}Physics department, Faculty of science, University of Zagreb, Zagreb, Croatia
\item \Idef{org98}Physics Department, Panjab University, Chandigarh, India
\item \Idef{org99}Physics Department, University of Jammu, Jammu, India
\item \Idef{org100}Physics Department, University of Rajasthan, Jaipur, India
\item \Idef{org101}Physikalisches Institut, Eberhard-Karls-Universit\"{a}t T\"{u}bingen, T\"{u}bingen, Germany
\item \Idef{org102}Physikalisches Institut, Ruprecht-Karls-Universit\"{a}t Heidelberg, Heidelberg, Germany
\item \Idef{org103}Physik Department, Technische Universit\"{a}t M\"{u}nchen, Munich, Germany
\item \Idef{org104}Research Division and ExtreMe Matter Institute EMMI, GSI Helmholtzzentrum f\"ur Schwerionenforschung GmbH, Darmstadt, Germany
\item \Idef{org105}Rudjer Bo\v{s}kovi\'{c} Institute, Zagreb, Croatia
\item \Idef{org106}Russian Federal Nuclear Center (VNIIEF), Sarov, Russia
\item \Idef{org107}Saha Institute of Nuclear Physics, Homi Bhabha National Institute, Kolkata, India
\item \Idef{org108}School of Physics and Astronomy, University of Birmingham, Birmingham, United Kingdom
\item \Idef{org109}Secci\'{o}n F\'{\i}sica, Departamento de Ciencias, Pontificia Universidad Cat\'{o}lica del Per\'{u}, Lima, Peru
\item \Idef{org110}Shanghai Institute of Applied Physics, Shanghai, China
\item \Idef{org111}St. Petersburg State University, St. Petersburg, Russia
\item \Idef{org112}Stefan Meyer Institut f\"{u}r Subatomare Physik (SMI), Vienna, Austria
\item \Idef{org113}SUBATECH, IMT Atlantique, Universit\'{e} de Nantes, CNRS-IN2P3, Nantes, France
\item \Idef{org114}Suranaree University of Technology, Nakhon Ratchasima, Thailand
\item \Idef{org115}Technical University of Ko\v{s}ice, Ko\v{s}ice, Slovakia
\item \Idef{org116}Technische Universit\"{a}t M\"{u}nchen, Excellence Cluster 'Universe', Munich, Germany
\item \Idef{org117}The Henryk Niewodniczanski Institute of Nuclear Physics, Polish Academy of Sciences, Cracow, Poland
\item \Idef{org118}The University of Texas at Austin, Austin, Texas, United States
\item \Idef{org119}Universidad Aut\'{o}noma de Sinaloa, Culiac\'{a}n, Mexico
\item \Idef{org120}Universidade de S\~{a}o Paulo (USP), S\~{a}o Paulo, Brazil
\item \Idef{org121}Universidade Estadual de Campinas (UNICAMP), Campinas, Brazil
\item \Idef{org122}Universidade Federal do ABC, Santo Andre, Brazil
\item \Idef{org123}University College of Southeast Norway, Tonsberg, Norway
\item \Idef{org124}University of Cape Town, Cape Town, South Africa
\item \Idef{org125}University of Houston, Houston, Texas, United States
\item \Idef{org126}University of Jyv\"{a}skyl\"{a}, Jyv\"{a}skyl\"{a}, Finland
\item \Idef{org127}University of Liverpool, Liverpool, United Kingdom
\item \Idef{org128}University of Science and Techonology of China, Hefei, China
\item \Idef{org129}University of Tennessee, Knoxville, Tennessee, United States
\item \Idef{org130}University of the Witwatersrand, Johannesburg, South Africa
\item \Idef{org131}University of Tokyo, Tokyo, Japan
\item \Idef{org132}University of Tsukuba, Tsukuba, Japan
\item \Idef{org133}Universit\'{e} Clermont Auvergne, CNRS/IN2P3, LPC, Clermont-Ferrand, France
\item \Idef{org134}Universit\'{e} de Lyon, Universit\'{e} Lyon 1, CNRS/IN2P3, IPN-Lyon, Villeurbanne, Lyon, France
\item \Idef{org135}Universit\'{e} de Strasbourg, CNRS, IPHC UMR 7178, F-67000 Strasbourg, France, Strasbourg, France
\item \Idef{org136} Universit\'{e} Paris-Saclay Centre d¿\'Etudes de Saclay (CEA), IRFU, Department de Physique Nucl\'{e}aire (DPhN), Saclay, France
\item \Idef{org137}Universit\`{a} degli Studi di Foggia, Foggia, Italy
\item \Idef{org138}Universit\`{a} degli Studi di Pavia, Pavia, Italy
\item \Idef{org139}Universit\`{a} di Brescia, Brescia, Italy
\item \Idef{org140}Variable Energy Cyclotron Centre, Homi Bhabha National Institute, Kolkata, India
\item \Idef{org141}Warsaw University of Technology, Warsaw, Poland
\item \Idef{org142}Wayne State University, Detroit, Michigan, United States
\item \Idef{org143}Westf\"{a}lische Wilhelms-Universit\"{a}t M\"{u}nster, Institut f\"{u}r Kernphysik, M\"{u}nster, Germany
\item \Idef{org144}Wigner Research Centre for Physics, Hungarian Academy of Sciences, Budapest, Hungary
\item \Idef{org145}Yale University, New Haven, Connecticut, United States
\item \Idef{org146}Yonsei University, Seoul, Republic of Korea
\end{Authlist}
\endgroup

\end{document}